\documentclass[twoside,11pt]{article}

\usepackage{jmlr2e}

\usepackage{times,url,mathrsfs}
\usepackage{amsmath,wrapfig,color}
\usepackage{amsfonts}
\usepackage{graphicx}
\usepackage{subfigure}

\jmlrheading{}{}{}{}{}{}{}

\ShortHeadings{Sign-Full Random Projections}{Ping Li}
\firstpageno{1}

\begin{document}

\title{Sign-Full Random Projections}

\author{\name Ping Li, \hspace{0.2in} Baidu Research USA, \email pingli98@gmail.com}

\editor{}

\maketitle

\begin{abstract}
\noindent The\footnote{The work was first presented in Stanford Statistics Department.} method of 1-bit (``sign-sign'') random projections  has been a popular tool for efficient search and machine learning on large datasets. Given two $D$-dim data vectors $u$, $v\in\mathbb{R}^D$, one can generate $x = \sum_{i=1}^D u_i r_i$, and $y = \sum_{i=1}^D v_i r_i$, where $r_i\sim N(0,1)$ iid. The ``collision probability'' is ${Pr}\left(sgn(x)=sgn(y)\right) = 1-\frac{\cos^{-1}\rho}{\pi}$, where $\rho = \rho(u,v)$ is the cosine similarity.\\

\noindent We develop ``sign-full'' random projections by estimating $\rho$ from (e.g.,) the expectation $E(sgn(x)y)=\sqrt{\frac{2}{\pi}} \rho$, which can be further substantially improved by  normalizing  $y$. For nonnegative data, we  recommend  an interesting estimator based on $E\left(y_- 1_{x\geq 0} + y_+ 1_{x<0}\right)$ and its normalized version.  The recommended estimator almost matches the accuracy of the (computationally  expensive) maximum likelihood estimator. At high similarity ($\rho\rightarrow1$), the asymptotic variance of recommended estimator is only $\frac{4}{3\pi} \approx 0.4$ of the estimator for sign-sign projections. At small $k$ and high similarity, the improvement would be even much more substantial.\\

\noindent In applications such as near neighbor search, duplicate detection, knn-classification, etc, the training data  are first transformed via random projections and then only the signs of the projected data points are stored (i.e., the $sgn(x)$). The original training data  are discarded. When a new data point arrives, we apply random projections but we do not necessarily need to quantize the projected data (i.e., the $y$) to 1-bit. Therefore, sign-full random projections can be practically  useful. Roughly speaking, compared to the classical sign-sign random projections, sign-full random projections  can reduce  storage (and number of projections) by a factor of 2. This gain essentially comes at no additional cost. {\bf(All technical proofs are in the Appendix.)}

\end{abstract}

\section{Introduction}

Consider two high-dimensional data vectors, $u, v\in\mathbb{R}^D$. Suppose we generate a $D$-dim random vector whose entries are iid standard normal, i.e., $r_i \sim N(0,1)$, and compute
\begin{align}\notag
x = \sum_{i=1}^D u_i r_i,\hspace{0.3in} y = \sum_{i=1}^D v_i r_i
\end{align}
We have in  expectation $E(xy) = \left<u,v\right>=\sum_{i=1}^D u_i v_i$.  If we generate  $x$ and $y$ independently for $k$ times, then  $\frac{1}{k}\sum_{j=1}^k x_j y_j\approx E(xy) = \left<u,v\right>$, and the quality of approximation improves as $k$ increases. This  idea of random projections  has been widely used for large-scale search and machine learning~\citep{Article:JL84,Book:Vempala,Proc:Papadimitriou_PODS98,Proc:Dasgupta_FOCS99,Proc:Datar_SCG04,Proc:Li_Hastie_Church_COLT06}.

\vspace{-0.05in}
\subsection{Sign-Sign (1-Bit) Random Projections}

A  popular variant is the  ``1-bit'' random projections, which we refer to as ``sign-sign'' random projections, based on the following result of ``collision probability''
\begin{align}\label{eqn_1bitProb}
{Pr}\left(sgn(x) = sgn(y)\right) = 1-\frac{\cos^{-1}\rho}{\pi}
\end{align}
where $\rho = \rho(u,v) =  \frac{\sum_{i=1}^D u_i v_i}{\sqrt{\sum_{i=1}^D u_i^2}\sqrt{\sum_{i=1}^D v_i^2}}$
 is the ``cosine'' similarity between the two original data vectors $u$ and $v$. Note that by using only the signs of the projected data, we will lose the information about the norms of the original vectors. Thus, in this context, with no loss of generality, we will assume that the original data vectors are normalized, i.e., $\sum_{i=1}^D u_i^2 = \sum_{i=1}^D v_i^2=1$, just for notational convenience. In other words, without loss of generality, we can assume that $x \sim N(0,1)$ and $y\sim N(0,1)$.

The result~(\ref{eqn_1bitProb}) was seen in~\citep{Article:Goemans_JACM95} and~\citep{Proc:Charikar}. The method of sign-sign random projections has become popular, for example in web search~\citep{Proc:Henzinger_SIGIR06,Proc:Manku_WWW07,Proc:Grimes_WWW08}. It is known that the method is  effective in the high-similarity region ($\rho\rightarrow 1$).

In this paper, we  take advantage of $E(sgn(x)y)$ and  several variants to considerably improve 1-bit random projections. This gain  essentially comes at no additional cost. Basically, the training data  after  projections are stored using  signs (e.g., $sgn(x)$). When a new data vector arrives, however, we need to generate its random projections ($y$) but  do not necessarily have to quantize them. This is the motivation.

\vspace{-0.05in}
\subsection{Estimators Based on Full Information}

In this context, since we are only concerned with estimating the cosine $\rho$, we can without loss of generality assume that the original data are normalized, i.e., $\|u\|=\|v\|=1$.  The  projected data  thus follow a bi-variant normal distribution:\footnote{Even if the data are not normalized, the results presented in this paper remain essentially the same. For un-normalized estimators there will be a scaling factor. For example, $E(sgn(x)y)=\sqrt{\frac{2}{\pi}}\rho\|v\|$ instead of $E(sgn(x)y)=\sqrt{\frac{2}{\pi}}\rho$.}
\begin{align}\notag
\left[\begin{array}{c}x_j\\ y_j\end{array} \right] \sim N
\left(
\left[\begin{array}{c}0\\ 0\end{array} \right],\
\left[\begin{array}{cc}1 &\rho\\ \rho &1
\end{array} \right]
\right), \text{ iid}\hspace{0.25in} j = 1, 2, ..., k.
\end{align}
where $\rho = \sum_{i=1}^D u_iv_i$.  The obvious estimator for $\rho$ is based on the inner product of random projections:
\begin{align}\notag
&\hat{\rho}_f = \frac{1}{k}\sum_{j=1}^k x_j y_j,\hspace{0.3in}
E\left(\hat{\rho}_f\right) = \rho\\\notag
&Var\left(\hat{\rho}_f\right) =\frac{V_f}{k},\hspace{0.2in} V_f= 1+\rho^2
\end{align}
See the derivation of variance ($V_f$) in~\citep{Proc:Li_Hastie_Church_COLT06}. Note that  $Var(\hat{\rho}_f)$ is the largest when $|\rho|=1$. This is disappointing, because when two  data vectors are  identical, we ought to be able to estimate their similarity with no error.

One can  improve the estimator by simply normalizing the projected data. See~\citep{Book:Anderson03} for the derivation.
\begin{align}\notag
&\hat{\rho}_{f,n} = \frac{\sum_{j=1}^k x_j y_j}{\sqrt{\sum_{j=1}^k x_j^2}\sqrt{\sum_{j=1}^k y_j^2}},\hspace{0.2in} E\left(\hat{\rho}_{f,n}\right) = \rho + O\left(\frac{1}{k}\right)\\\notag
&Var\left(\hat{\rho}_{f,n}\right) =\frac{V_{f,n}}{k} + O\left(\frac{1}{k^2}\right),\hspace{0.2in} V_{f,n}= \left(1-\rho^2\right)^2
\end{align}
In particular, $V_{f,n} = 0$ when $|\rho|=1$, as desired.

One can further improve $\hat{\rho}_{f,n}$ but not too much. The theoretical limit (i.e., the Cram\'er-Rao bound) of the asymptotic variance~\citep{Book:Lehmann_Casella} can be obtained by the maximum likelihood estimator (MLE), which is the solution of the following cubic equation:
\begin{align}\notag
&\rho^3-\rho^2\sum_{j=1}^k x_jy_j
+\rho\left(-1+\sum_{i=1}^k x_j^2 + \sum_{j=1}^k y_j^2\right) -\sum_{j=1}^k x_jy_j=0
\end{align}
This cubic equation can have multiple real roots with a small probability~\citep{Proc:Li_Hastie_Church_COLT06}, which decreases exponentially fast with increasing  $k$. The MLE is asymptotically unbiased and its asymptotic variance becomes:
\begin{align}\notag
&E\left(\hat{\rho}_{f,m}\right) = \rho + O\left(\frac{1}{k}\right)\\\notag
&Var\left(\hat{\rho}_{f,m}\right) =\frac{V_{f,m}}{k} + O\left(\frac{1}{k^2}\right),\hspace{0.2in} V_{f,m}= \frac{\left(1-\rho^2\right)^2}{1+\rho^2}
\end{align}

\subsection{Estimator Based on Sign-Sign Random Projections}

\begin{figure}[h!]
\begin{center}
\includegraphics[width=3in]{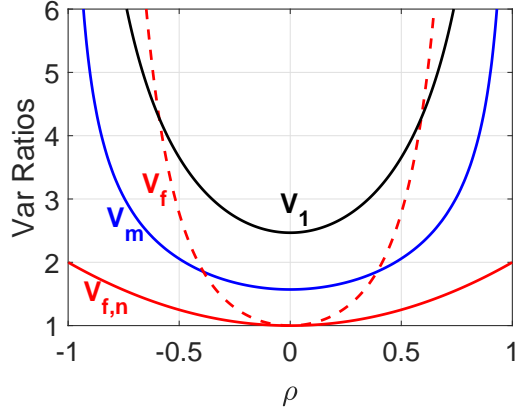}
\end{center}
\vspace{-0.2in}
\caption{Ratios of  variance factors: $\frac{V_1}{V_{f,m}}$, $\frac{V_f}{V_{f,m}}$, $\frac{V_m}{V_{f,m}}$, $\frac{V_{f,n}}{V_{f,m}}$. Because $V_{f,m}$ is the theoretically smallest variance factor, the ratios are always larger than 1, and we can use them to compare estimators (lower the better). Note that $V_m$ is the variance factor for the MLE of sign-full random projections (see Section~\ref{sec_estimators}).}\label{fig_Vf}\vspace{-0.1in}
\end{figure}

From  ${Pr}\left({sgn}(x_j) = {sgn}(y_j)\right)=1 - \frac{1}{\pi}\cos^{-1}\rho$, we have an asymptotically unbiased estimator and its variance:
\begin{align}\notag
&\hat{\rho}_1 = \cos\pi\left(1-\frac{1}{k}\sum_{j=1}^k1_{sgn(x_j)=sgn(y_j)}\right),\\\notag
&E\left(\hat{\rho}_1\right) = \rho + O\left(\frac{1}{k}\right),\hspace{0.2in}
Var\left(\hat{\rho}_1\right) = \frac{V_1}{k} + O\left(\frac{1}{k^2}\right),\\\notag
&V_1 = \cos^{-1}\rho\left(\pi-\cos^{-1}\rho\right)(1-\rho^2)
\end{align}
As later will be shown in Lemma~\ref{lem_V}, we have when $|\rho|\rightarrow1$,
\begin{align}\notag
&V_1 = 2\sqrt{2}\pi\left(1-|\rho|\right)^{3/2} + o\left(\left(1-|\rho|\right)^{3/2}\right),
\end{align}
This rate is slower than $O\left((1-|\rho|)^2\right)$, which is the rate at which $V_{f,n}$ and $V_{f,m}$ approach 0. Figure~\ref{fig_Vf} compares the estimators in terms of $\frac{V_1}{V_{f,m}}$, $\frac{V_f}{V_{f,m}}$, and $\frac{V_{f,n}}{V_{f,m}}$.  Basically, $V_{f,n}<V_{f}$ always which means we should always use the normalized estimator. Note that $V_1<V_{f}$ if $|\rho|>0.5902$.

\section{Estimators {for} Sign-Full Random Projections}\label{sec_estimators}

In many practical scenarios such as near-neighbor search and near-neighbor classification, we can store signs of the projected data (i.e., $sgn(x_j)$) and discard the original high-dimensional data. When a new data vector arrives, we generate its projected vector (i.e., $y$). At this point we actually have the option to choose whether we would like to use the full information or just the signs (i.e., $sgn(y_j)$) to estimate the similarity  $\rho$. If we are able to find a better (more accurate) estimator by using the full information of $y_j$, there is no reason why we have to only use the sign of $y_j$.

We first examine the maximum likelihood estimator (MLE), to understand the theoretical limit of sign-full projections.
\begin{theorem}\label{thm_mle}
Given $k$ iid samples $(sign(x_j),\ y_j)$, $j=1, 2, ..., k$, with $x_j\sim N(0,1)$, $y_j\sim N(0,1)$,  $E(x_jy_j)=\rho$, the maximum likelihood estimator (MLE, denoted by $\hat{\rho}_m$) is the solution to the following equation:
\begin{align}\label{eqn_mle}
\sum_{j=1}^k\frac{\phi\left(\frac{\rho}{\sqrt{1-\rho^2}}{sgn}(x_j) y_j\right)}{\Phi\left(\frac{\rho}{\sqrt{1-\rho^2} }{sgn}(x_j)y_j\right)}{sgn}(x_j)y_j = 0
\end{align}
where $\phi(t)= \frac{1}{\sqrt{2\pi}}e^{-t^2/2}$ and $\Phi(t) = \int_{-\infty}^t \phi(t)dt$ are respectively the pdf and cdf of the standard normal.
\begin{align}\notag
&E\left(\hat{\rho}_m\right) = \rho+O\left(\frac{1}{k}\right)\\\notag
&Var\left(\hat{\rho}_m\right) = \frac{V_m}{k}+O\left(\frac{1}{k^2}\right)
\end{align}
where
\begin{align}\notag
\frac{1}{V_m}=&E\left\{ \frac{\rho}{(1-\rho^2)^{7/2}}\frac{\phi\left(\frac{\rho}{\sqrt{1-\rho^2}}{sgn}(x_j) y_j\right)
}{\Phi\left(\frac{\rho}{\sqrt{1-\rho^2}}{sgn}(x_j) y_j\right)} {sgn}(x_j)y_j^3\right\}\\\label{eqn_Vm}
 +&E\left\{
\frac{1}{(1-\rho^2)^{3}}\frac{\phi^2\left(\frac{\rho}{\sqrt{1-\rho^2}}{sgn}(x_j) y_j\right)
}{\Phi^2\left(\frac{\rho}{\sqrt{1-\rho^2}}{sgn}(x_j) y_j\right)} y_j^2\right\}\\\notag
-&E\left\{
\frac{3\rho}{(1-\rho^2)^{5/2}}\frac{\phi\left(\frac{\rho}{\sqrt{1-\rho^2}}{sgn}(x_j) y_j\right)}{\Phi\left(\frac{\rho}{\sqrt{1-\rho^2}}{sgn}(x_j) y_j\right)}
{sgn}(x_j)y_j\right\}
\end{align}

\noindent\textbf{Proof:}\hspace{0.2in} See Appendix A.$\hfill\Box$.
\end{theorem}

As the MLE equation (\ref{eqn_mle}) is quite sophisticated, we study this estimator  mainly for theoretical interest, for example, for evaluating other estimators. We can evaluate the expectations in (\ref{eqn_Vm}) by simulations. Figure~\ref{fig_Vf} already plots $\frac{V_m}{V_{f,m}}$, to compare $\hat{\rho}_m$ with   three estimators: $\hat{\rho}_1$, $\hat{\rho}_{f}$,  $\hat{\rho}_{f,n}$. The figure shows that $\hat{\rho}_m$ indeed substantially improves $\hat{\rho}_1$.

Next, we seek estimators which are much simpler than $\hat{\rho}_m$. Ideally, we look for estimators which can be written as ``inner products''. In this paper, we propose \textbf{four} such estimators. We first present a Lemma which will be needed for deriving these estimators and proving their properties.
\begin{lemma}\label{lem_int}
\begin{align}
&\int_0^\infty t e^{-t^2/2}\Phi\left(\frac{\rho t}{\sqrt{1-\rho^2}}\right) dt  = \frac{1+\rho}{2}\\
&\int_0^\infty t^3 e^{-t^2/2}\Phi\left(\frac{\rho t}{\sqrt{1-\rho^2}}\right) dt  = \frac{1}{2}\left(2+3\rho-\rho^3\right)\\
&\int_0^\infty t^2 e^{-t^2/2}\Phi\left(\frac{\rho t}{\sqrt{1-\rho^2}}\right) dt
= 1_{\rho\geq0}\sqrt{\frac{\pi}{2}}-\sqrt{\frac{1}{2\pi}}\left(\tan^{-1}\frac{\sqrt{1-\rho^2}}{\rho}-\rho\sqrt{1-\rho^2}\right)
\end{align}
where we denote that $\tan^{-1}\left(\frac{1}{0}\right)  = \tan^{-1}\left(\frac{1}{0+}\right) =\frac{\pi}{2}$.

\noindent\textbf{Proof:}\hspace{0.2in} See Appendix B.$\hfill\Box$.
\end{lemma}

\newpage

The first estimator we present is based on the theoretical moments of $(sgn(x_j)y_j)$ as shown in Theorem~\ref{thm_g}.
\begin{theorem}\label{thm_g}
\begin{align}
&E(sgn(x_j)y_j) = \sqrt{\frac{2}{\pi}}\rho\\
&E\left((sgn(x_j)y_j)^3\right)=\frac{1}{\sqrt{2\pi}}\left(6\rho-2\rho^3\right)\\
&E\left((sgn(x_j)y_j)^2\right)=1,\hspace{0.2in} E\left((sgn(x_j)y_j)^4\right)=3
\end{align}
\noindent\textbf{Proof:}\hspace{0.2in} See Appendix C.$\hfill\Box$.
\end{theorem}

Theorem~\ref{thm_g} leads to a simple estimator $\hat{\rho}_g$ and its variance:
\begin{align}
&\hat{\rho}_g = \frac{1}{k}\sum_{j=1}^k \sqrt{\frac{\pi}{2}}{sgn}(x_j) y_j,\hspace{0.2in}
{E}\left(\hat{\rho}_g\right) = \rho\\\label{eqn_Vg}
&{Var}\left(\hat{\rho}_g\right) = \frac{V_g}{k},\hspace{0.2in} V_g = \frac{\pi}{2}-\rho^2
\end{align}
The variance does not vanish when $|\rho|\rightarrow1$. Interestingly, the variance can be substantially reduced by applying a normalization step on $y_j$, as shown in Theorem~\ref{thm_gn}.

\begin{theorem}\label{thm_gn}
As $k\rightarrow\infty$, the following asymptotic normality holds:
\begin{align}
&\sqrt{k}\left(\frac{\sum_{j=1}^k {sgn}(x_j) y_j}{\sqrt{k}\sqrt{\sum_{j=1}^k y_j^2}} - \sqrt{\frac{2}{\pi}}\rho\right)\overset{D}{\Longrightarrow}
N\left(0,\ V_{g,n}\right)\\\label{eqn_Vgn}
&V_{g,n} = V_g - \rho^2\left(3/2-\rho^2\right)
\end{align}
where $V_g=\frac{\pi}{2}-\rho^2$ as in (\ref{eqn_Vg}).\\

\noindent\textbf{Proof:}\hspace{0.2in} See Appendix D.$\hfill\Box$.
\end{theorem}

Theorem~\ref{thm_gn} leads to the following estimator $\hat{\rho}_{g,n}$:
\begin{align}
&\hat{\rho}_{g,n} = \sqrt{\frac{\pi}{2}}\left(\frac{\sum_{j=1}^k {sgn}(x_j) y_j}{\sqrt{k}\sqrt{\sum_{j=1}^k y_j^2}}\right)\\\notag
&{E}\left(\hat{\rho}_{g,n}\right) = \rho + O\left(\frac{1}{k}\right)\\\notag
&{Var}\left(\hat{\rho}_{g,n}\right) = \frac{V_{g,n}}{k} + O\left(\frac{1}{k^2}\right)
\end{align}
While this normalization  always helps (since $V_{g,n} \leq V_g$), the estimator still does not have the desired property that the variance should approach 0 as $|\rho|\rightarrow1$.

\newpage

It turns out that we can improve $\hat{\rho}_{g,n}$ at least for nonnegative data ($\rho\geq 0$), based on the results in Theorem~\ref{thm_s}.
\begin{theorem}\label{thm_s}
\begin{align}
&{E}\left(y_- 1_{x<0} + y_+1_{x\geq0}\right)=\frac{1+\rho}{\sqrt{2\pi}}\\
&{E}\left(y_- 1_{x<0} + y_+1_{x\geq0}\right)^2
=1_{\rho\geq0}-\frac{1}{\pi}\left(\tan^{-1}\left(\frac{\sqrt{1-\rho^2}}{\rho}\right)-\rho\sqrt{1-\rho^2}\right)\\
&{E}\left(y_- 1_{x\geq0} + y_+1_{x<0}\right)=\frac{1-\rho}{\sqrt{2\pi}}\\
&{E}\left(y_- 1_{x\geq0} + y_+1_{x<0}\right)^2
=1_{\rho<0}+\frac{1}{\pi}\left(\tan^{-1}\left(\frac{\sqrt{1-\rho^2}}{\rho}\right)-\rho\sqrt{1-\rho^2}\right)
\end{align}
\noindent\textbf{Proof:}\hspace{0.2in} See Appendix E.$\hfill\Box$.\\
\end{theorem}

This leads to another estimator, denoted by $\hat{\rho}_s$:
\begin{align}
&\hat{\rho}_s =1- \frac{\sqrt{2\pi}}{k}\sum_{j=1}^k \left[y_{j-}1_{x_j\geq0} +y_{j+}1_{x_j<0} \right] \\\notag
&E\left(\hat{\rho}_s\right) = \rho,\hspace{0.2in}
{Var}\left(\hat{\rho}_s\right) = \frac{V_s}{k}\\ \label{eqn_Vs} &V_s={2\pi}
\left[1_{\rho<0}+\frac{1}{\pi}\left(\tan^{-1}\left(\frac{\sqrt{1-\rho^2}}{\rho}\right)-\rho\sqrt{1-\rho^2}\right)-\frac{(1-\rho)^2}{2\pi}\right]
\end{align}
Recall that we denote $\tan^{-1}\left(\frac{1}{0}\right)  = \tan^{-1}\left(\frac{1}{0+}\right) =\frac{\pi}{2}$.

The variance of $\hat{\rho}_s$ has the desired property that it approaches zero as $\rho\rightarrow1$. However, when $\rho\rightarrow-1$, the variance becomes large. In fact, even at $\rho=0$, the variance is already fairly large as we will soon show. Thus, we still hope to be able to reduce the variance by normalizing $y$.

\begin{theorem}\label{thm_sn}
\begin{align}
&\sqrt{k}\left(\frac{\sum_{j=1}^k y_{j-}1_{x_j\geq0} + y_{j+}1_{x_j<0}}{\sqrt{k}\sqrt{\sum_{j=1}^k y_j^2}}-\frac{1-\rho}{\sqrt{2\pi}}\right)
\overset{D}{\Longrightarrow}
N\left(0,\ V_{s,n}\right)\\\label{eqn_Vsn}
&V_{s,n}=V_s -\frac{(1-\rho)^2}{{4\pi}}\left(1-2\rho-2\rho^2\right)
\end{align}
where $V_s$ is in (\ref{eqn_Vs}).\\

\noindent\textbf{Proof:}\hspace{0.2in} See Appendix F.$\hfill\Box$.
\end{theorem}

\begin{figure}[h!]
\begin{center}
\includegraphics[width=3.5in]{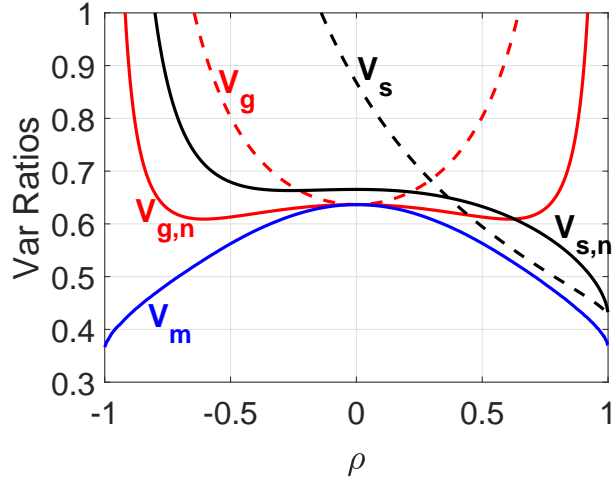}
\end{center}
\vspace{-0.2in}
\caption{Variance factor ratios: $\frac{V_m}{V_1}$,  $\frac{V_g}{V_1}$,  $\frac{V_{g,n}}{V_1}$,  $\frac{V_s}{V_1}$,  $\frac{V_{s,n}}{V_1}$, to compare the five estimators developed for sign-full random projections, in terms of the relative improvement with respect to the 1-bit estimator $\hat{\rho}_1$. The MLE ($\hat{\rho}_{m}$, solid blue curve) achieves the lowest asymptotic variance. When $\rho=0$, $\frac{V_m}{V_1}=\frac{V_g}{V_1}=\frac{V_{g,n}}{V_1}=\frac{2}{\pi}\approx 0.6366$, $\frac{V_{s}}{V_1} = \frac{4}{\pi} -\frac{4}{\pi^2}\approx 0.8680$, $\frac{V_{s,n}}{V_1} = \frac{4}{\pi}-\frac{6}{\pi^2} \approx 0.6653$. When $\rho\rightarrow1$, $\frac{V_s}{V_1} = \frac{V_{s,n}}{V_1}=\frac{4}{3\pi}\approx 0.4244 $. However, $\frac{V_g}{V_1}=\infty$ and $\frac{V_{g,n}}{V_1}=\infty$ when $\rho\rightarrow1$, indicating  that $\hat{\rho}_{g}$ and $\hat{\rho}_{g,n}$  are poor estimators for the high similarity region. Overall, $\hat{\rho}_{s,n}$ is a very good estimator, at least for nonnegative data.}
\label{fig_Vsn}
\end{figure}

This leads to the following estimator:
\begin{align}
&\hat{\rho}_{s,n} =1- \frac{\sum_{j=1}^k \sqrt{2\pi}\left[y_{j-}1_{x_j\geq0} +y_{j+}1_{x_j<0} \right]}{\sqrt{k}\sqrt{\sum_{j=1}^ky_j^2}} \\\notag
&E\left(\hat{\rho}_{s,n}\right) = \rho+O\left(\frac{1}{k}\right),\ {Var}\left(\hat{\rho}_{s,n}\right) = \frac{V_{s,n}}{k} +O\left(\frac{1}{k^2}\right)
\end{align}
where $V_{s,n}$ is in (\ref{eqn_Vsn}).
The resultant estimator $\hat{\rho}_{s,n}$ still has the property that the variance approaches 0 as $\rho\rightarrow 1$. The normalization step however does not always help. From (\ref{eqn_Vsn}), we have  $V_s\geq V_{s,n}$ if $\rho\leq \frac{\sqrt{3}-1}{2}\approx 0.3660$. On the other hand, as shown in Figure~\ref{fig_Vsn}, the normalization step only increases the variance slightly if $\rho> 0.3660$.\\

Figure~\ref{fig_Vsn} plots the rations: $\frac{V_m}{V_1}$,  $\frac{V_g}{V_1}$,  $\frac{V_{g,n}}{V_1}$,  $\frac{V_s}{V_1}$,  $\frac{V_{s,n}}{V_1}$, to compare those five estimators in terms of their improvements relative to the 1-bit estimator $\hat{\rho}_1$. As expected, the MLE $\hat{\rho}_m$ achieves the smallest asymptotic variance and $\frac{V_m}{V_1} = \frac{2}{\pi}$ at $\rho=0$ and $\frac{V_m}{V_1} \approx 0.36$ at $|\rho|\rightarrow1$. This means in the high similarity region, using $\hat{\rho}_m$ can roughly reduce the required number of samples ($k$) by a factor of 3. Overall, $\hat{\rho}_{s,n}$ is the recommended estimator for practical use, because $\hat{\rho}_{s,n}$ is computationally simple and its variance is very close to the variance of the MLE, at least for $\rho\geq -0.4$. \vspace{-0.15in}

We summarize some numerical values in Lemma~\ref{lem_V}.
\begin{lemma}\label{lem_V}
At $\rho=0$,
\begin{align}
&\frac{V_m}{V_1}=\frac{V_g}{V_1}=\frac{V_{g,n}}{V_1}=\frac{2}{\pi}\approx 0.6366,\\
&\frac{V_{s}}{V_1} = \frac{4}{\pi} -\frac{4}{\pi^2}\approx 0.8680, \\
&\frac{V_{s,n}}{V_1} = \frac{4}{\pi}-\frac{6}{\pi^2} \approx 0.6653
\end{align}
As $|\rho|  \rightarrow 1$,
\begin{align}\label{eqn_V1_rate}
&V_1 = 2\sqrt{2}\pi\left(1-|\rho|\right)^{3/2} + o\left(\left(1-|\rho|\right)^{3/2}\right)
\end{align}
As $\rho  \rightarrow 1$,
\begin{align}
&\frac{V_s}{V_1} = \frac{V_{s,n}}{V_1}=\frac{4}{3\pi}\approx 0.4244,\\
& \frac{V_g}{V_1}=\infty, \hspace{0.2in} \frac{V_{g,n}}{V_1}=\infty
\end{align}
\textbf{Proof:}\hspace{0.2in} See Appendix G.$\hfill\Box$.
\end{lemma}

Overall, $\hat{\rho}_{s,n}$ is recommended, at least for nonnegative data ($\rho\geq 0$, which is common in practice). Typical applications are often concerned with the high similarity region. At $\rho\rightarrow 1$, the asymptotic variance of $\hat{\rho}_{s,n}$ approaches zero at the same rate as the MLE ($\hat{\rho}_m$), in particular, $\frac{V_{s,n}}{V_m} \approx 1.18$.

\section{A Simulation Study}

In this section, we provide a simulation study to verify the theoretical properties of the proposed four estimators for sign-full random projections: $\hat{\rho}_g$, $\hat{\rho}_{g,n}$, $\hat{\rho}_s$,  $\hat{\rho}_{s,n}$, as well as the estimator for sign-sign projections: $\hat{\rho}_1$:
\begin{align}\notag
&\hat{\rho}_1 = \cos\pi\left(1-\frac{1}{k}\sum_{j=1}^k1_{sgn(x_j)=sgn(y_j)}\right),\\\notag
&\hat{\rho}_g = \frac{1}{k}\sum_{j=1}^k \sqrt{\frac{\pi}{2}}{sgn}(x_j) y_j,\\\notag
&\hat{\rho}_{g,n} = \sqrt{\frac{\pi}{2}}\left(\frac{\sum_{j=1}^k {sgn}(x_j) y_j}{\sqrt{k}\sqrt{\sum_{j=1}^k y_j^2}}\right),\\\notag
&\hat{\rho}_s =1- \frac{\sqrt{2\pi}}{k}\sum_{j=1}^k \left[y_{j-}1_{x_j\geq0} +y_{j+}1_{x_j<0} \right], \\\notag
&\hat{\rho}_{s,n} =1- \frac{\sum_{j=1}^k \sqrt{2\pi}\left[y_{j-}1_{x_j\geq0} +y_{j+}1_{x_j<0} \right]}{\sqrt{k}\sqrt{\sum_{j=1}^ky_j^2}}
\end{align}

For  a given $\rho$, we simulate $k$ standard bi-variate normal variables $(x_j, y_j)$ with $E(x_jy_j)=\rho$, $j=1, ..., k$. Then we choose an estimator $\hat{\rho}$ to estimate $\rho$. With $10^6$ simulations, we assume that the empirical bias and variance of $\hat{\rho}$ are close to the true values. We plot the empirical mean square errors (MSEs): $MSE(\hat{\rho}) = Bias^2(\hat{\rho}) + Var(\hat{\rho})$, together with the theoretical variance of $\hat{\rho}$. If the empirical MSE curve and the theoretical variance overlap, we  know that the estimator is unbiased and the theoretical variance formula is verified.

\begin{figure}[h!]
\begin{center}
\mbox{
\includegraphics[width=2.25in]{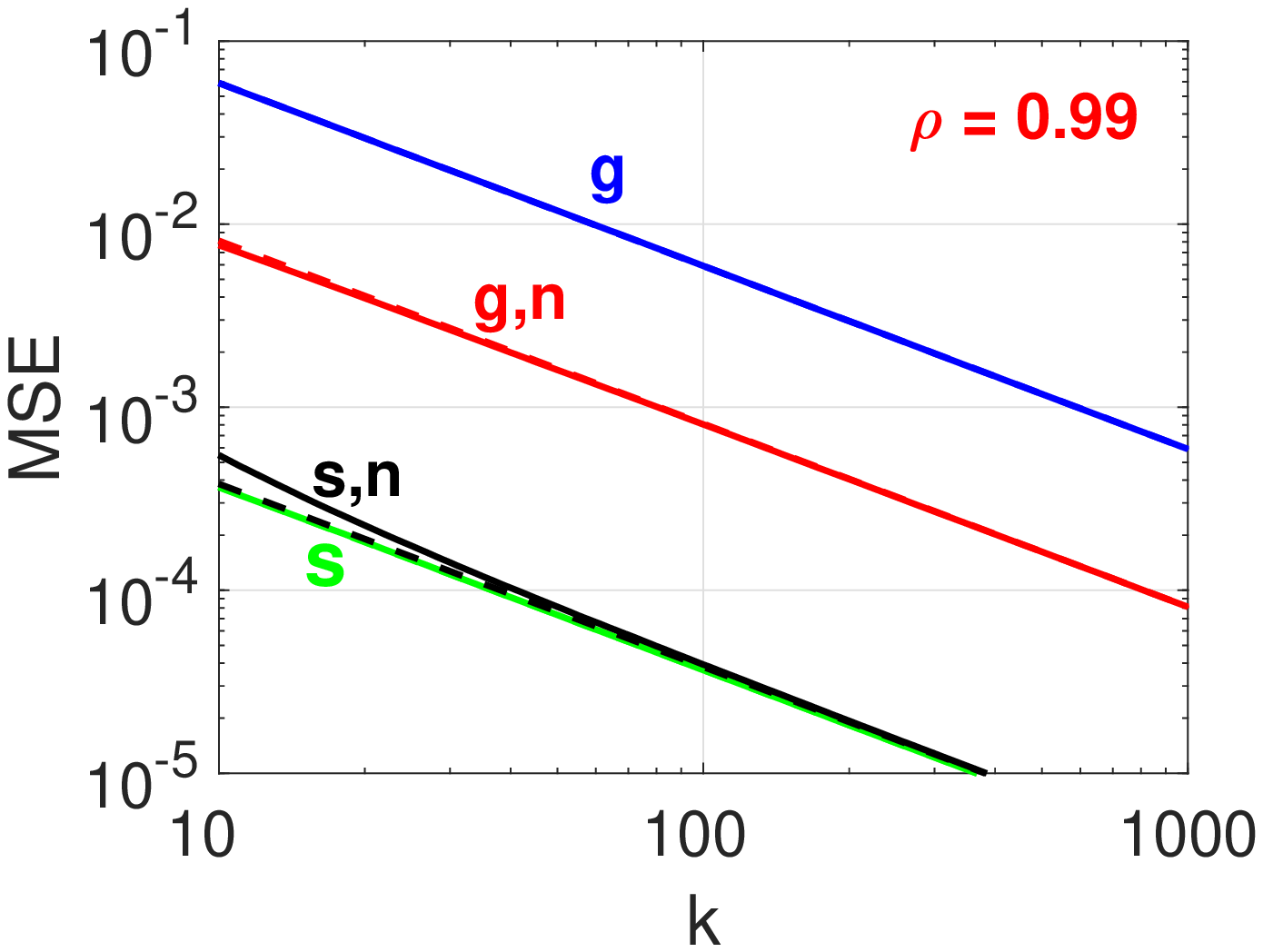}
\includegraphics[width=2.25in]{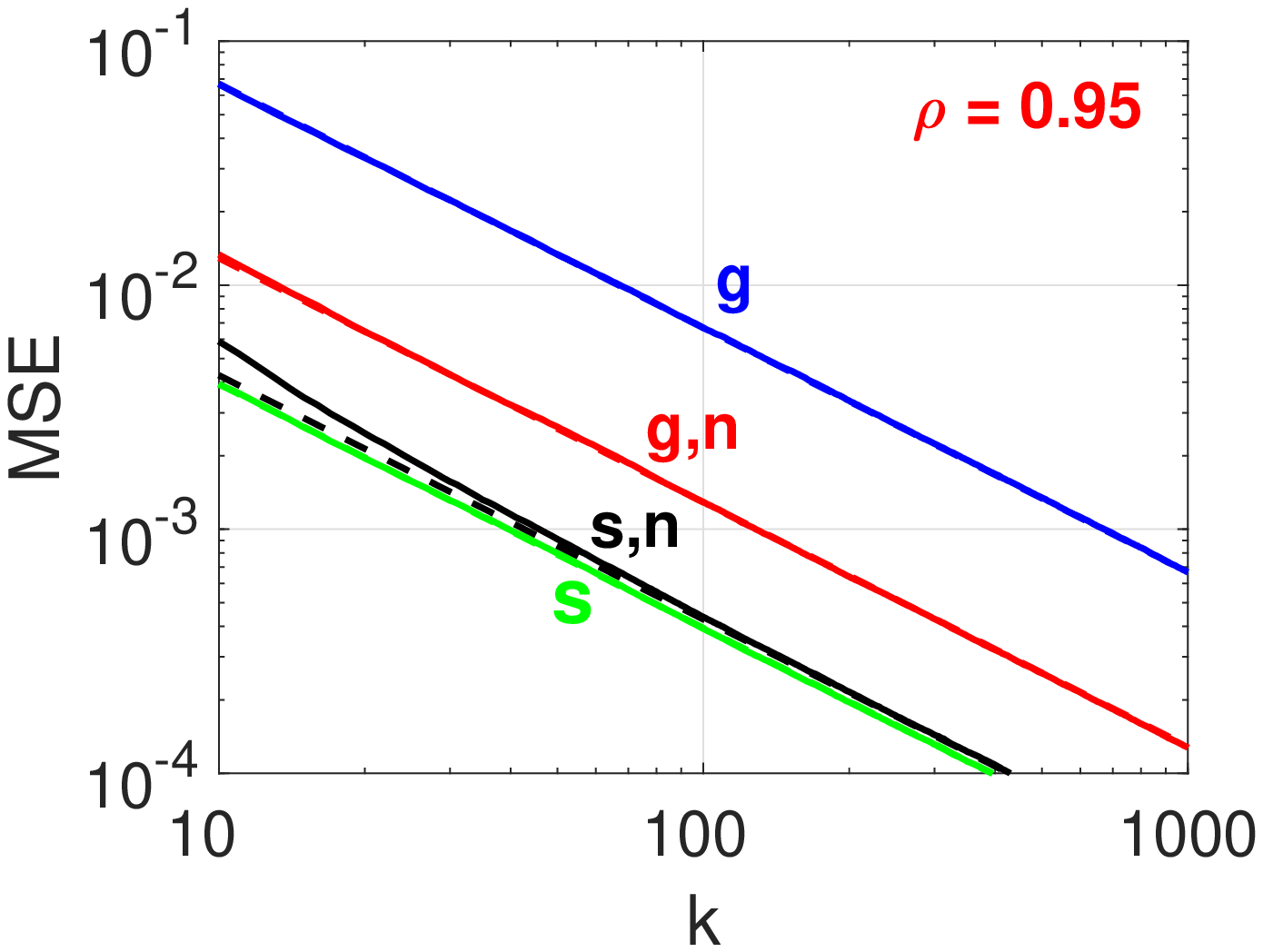}
}

\mbox{
\includegraphics[width=2.25in]{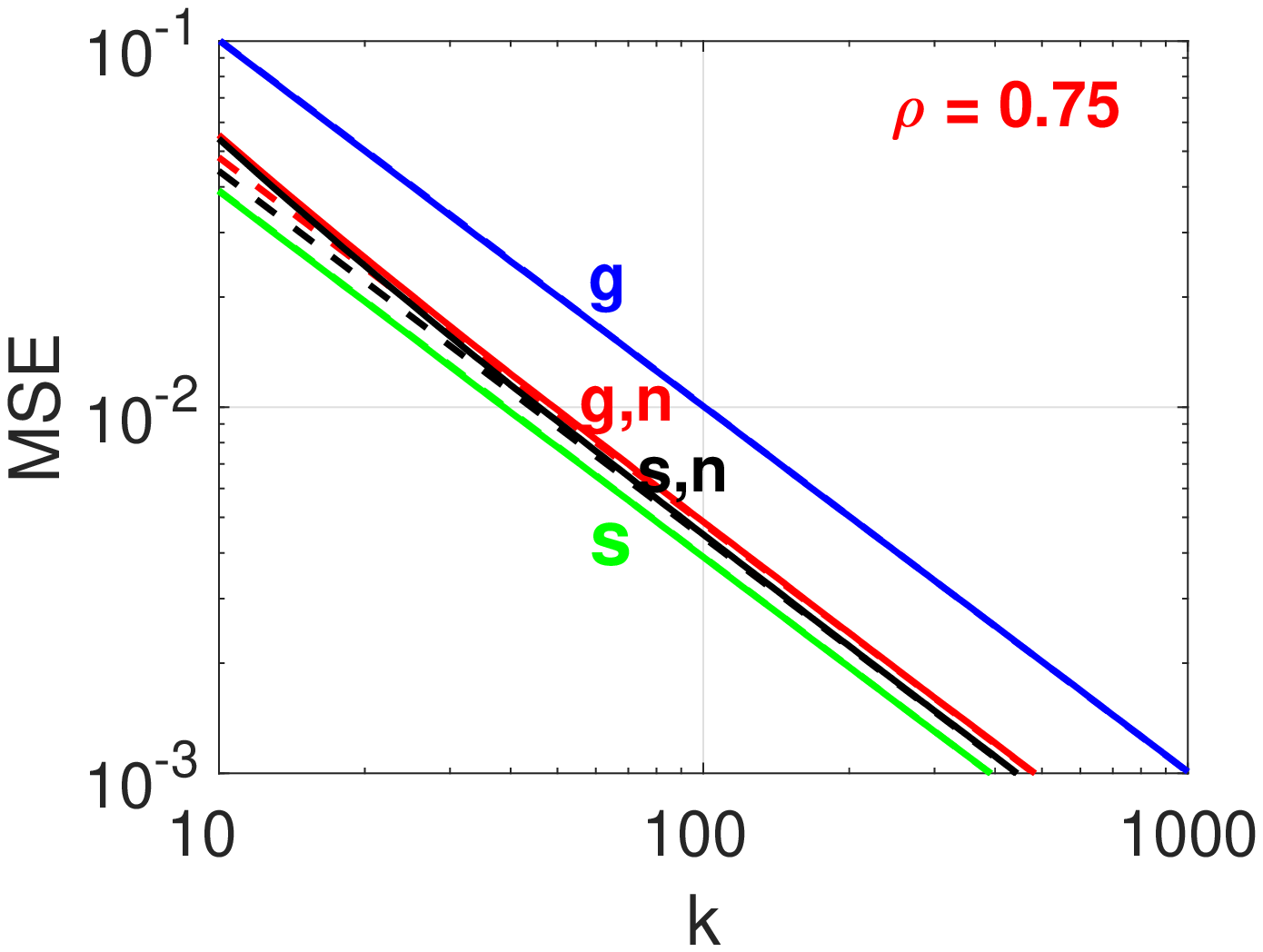}
\includegraphics[width=2.25in]{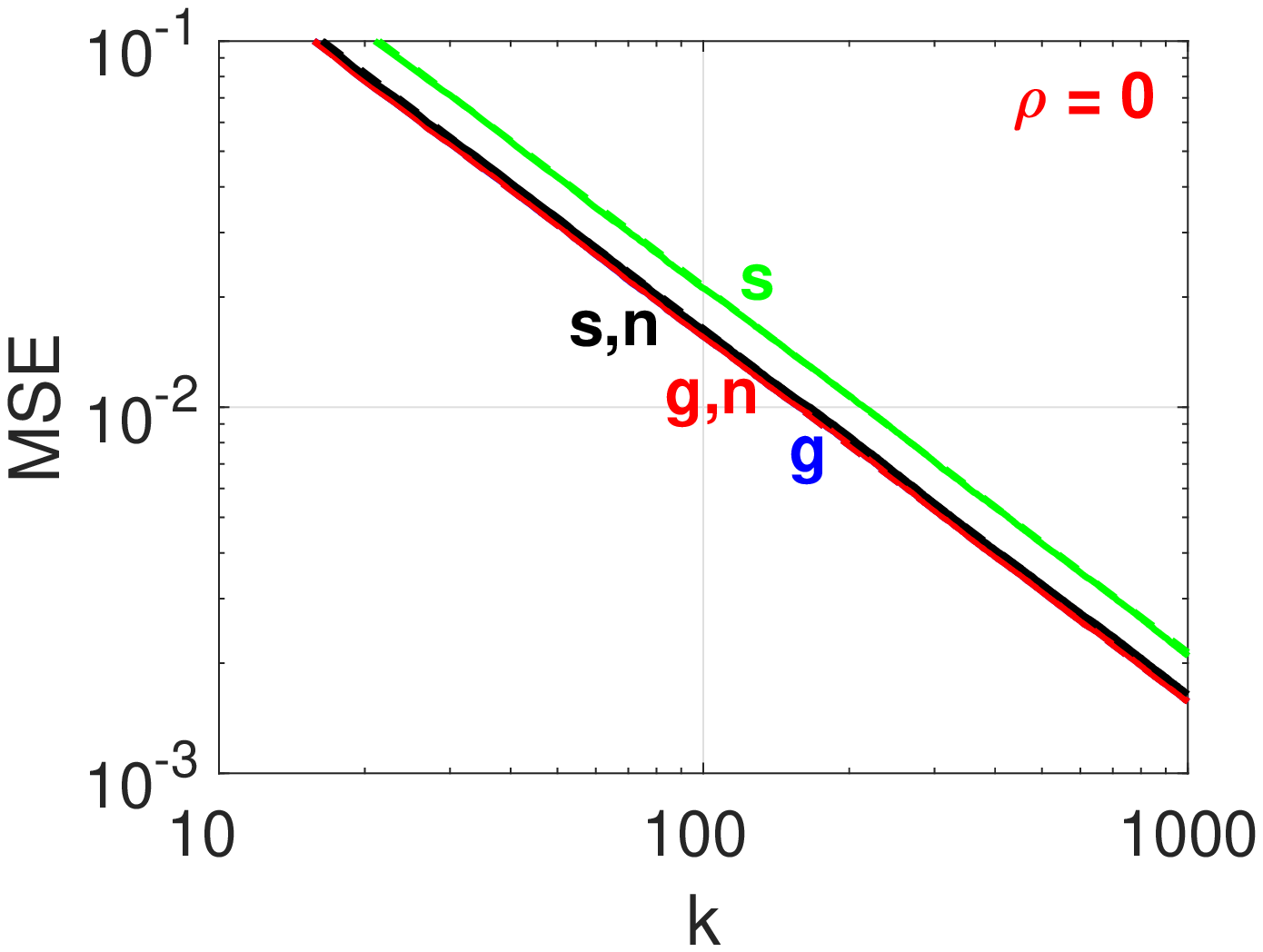}
}

\mbox{
\includegraphics[width=2.25in]{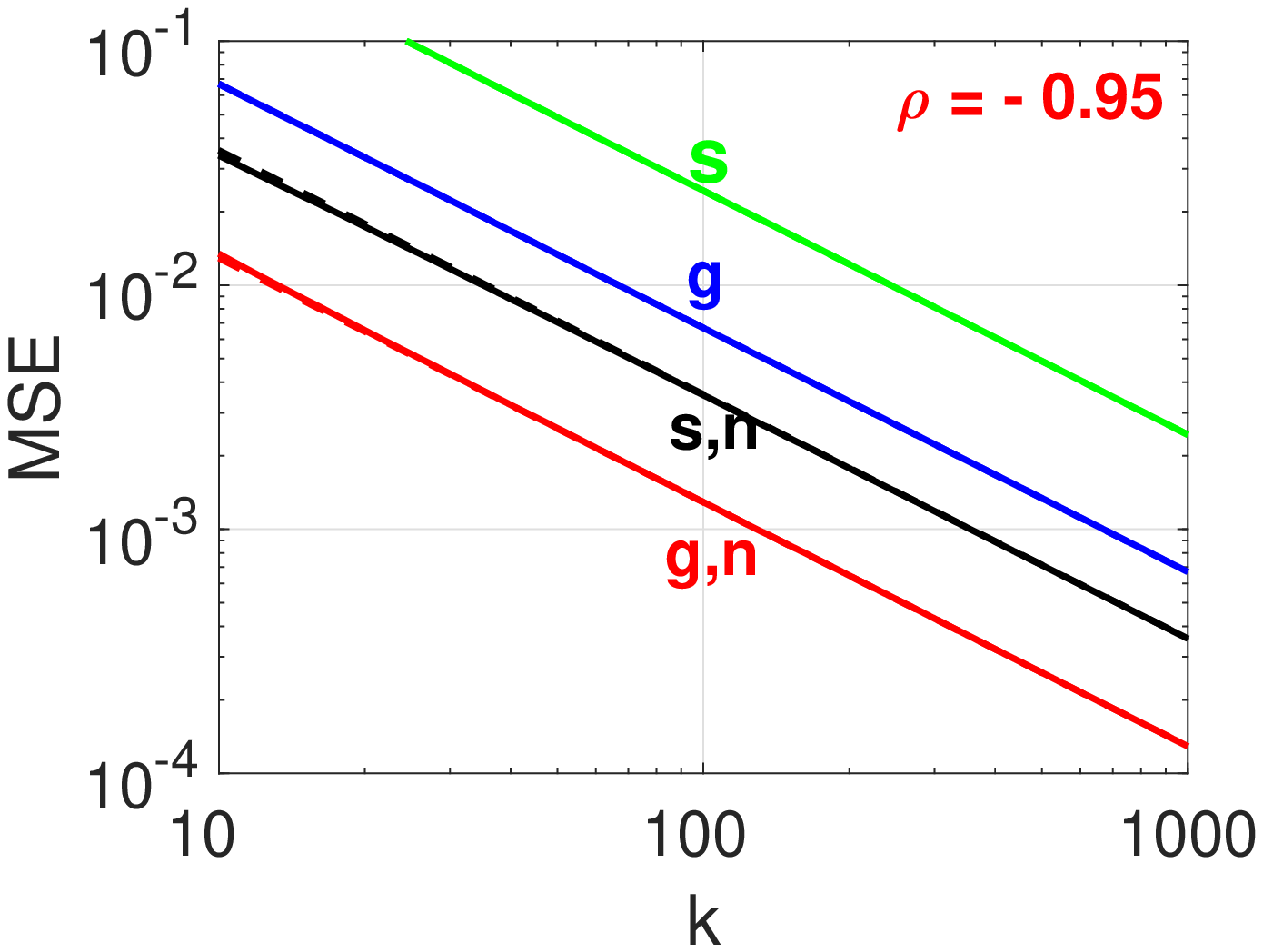}
\includegraphics[width=2.25in]{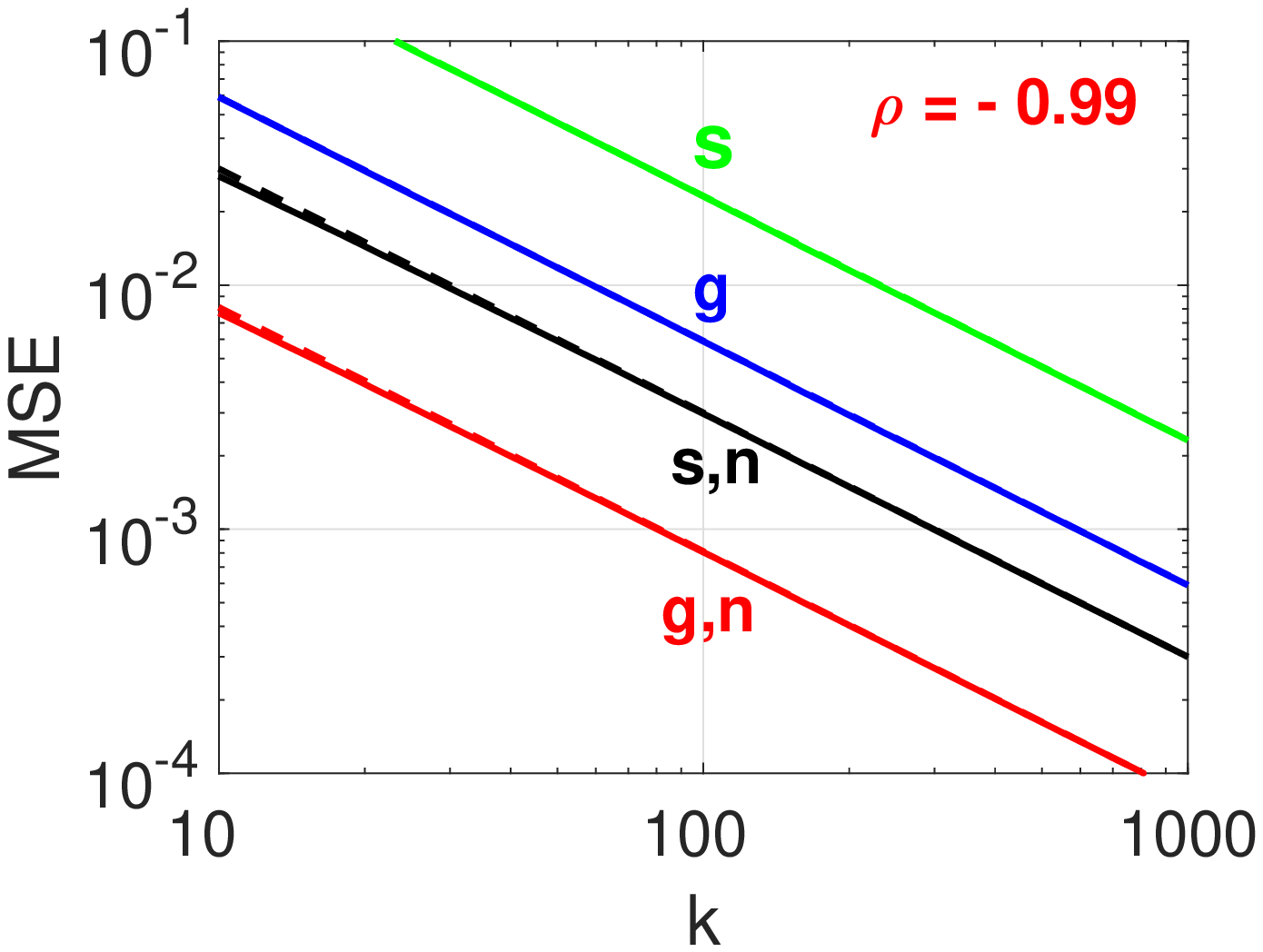}
}
\end{center}
\vspace{-0.2in}
\caption{Empirical MSEs (solid curves) for four proposed estimators, together with the theoretical (asymptotic) variances (dashed curves), for 6 selected $\rho$ values (one in each panel). For $\hat{\rho}_{g}$ and $\hat{\rho}_s$, the solid and dashed curves overlap, confirming that they are unbiased and the variance formulas are correct. For $\hat{\rho}_{g,n}$ and $\hat{\rho}_{s,n}$, the solid and dashed curves overlap when $k$ is not too small. }\label{fig_Est_MSE}
\end{figure}

Figure~\ref{fig_Est_MSE} presents the results for 6 selected $\rho$ values: $0.99, 0.95, 0.75 0, -0.95, -0.99$. Those simulations verify that both $\hat{\rho}_g$ and $\hat{\rho}_s$ are unbiased, while their normalized versions $\hat{\rho}_{g,n}$ and $\hat{\rho}_{s,n}$ are asymptotically (i.e., when $k$ is not too small) unbiased. The (asymptotic) variance formulas for these four estimators are verified since the solid and dashed curves overlap (when $k$ is not small).

\begin{figure}[h!]
\begin{center}
\mbox{
\includegraphics[width=2.25in]{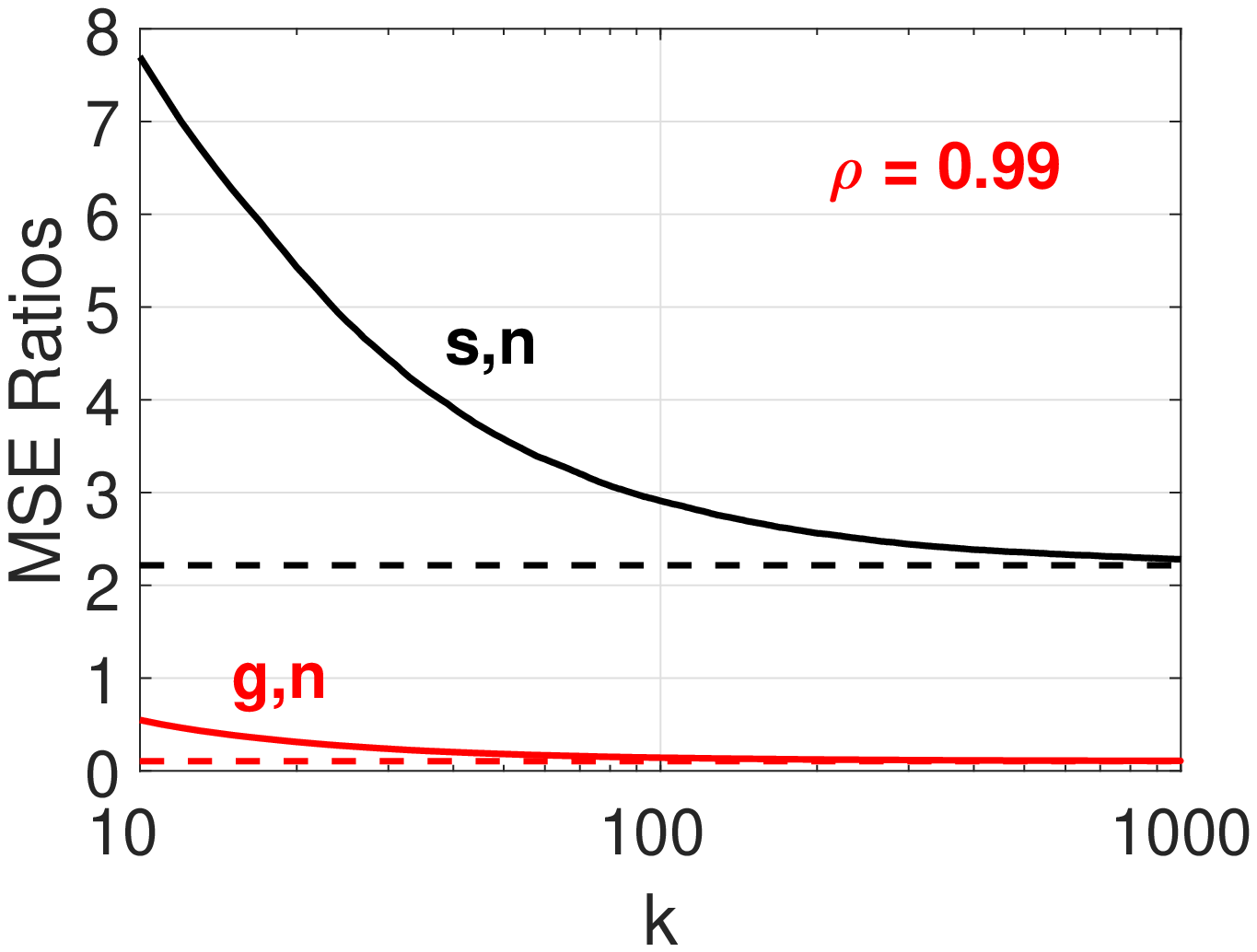}\hspace{-0.13in}
\includegraphics[width=2.25in]{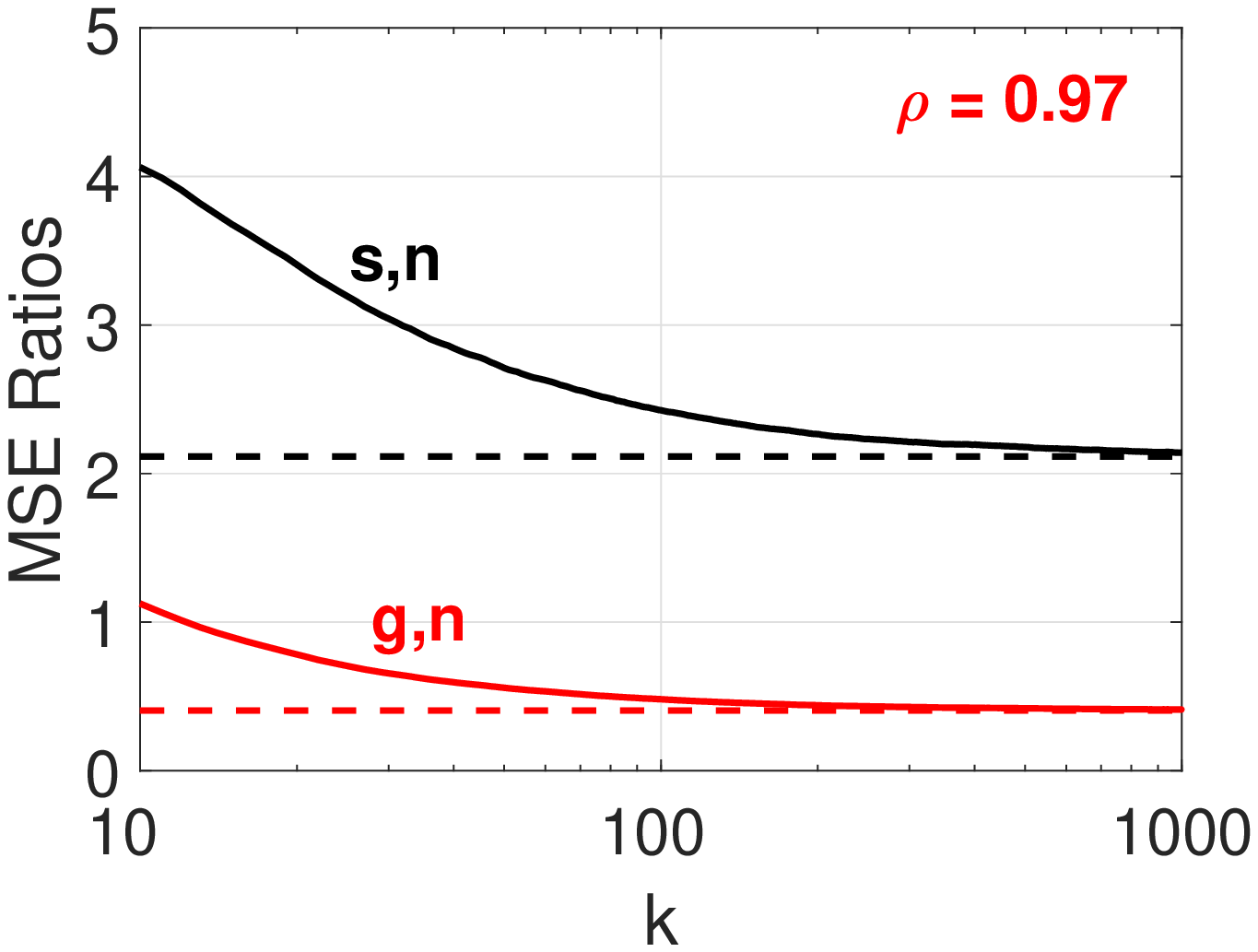}
}

\mbox{
\includegraphics[width=2.25in]{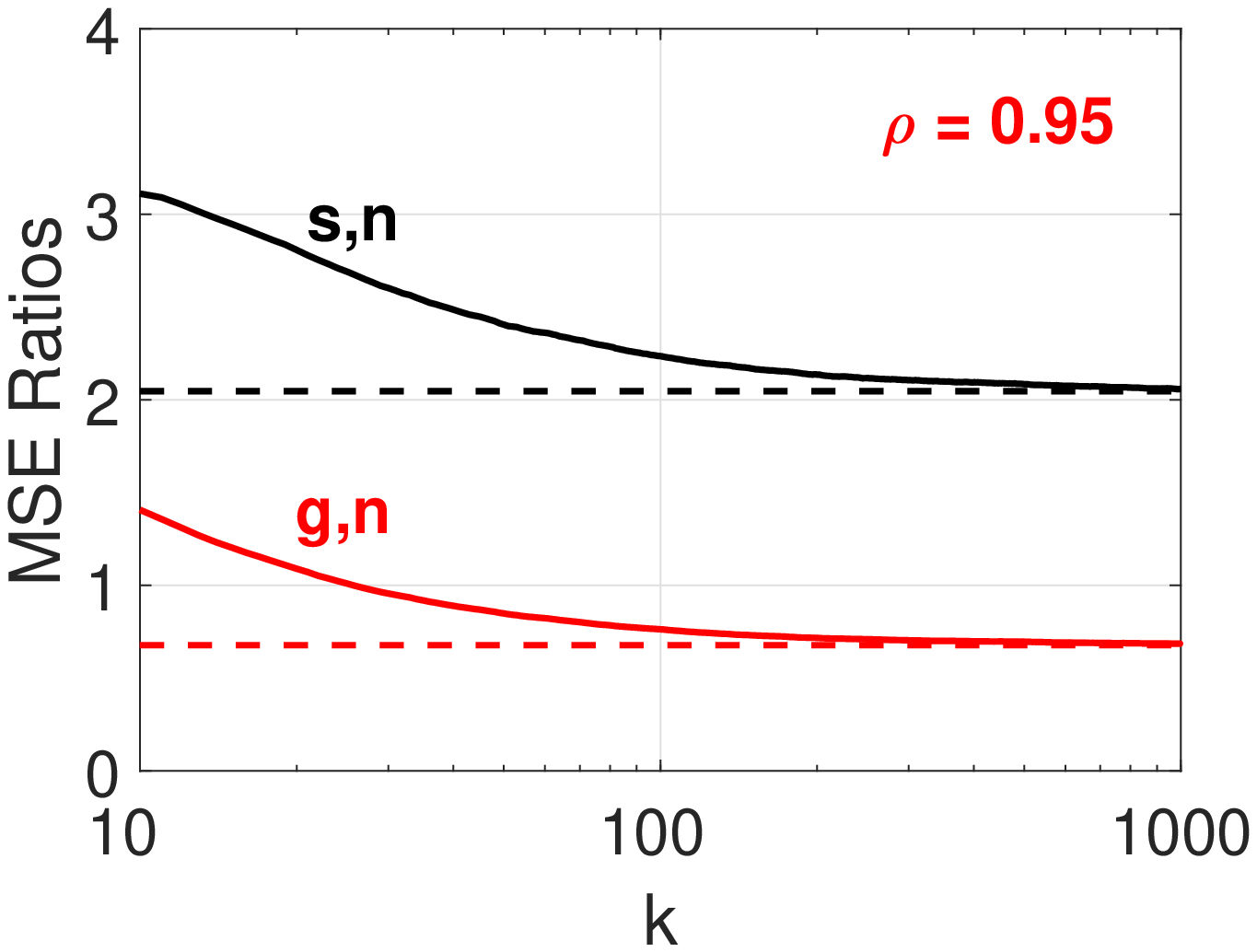}\hspace{-0.13in}
\includegraphics[width=2.25in]{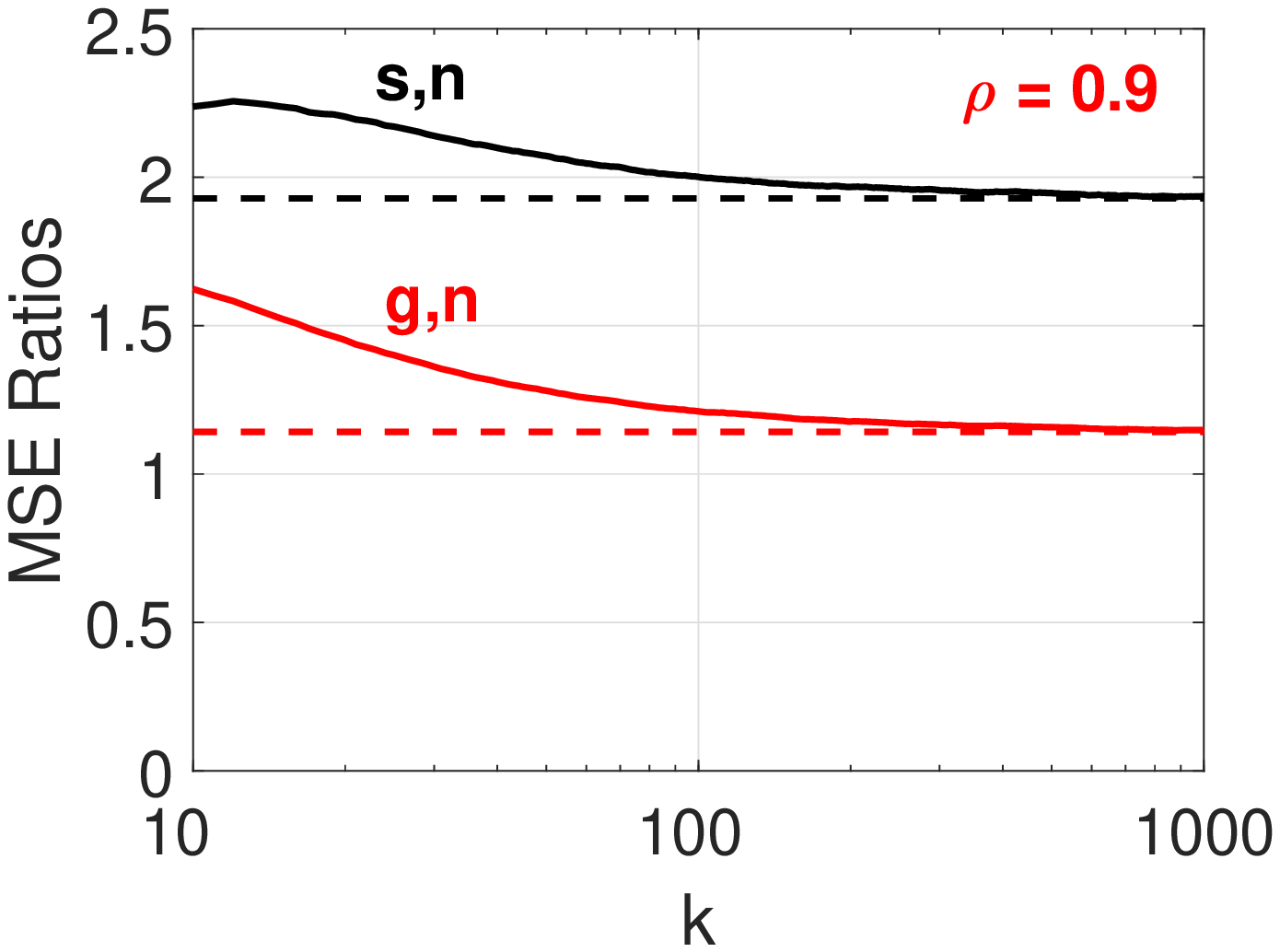}
}

\end{center}
\vspace{-0.2in}
\caption{Empirical MSE rations:  $\frac{MSE(\hat{\rho}_{1})}{MSE(\hat{\rho}_{s,n})}$ and $\frac{MSE(\hat{\rho}_{1})}{MSE(\hat{\rho}_{g,n})}$, together with the theoretical asymptotic variance ratios (dashed curves):  $\frac{V_{1}}{V_{s,n}}$ and $\frac{V_{1}}{V_{g,n}}$ (i.e., the reciprocal of those in Figure~\ref{fig_Vsn}). When $k$ is not small, the solid and dashed curves overlap. The results indicate that at high similarity and small $k$, the improvement from using sign-full random projections would be even much more substantial (e.g., the actual ratio can be as high as 8). }\label{fig_Est_Ratio}
\end{figure}

Figure~\ref{fig_Est_Ratio} presents the ratios of empirical MSEs (solid curves): $\frac{MSE(\hat{\rho}_{1})}{MSE(\hat{\rho}_{s,n})}$ and $\frac{MSE(\hat{\rho}_{1})}{MSE(\hat{\rho}_{g,n})}$, together with the theoretical asymptotic variance ratios (dashed curves):  $\frac{V_{1}}{V_{s,n}}$ and $\frac{V_{1}}{V_{g,n}}$ (i.e., the reciprocal of those in Figure~\ref{fig_Vsn}).  These curves again confirm the asymptotic variance formulas. In addition, they indicate that in the high similarity region, when the sample size $k$ is not too large, the improved gained from using $\hat{\rho}_{s,n}$ can be substantially more than what are predicted by theory. For example, when $\rho$ is close to 1 (e.g., $\rho=0.99$), theoretically $\frac{V_{1}}{V_{s,n}} = \frac{3}{4}\pi \approx 2.3562$, the actual improvement can be as much as a factor of 8 (at $k=10$). This is the additional advantage of $\hat{\rho}_{s,n}$.

\begin{figure}[h!]
\begin{center}
\mbox{
\includegraphics[width=2.25in]{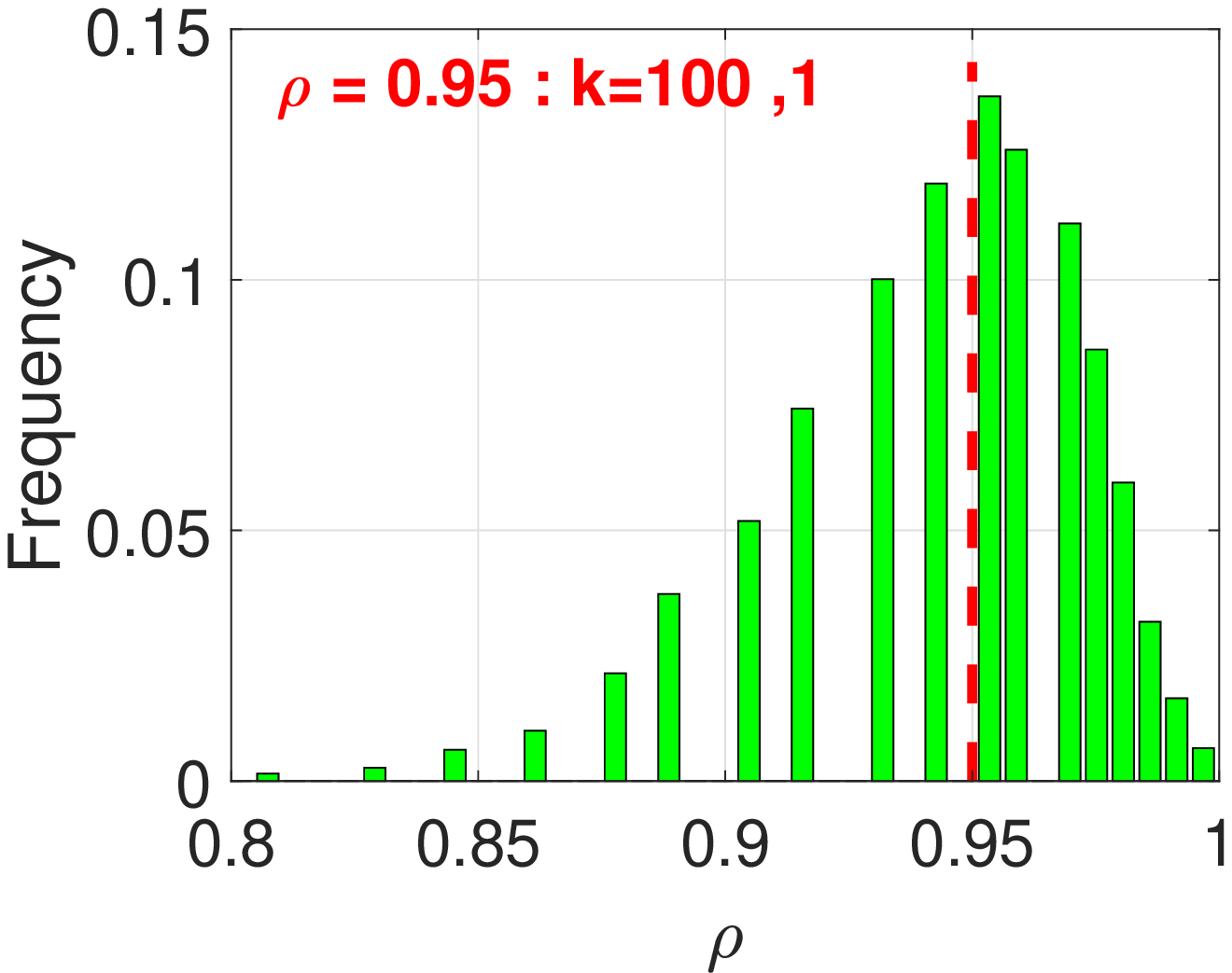}\hspace{-0.11in}
\includegraphics[width=2.25in]{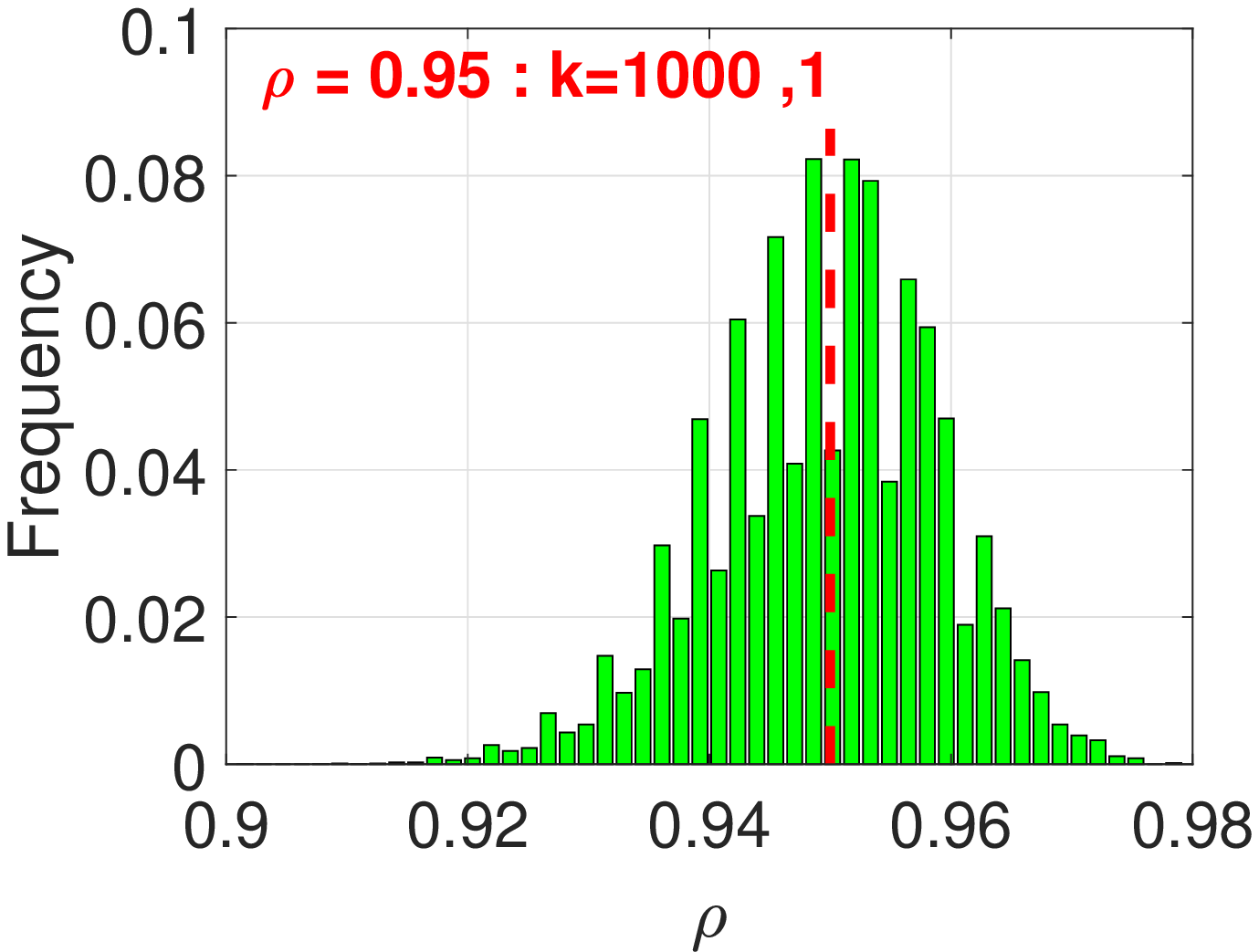}
}

\mbox{
\includegraphics[width=2.25in]{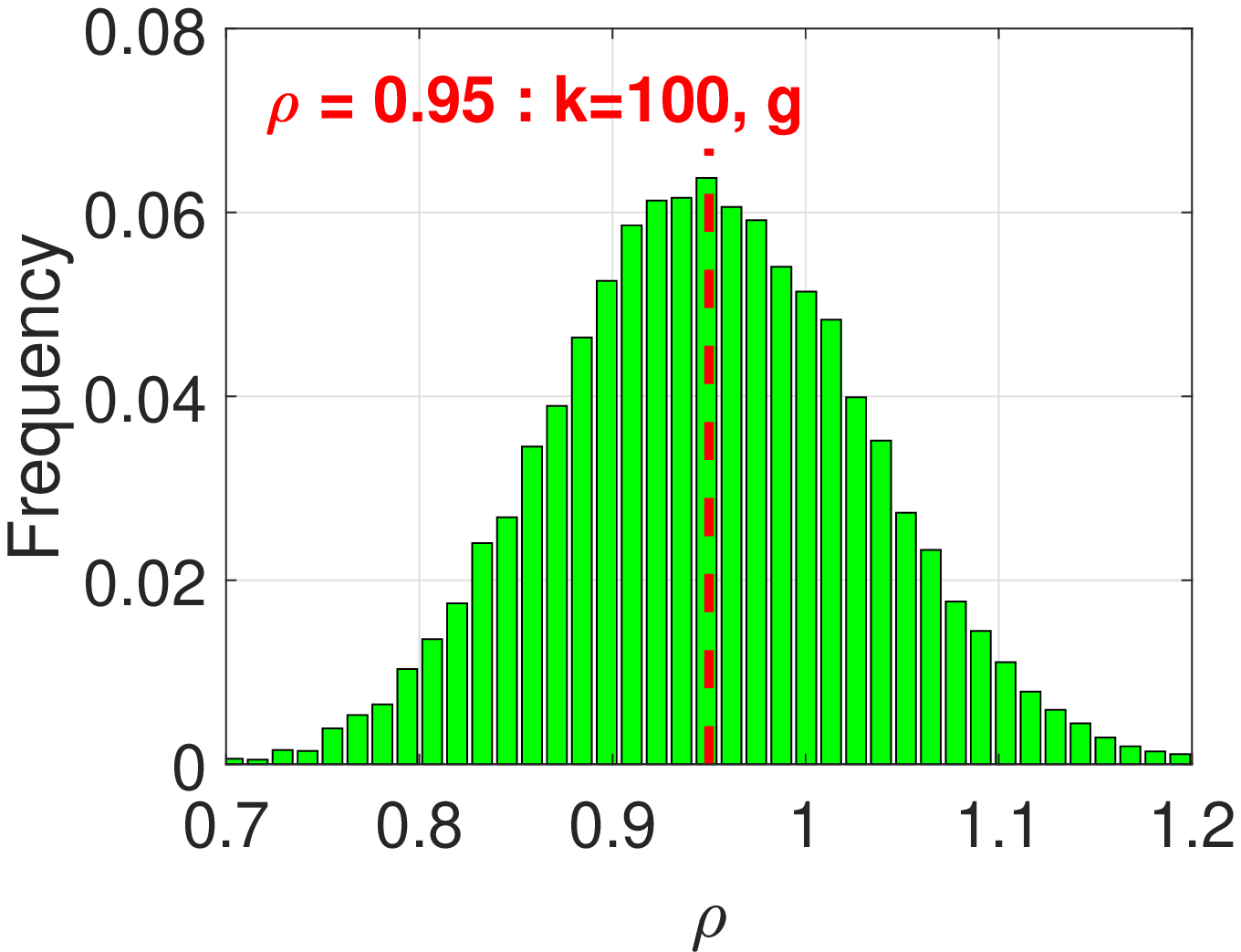}\hspace{-0.11in}
\includegraphics[width=2.25in]{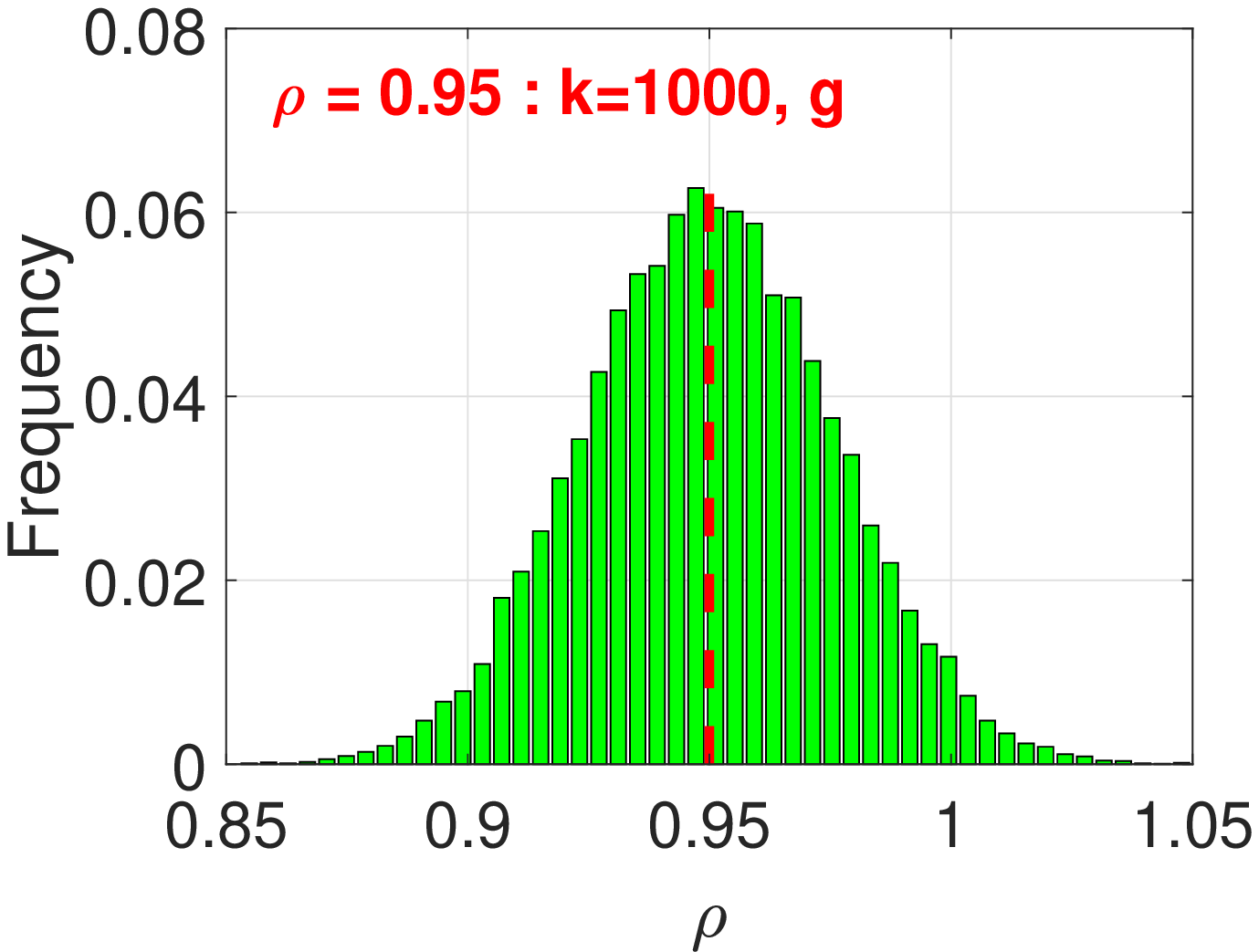}
}

\mbox{
\includegraphics[width=2.25in]{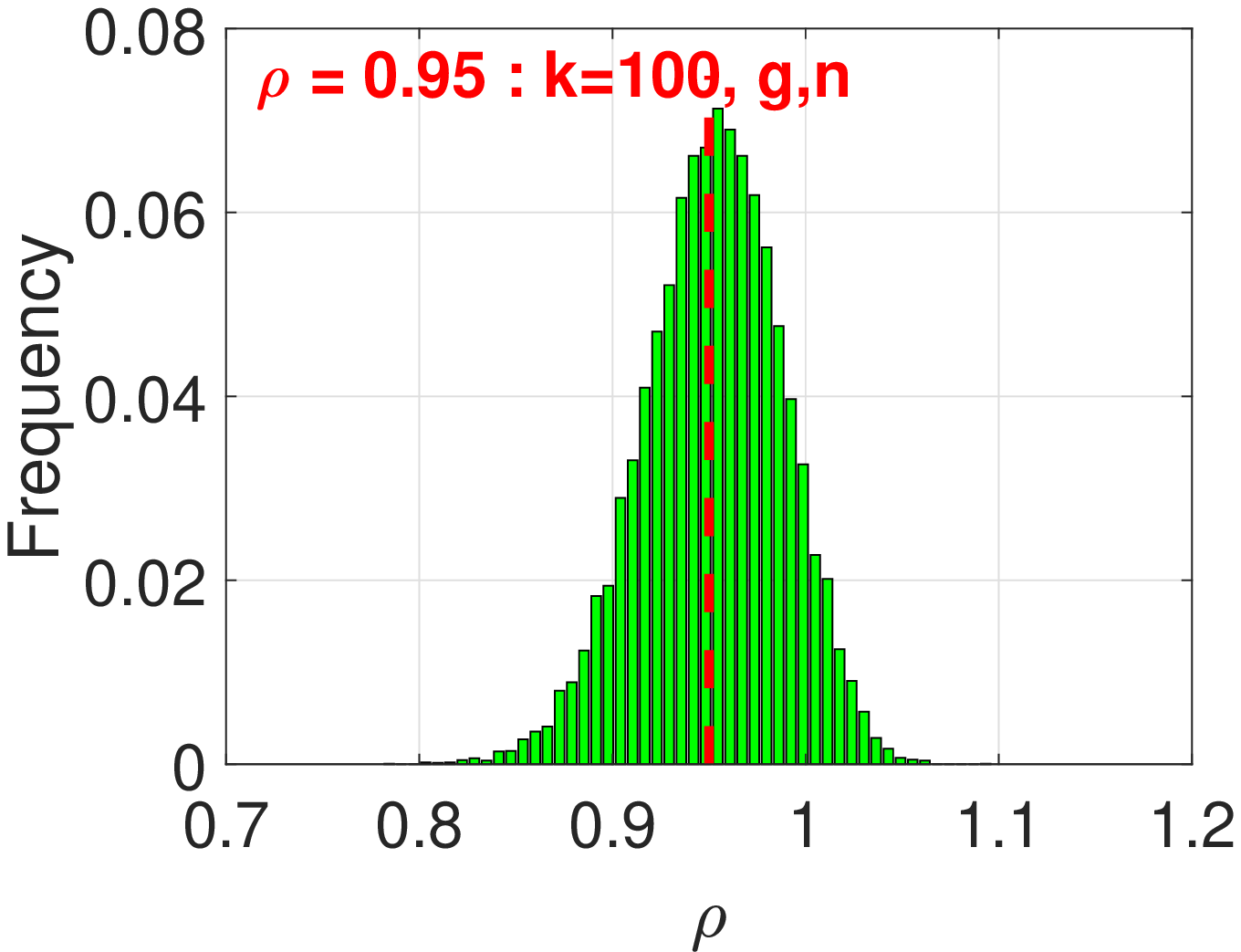}\hspace{-0.11in}
\includegraphics[width=2.25in]{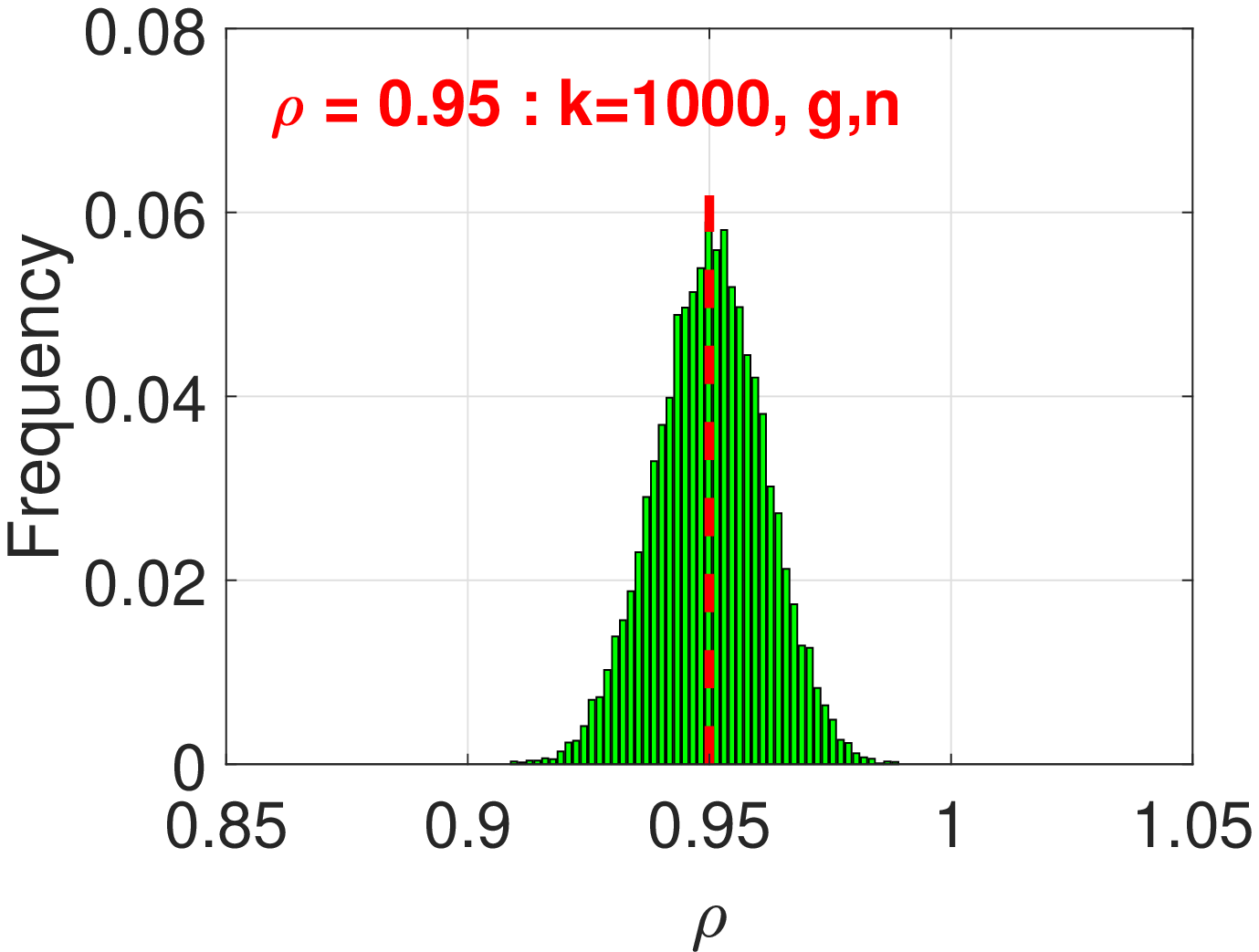}
}

\mbox{
\includegraphics[width=2.25in]{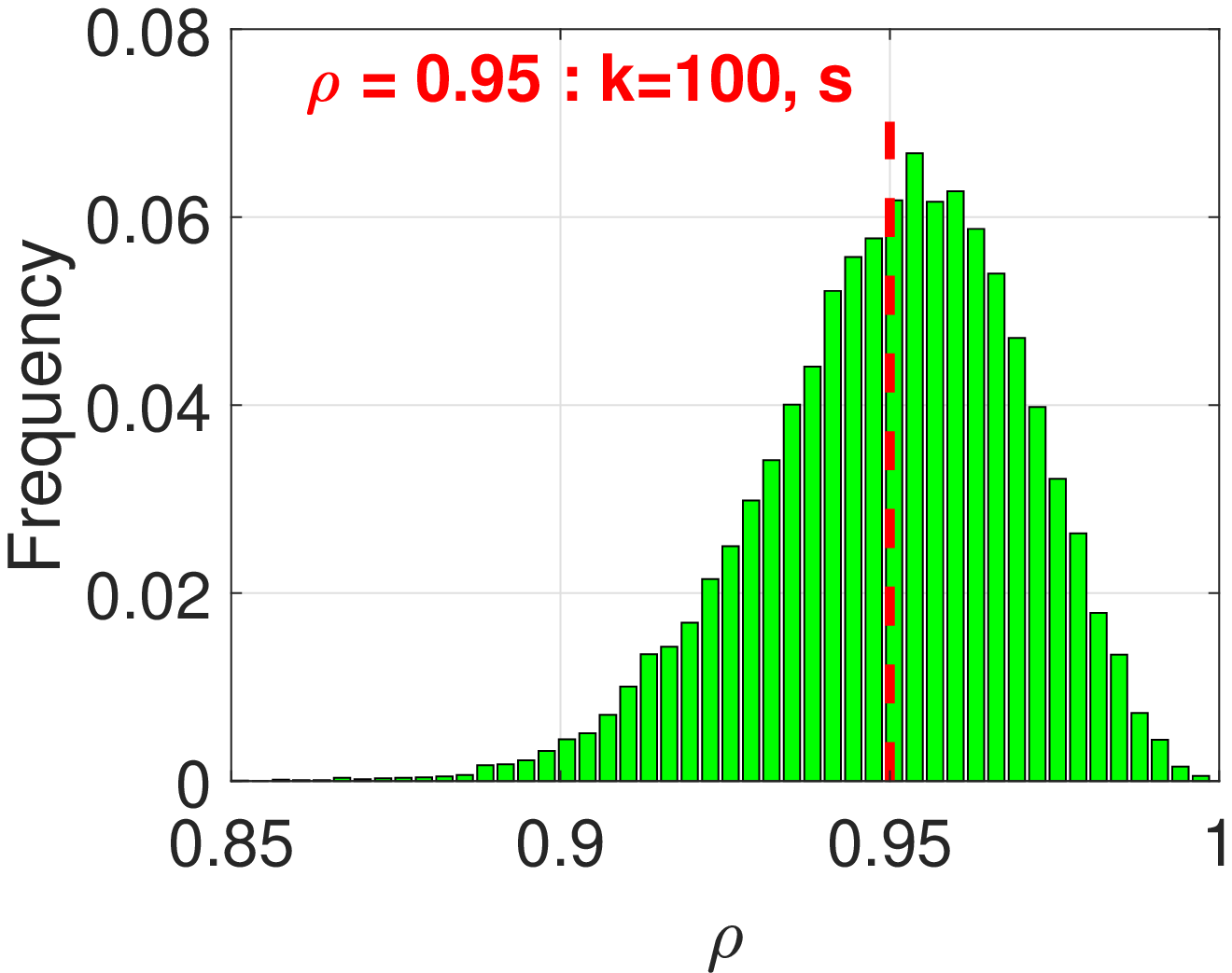}\hspace{-0.11in}
\includegraphics[width=2.25in]{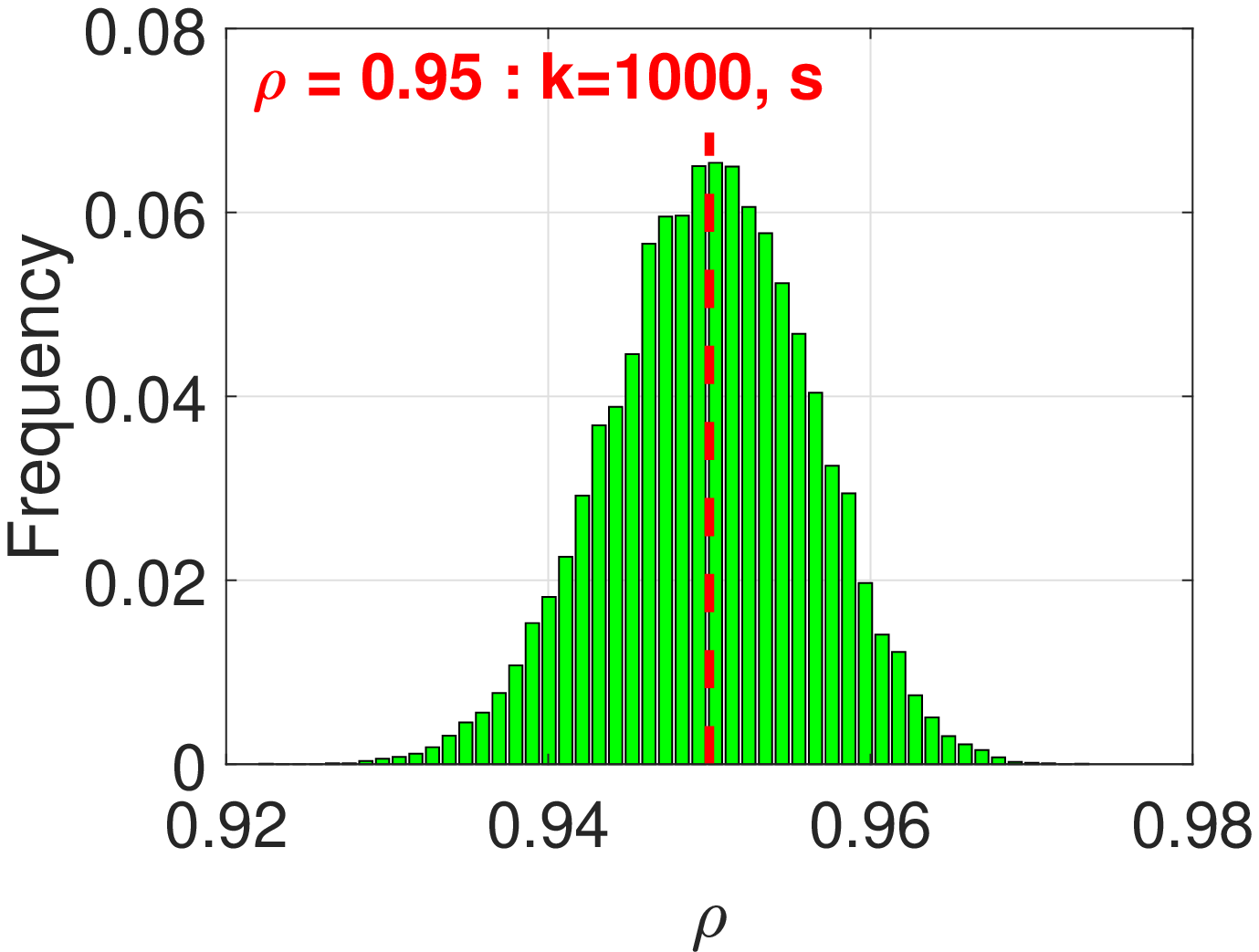}
}

\mbox{
\includegraphics[width=2.25in]{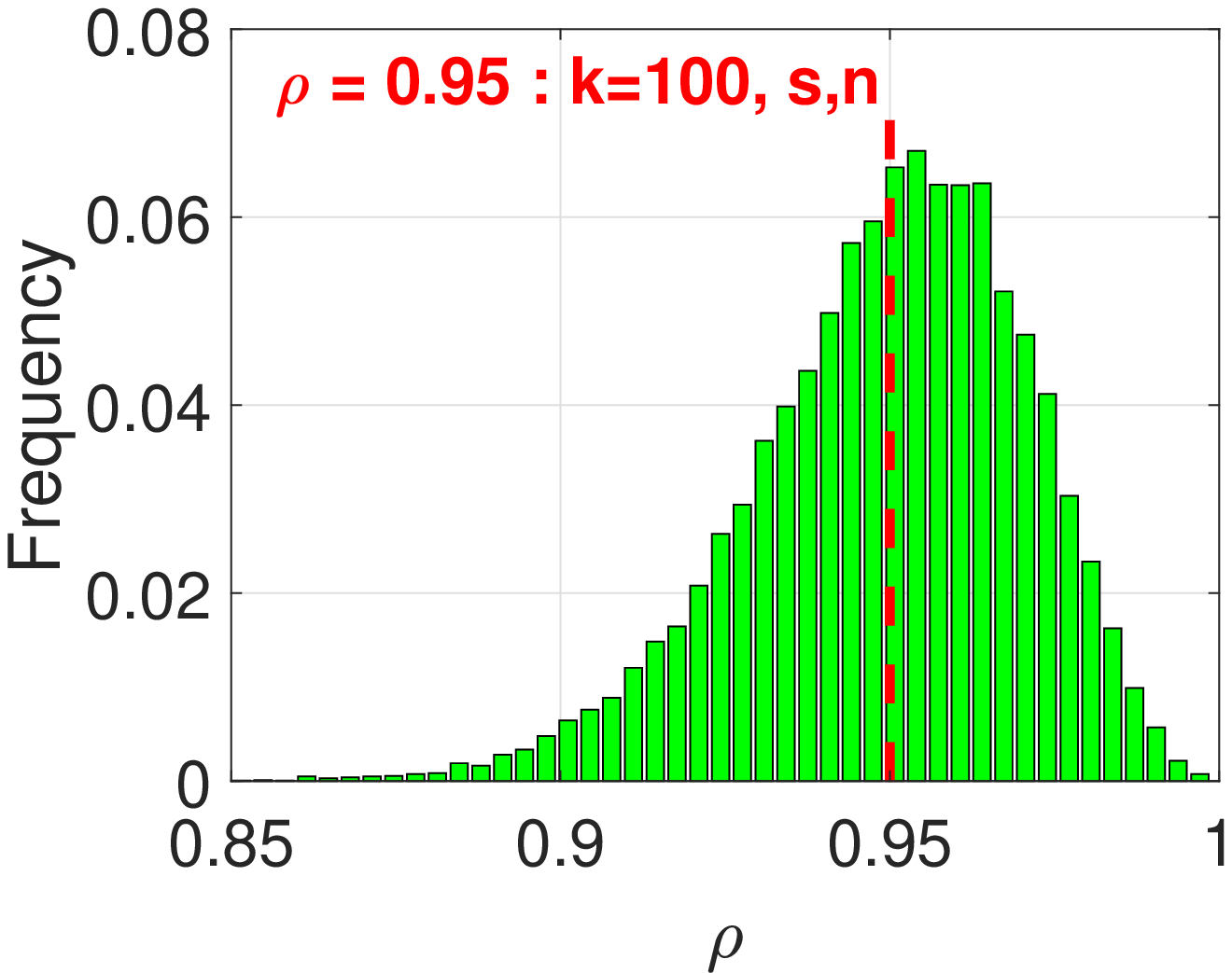}\hspace{-0.11in}
\includegraphics[width=2.25in]{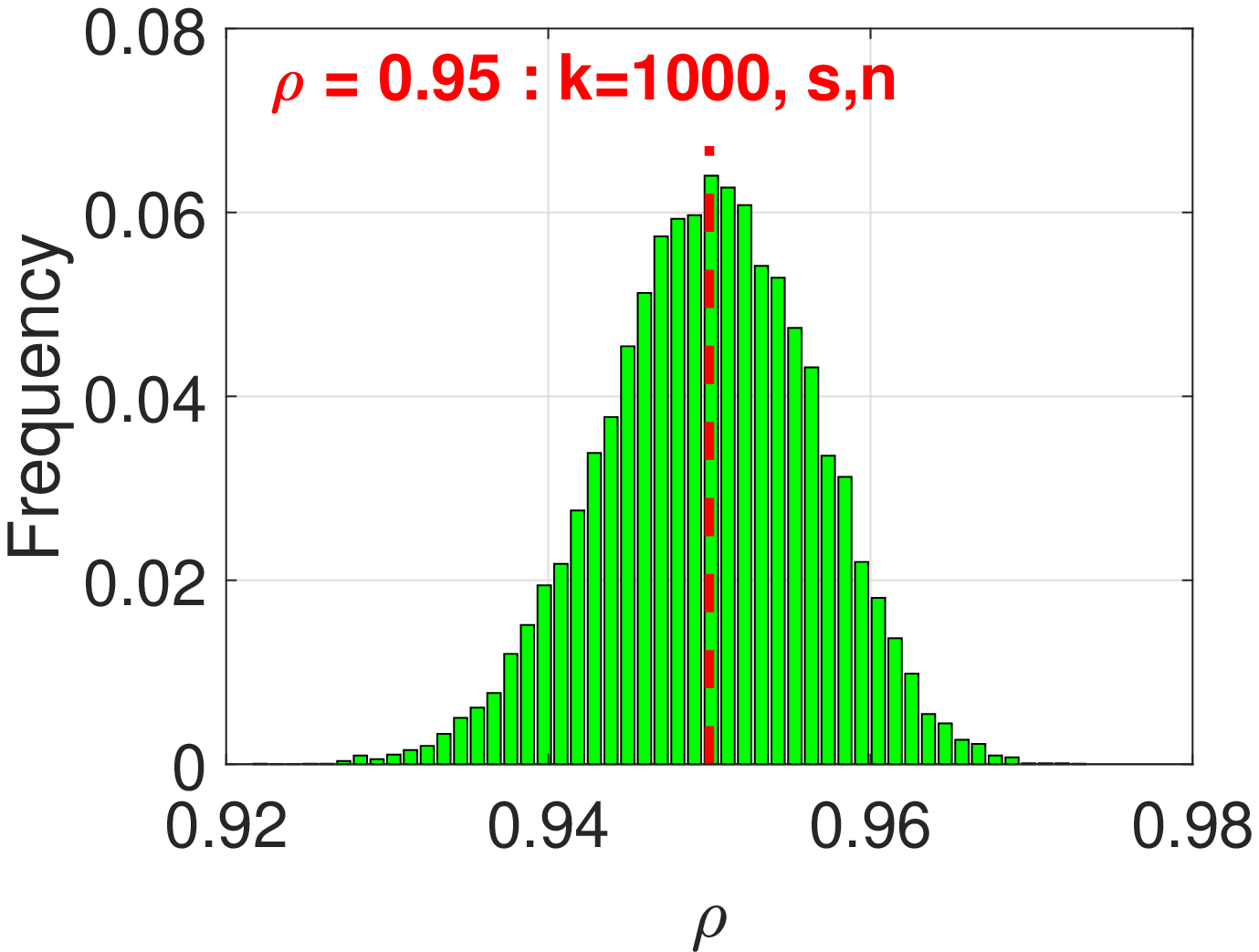}
}

\end{center}
\vspace{-0.2in}
\caption{Histograms of the estimates from five estimators: $\hat{\rho}_1$,  $\hat{\rho}_g$,  $\hat{\rho}_{g,n}$,  $\hat{\rho}_s$,  $\hat{\rho}_{s,n}$ (top to bottom), for $k = 100$ (left panels) and $1000$ (right panels).  $\hat{\rho}_s$ and $\hat{\rho}_{s,n}$ have the desired property that the estimates are smaller than 1. The truth $\rho=0.95$. }\label{fig_Hist_Est}
\end{figure}

Figure~\ref{fig_Hist_Est} provides the histograms of the estimates from five estimators, for $\rho=0.95$ and $k\in\{100,1000\}$.  In addition to showing the expected bell-shaped curves, the histograms reveal that $\hat{\rho}_g$ does not have another desired property that the estimates should be smaller than 1. The normalized version $\hat{\rho}_{g,n}$ helps but it is still not good enough. This figure once again confirms that $\hat{\rho}_{s,n}$ is an overall good estimator.

\newpage\clearpage

\section{An Experimental Study}

\begin{figure}[h!]
\begin{center}
\mbox{
\includegraphics[width=2.25in]{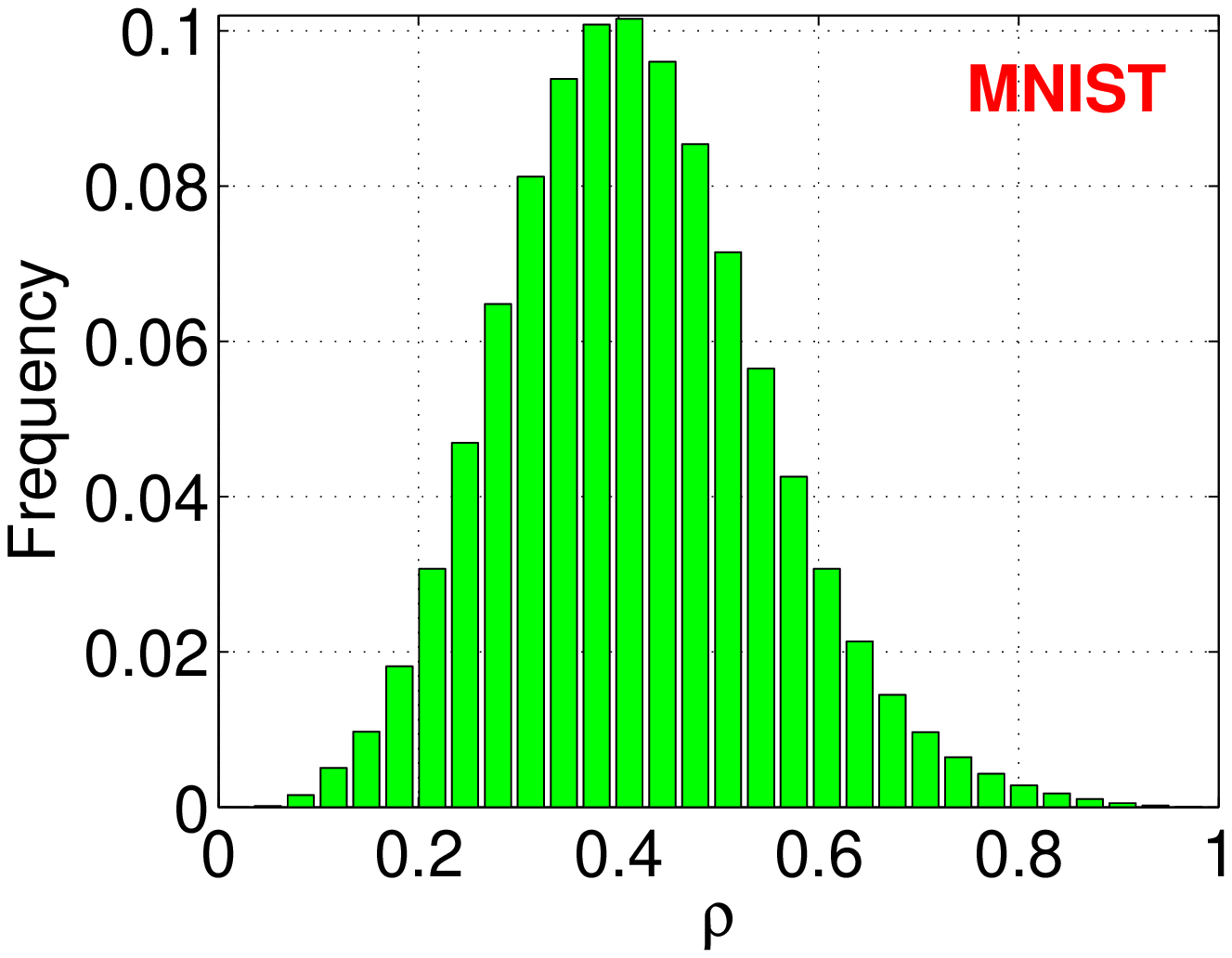}
\includegraphics[width=2.25in]{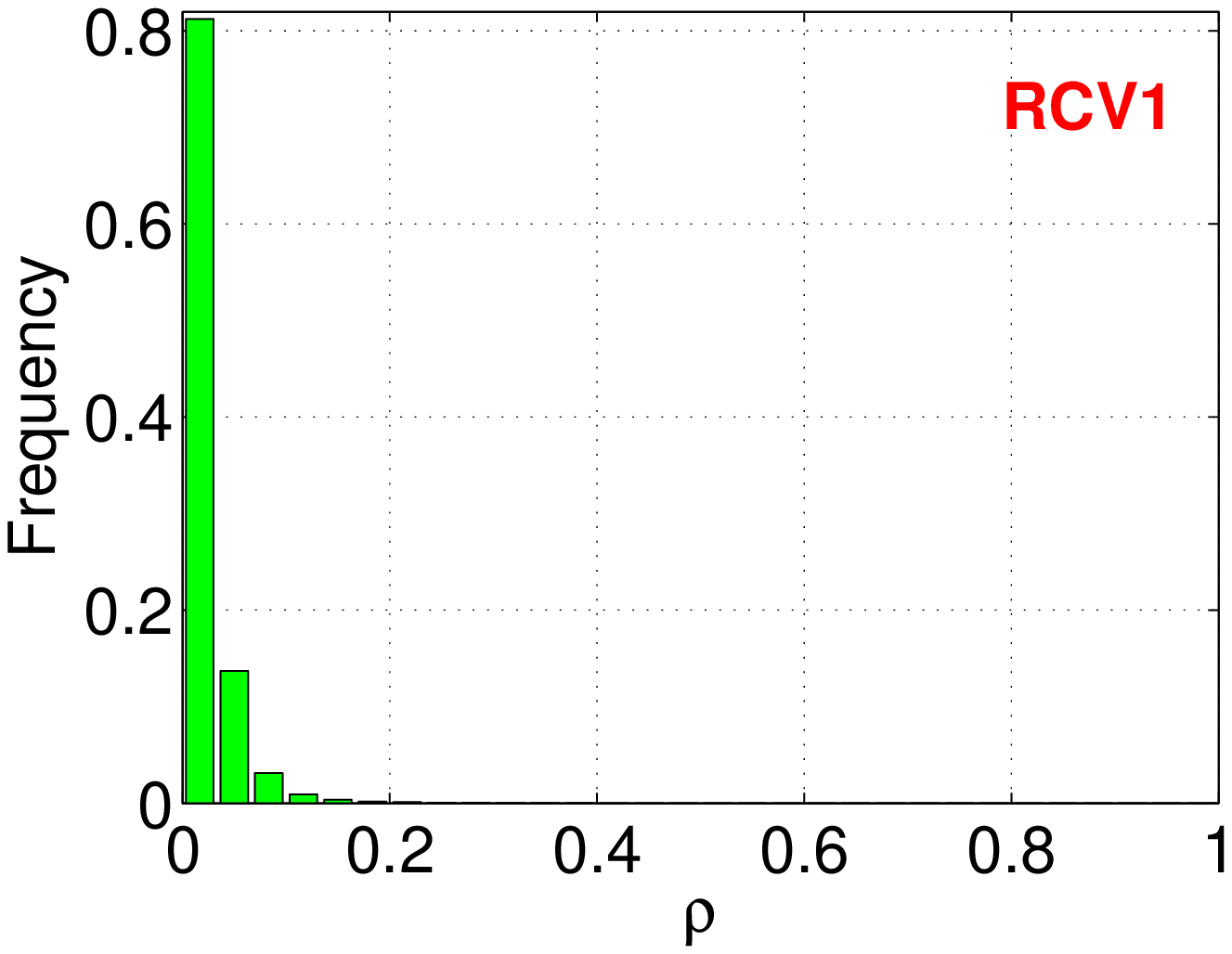}
}
\mbox{
\includegraphics[width=2.25in]{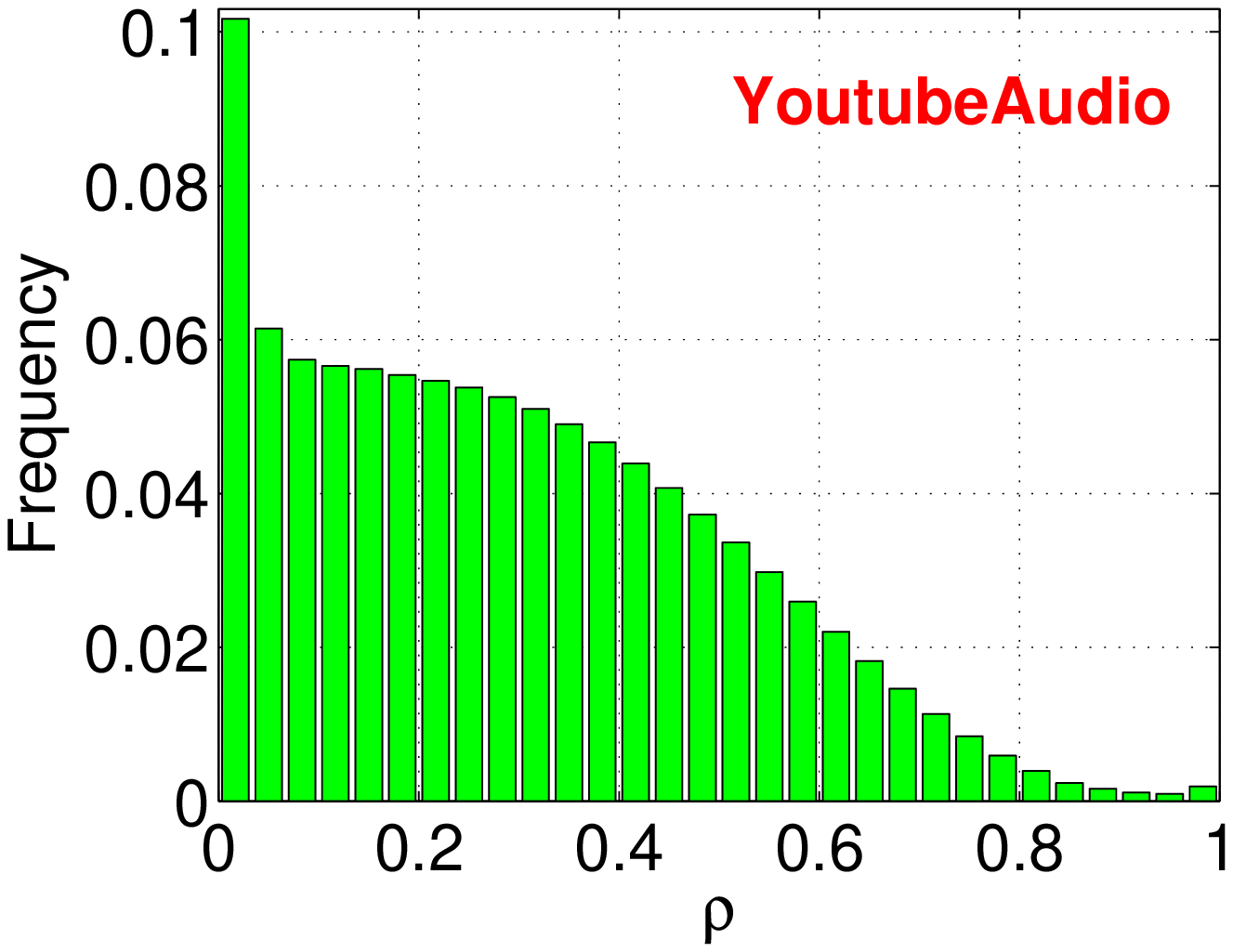}
\includegraphics[width=2.25in]{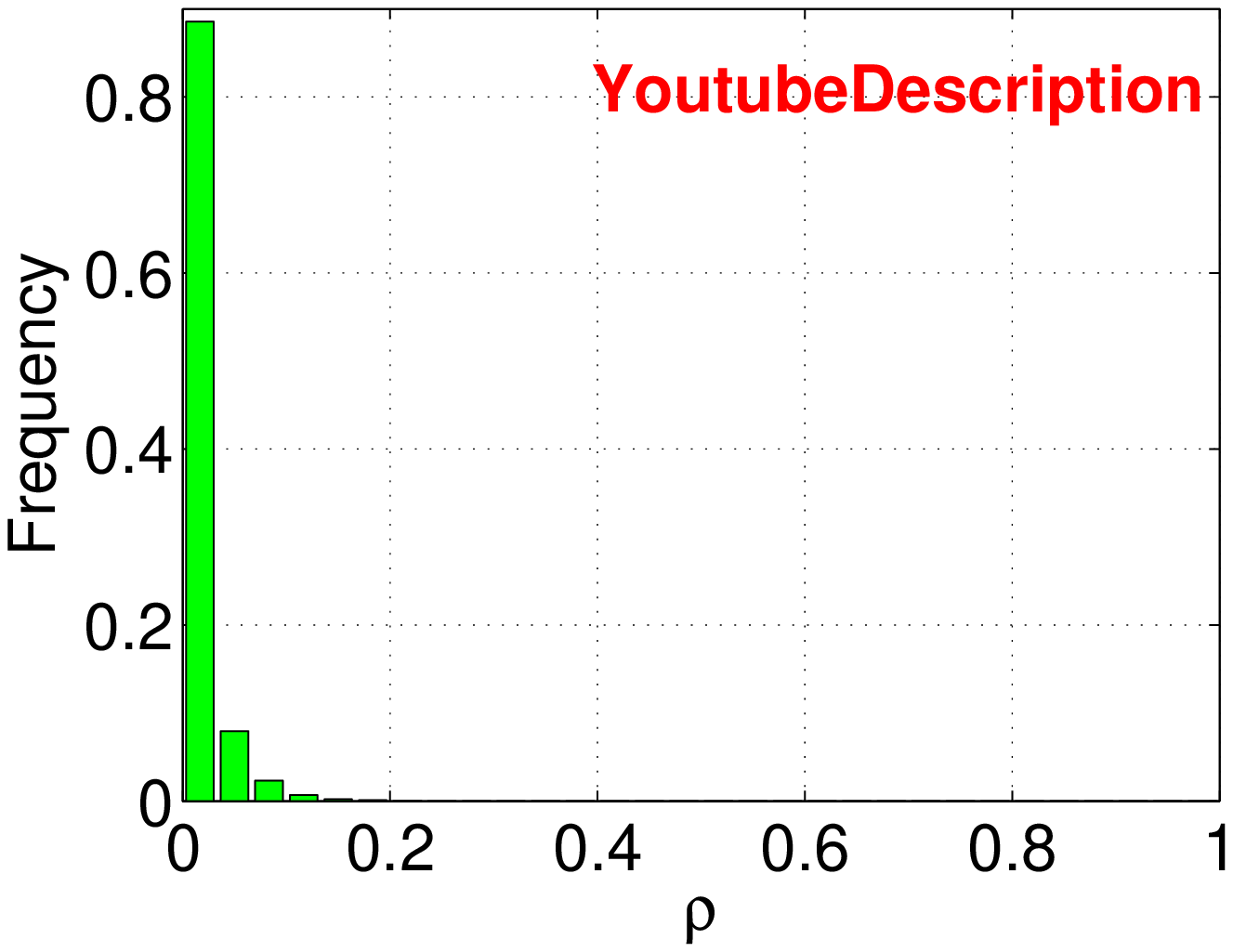}

}
\end{center}
\vspace{-0.2in}
\caption{Histograms of all pairwise $\rho$ values for the 4 datasets.  }\label{fig_hist_data}
\end{figure}

To further verify the theoretical results,  we  conduct an experimental study on the ranking task for near-neighbor search on 4 public datasets (see Table~\ref{tab_data} and Figure~\ref{fig_hist_data}).\vspace{-0.1in}

\begin{table}[h!]
\begin{center}
\caption{Information about the datasets}\vspace{0.03in}
\begin{tabular}{crrr} \hline
Dataset & \# Train &\# Query  & \# Dim\\ \hline
MNIST & 10,000 &10,000& 780\\
RCV1 & 10,000 &10,000& 47,236\\
YoutubeAudio & 10,000 &11,930 &2,000\\
YoutubeDescription & 10,000 &11,743 &12,183,626
\\ \hline
\end{tabular}\label{tab_data}
\end{center}\vspace{-0.1in}
\end{table}

These four datasets are downloaded from either the UCI repository or the LIBSVM website. When a dataset  contains significantly more than 10,000 training samples, we only use a random sample of it. The datasets represent  a wide range of application scenarios and data types. See Figure~\ref{fig_hist_data} for the frequencies of all pairwise  $\rho$ values.

For each data point in the query set, we estimate its similarity with every data point in the training set, using random projections. The goal is to return training data points with which the estimated similarities are larger than a pre-specified threshold $\rho_0$. For each query point, we rank all the (estimated) similarities and return top-$L$ points. We can then compute the precision and recall
\begin{align}\notag
&Precision = \frac{\# \text{ retrieved points with true similarities }\geq \rho_0}{L},\\\notag
&Recall = \frac{\# \text{ retrieved points with true similarities }\geq \rho_0}{\# \text{total points with true similarities }\geq \rho_0}
\end{align}
We report the averaged precision-recall values over all query data points. By varying $L$ from 1 to the number of training data points, we obtain a precision-recall curve. Therefore, for each $\rho_0$ and $k$, and each estimator ($\hat{\rho}_1$, $\hat{\rho}_{s,n}$, or $\hat{\rho}_{g,n}$), we report one precision-recall curve.\\

\begin{figure*}[h!]
\begin{center}
\mbox{
\includegraphics[width=2.2in]{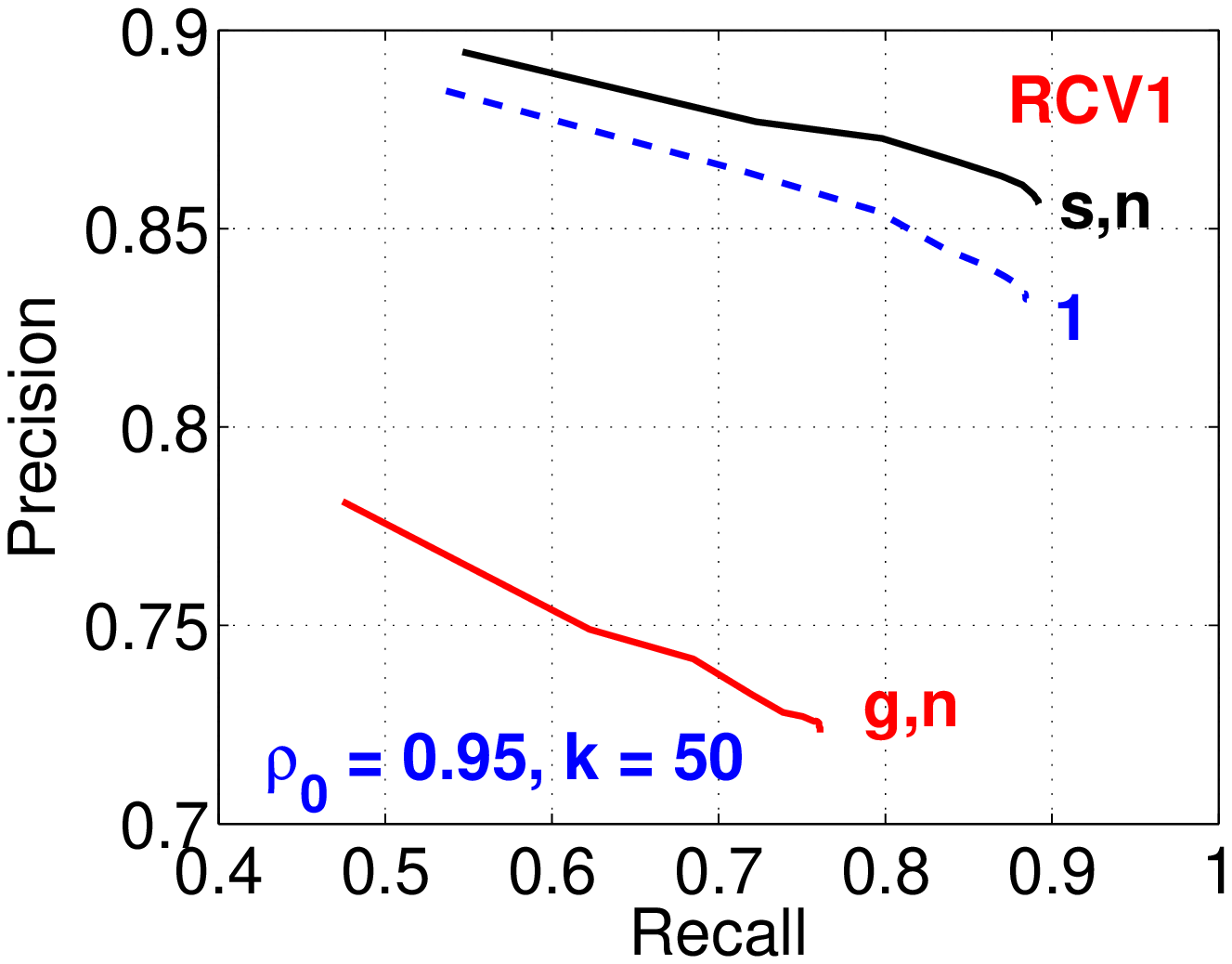}\hspace{-0.15in}
\includegraphics[width=2.2in]{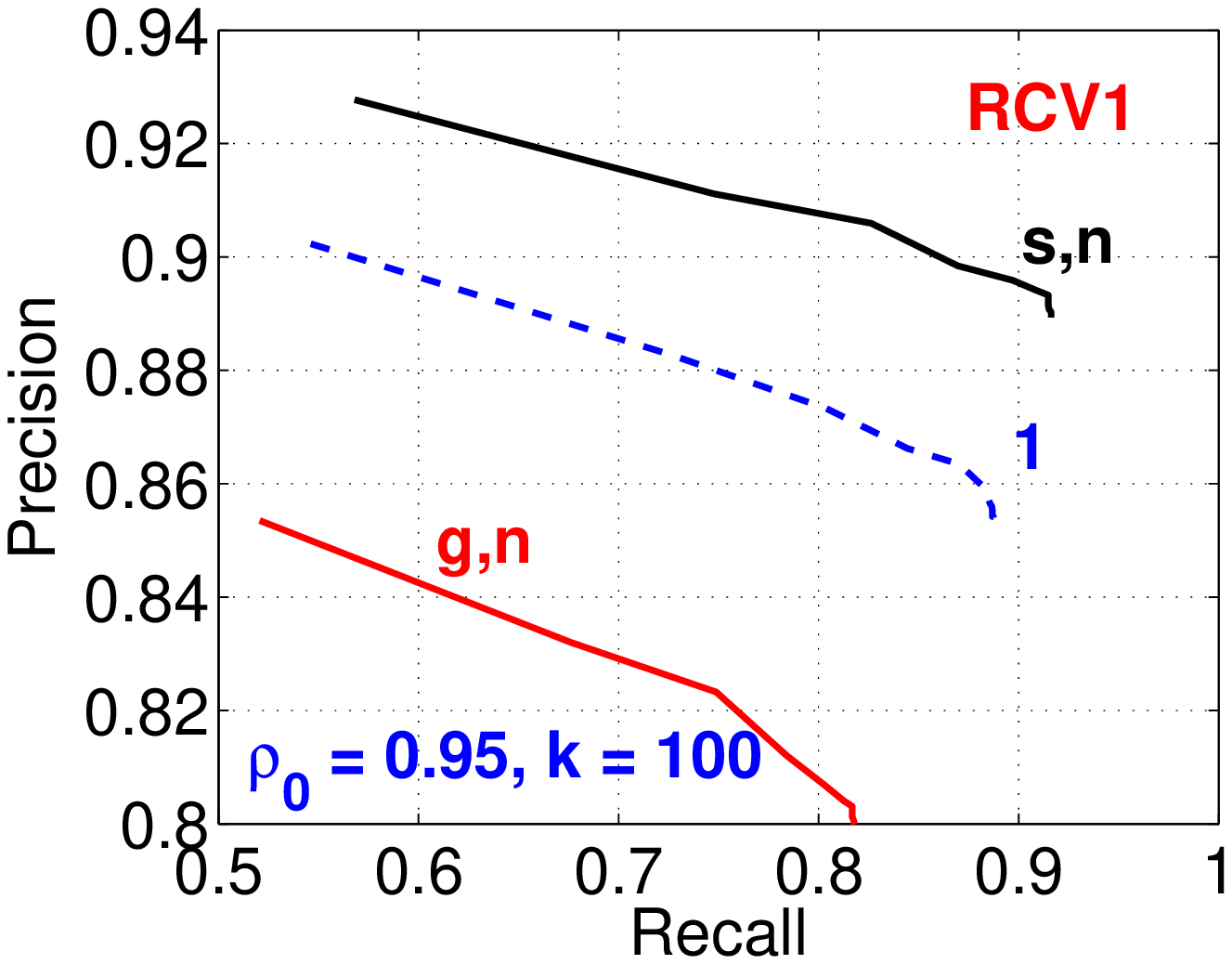}\hspace{-0.15in}
\includegraphics[width=2.2in]{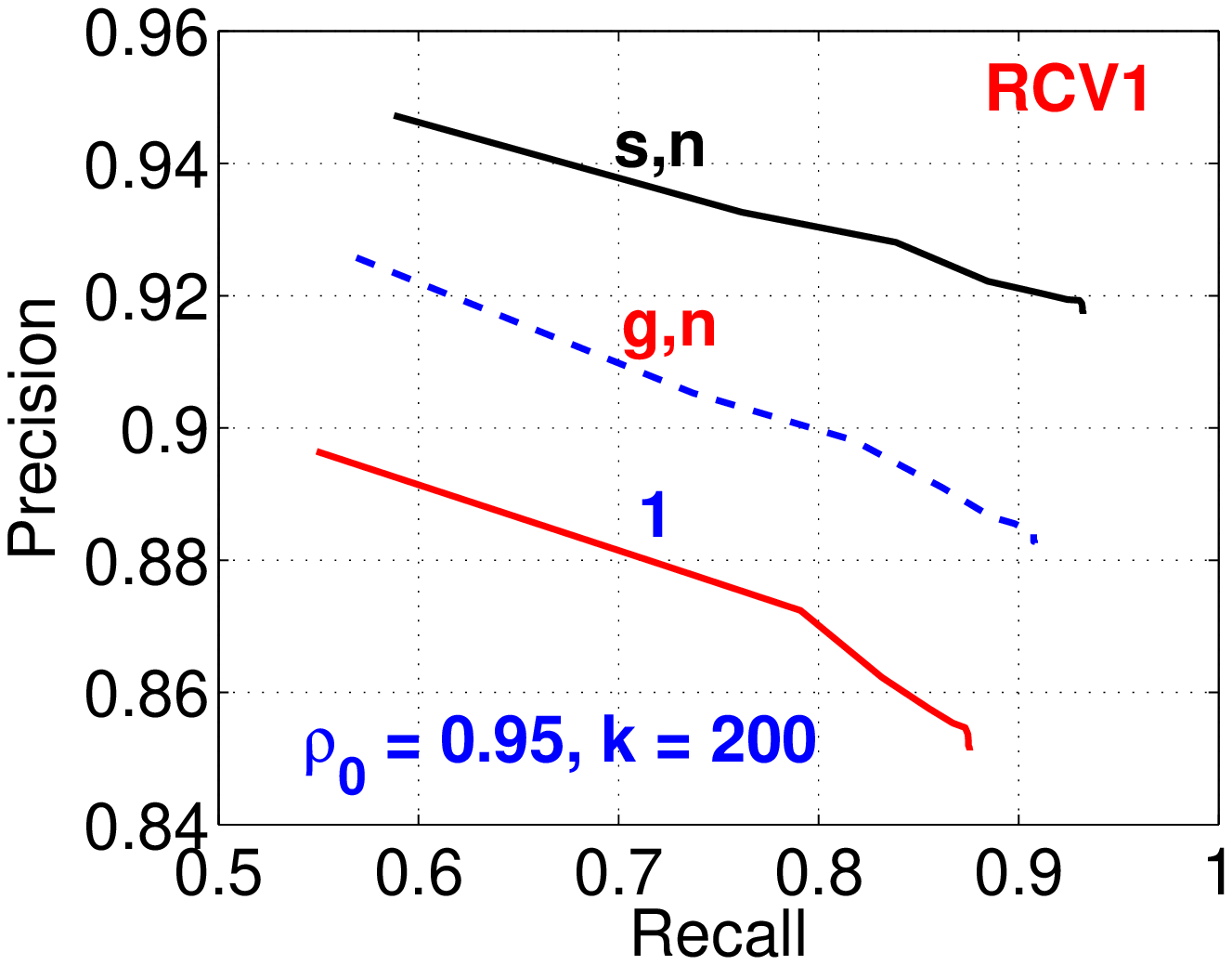}\hspace{-0.15in}
}

\mbox{
\includegraphics[width=2.2in]{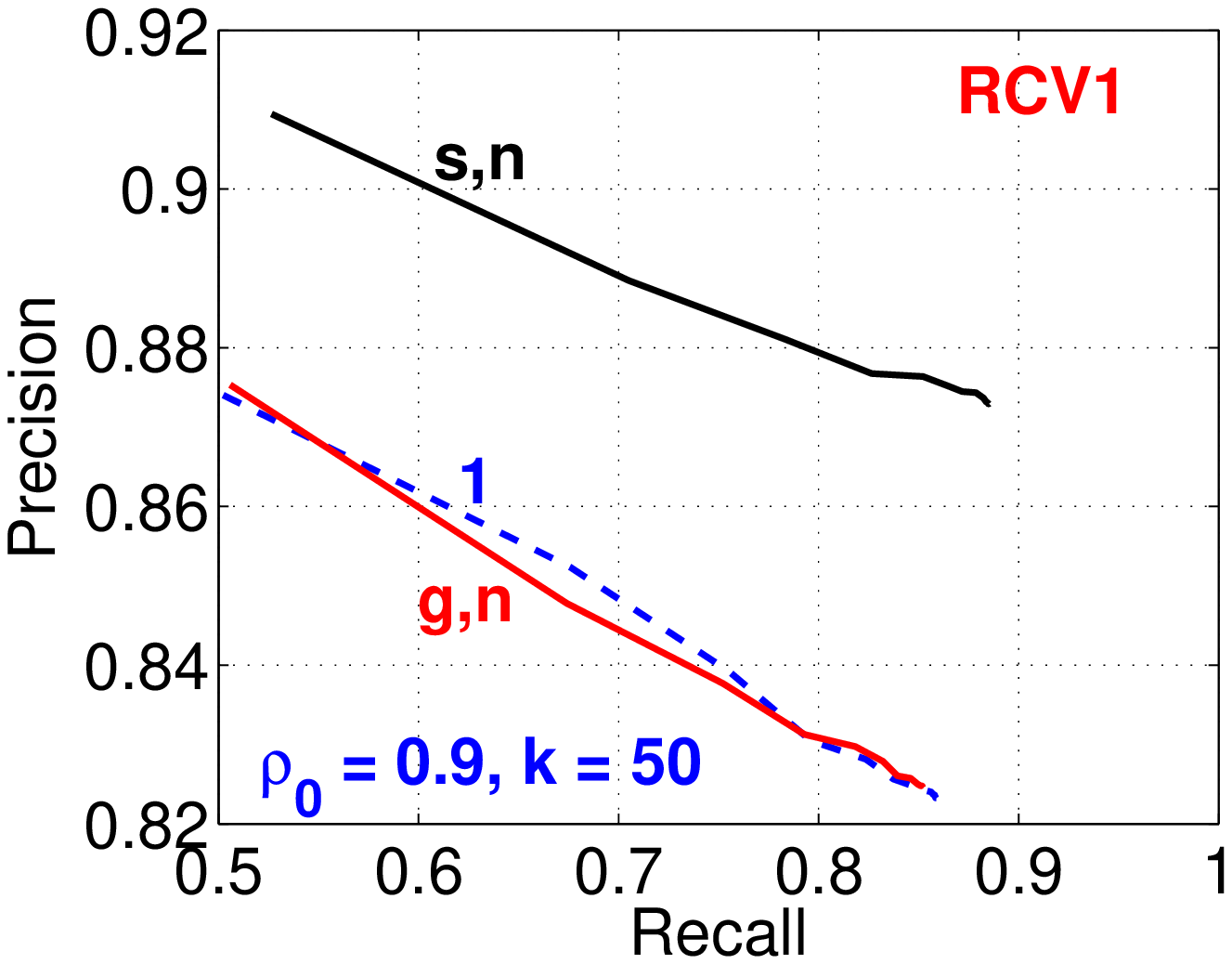}\hspace{-0.15in}
\includegraphics[width=2.2in]{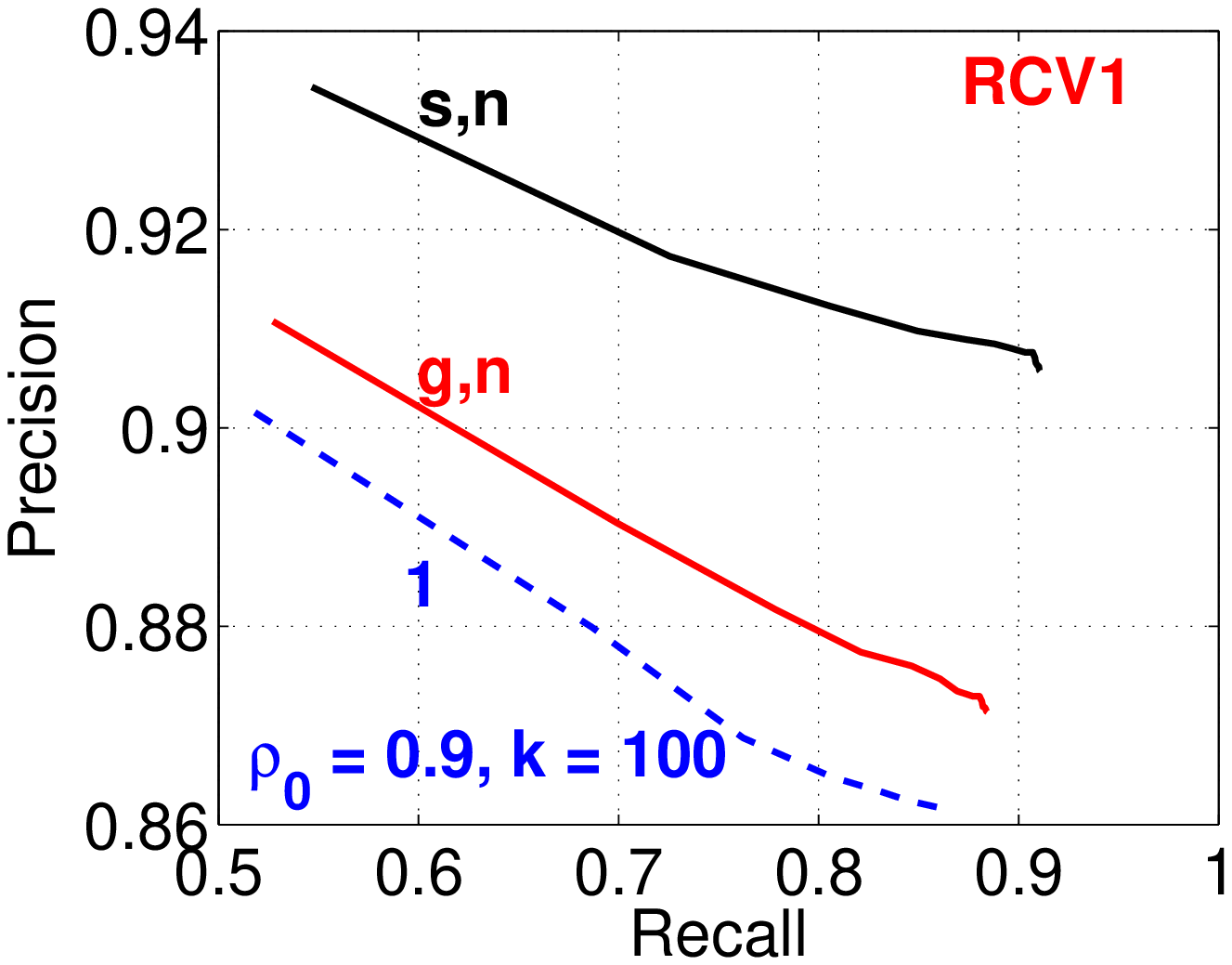}\hspace{-0.15in}
\includegraphics[width=2.2in]{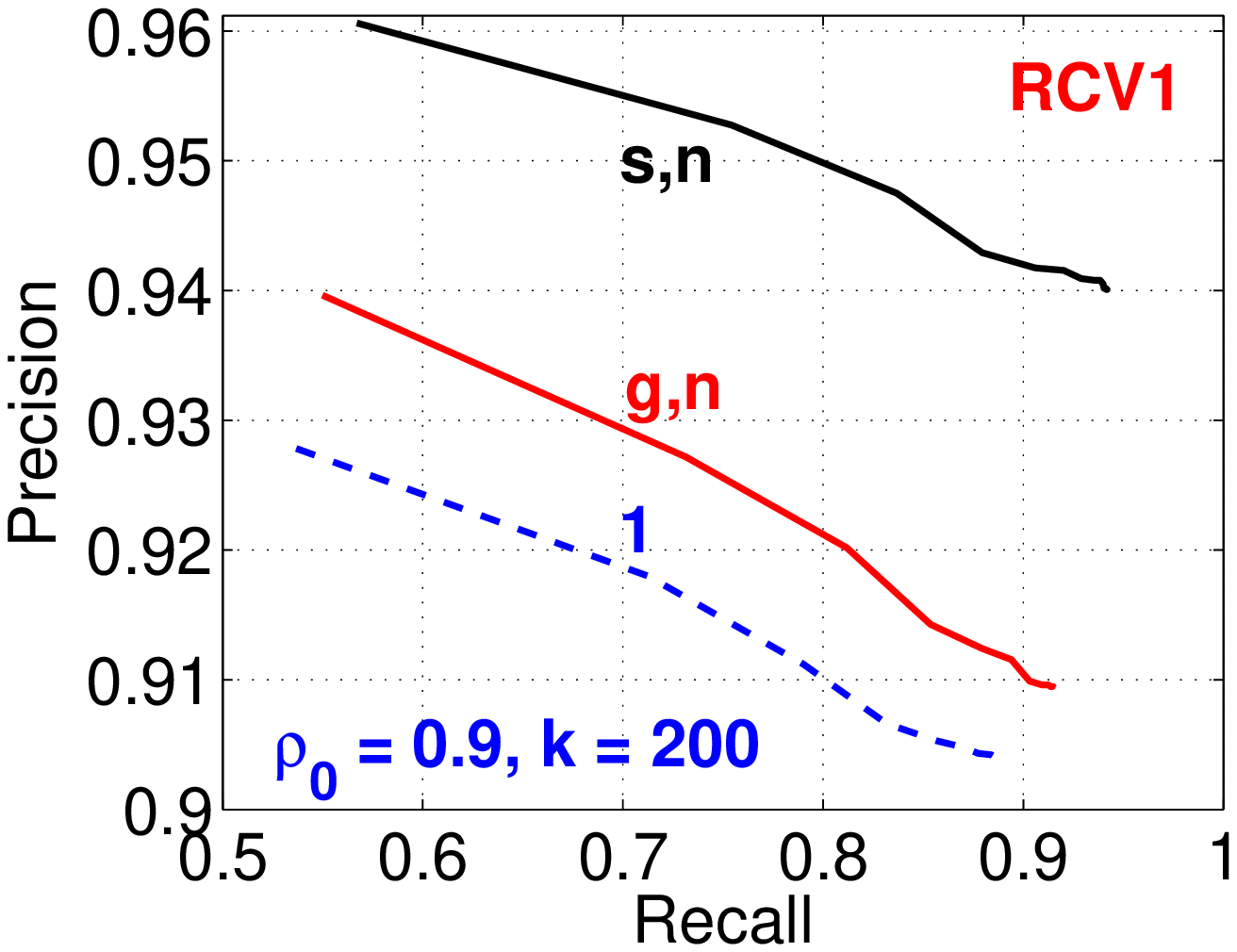}\hspace{-0.15in}
}

\mbox{
\includegraphics[width=2.2in]{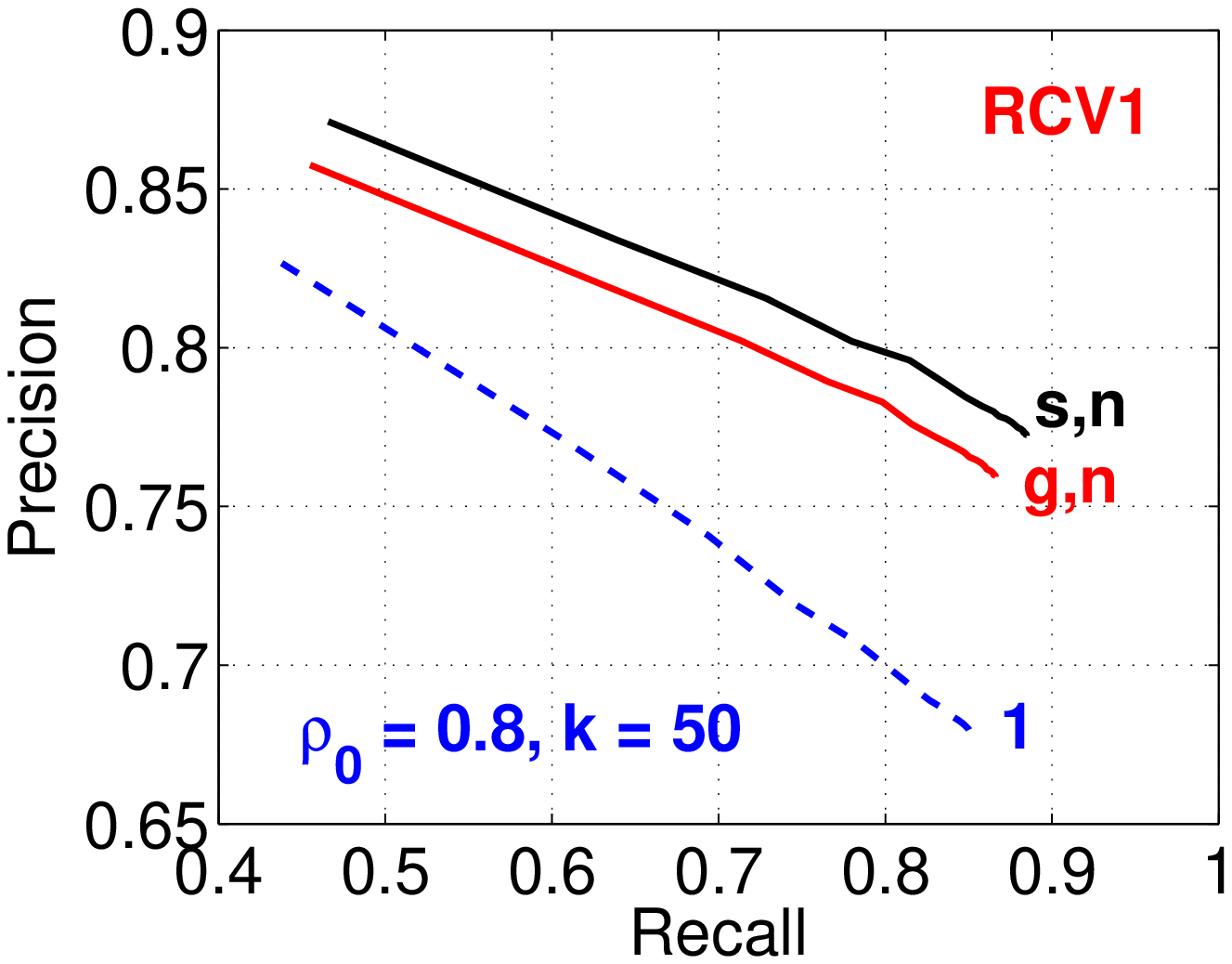}\hspace{-0.15in}
\includegraphics[width=2.2in]{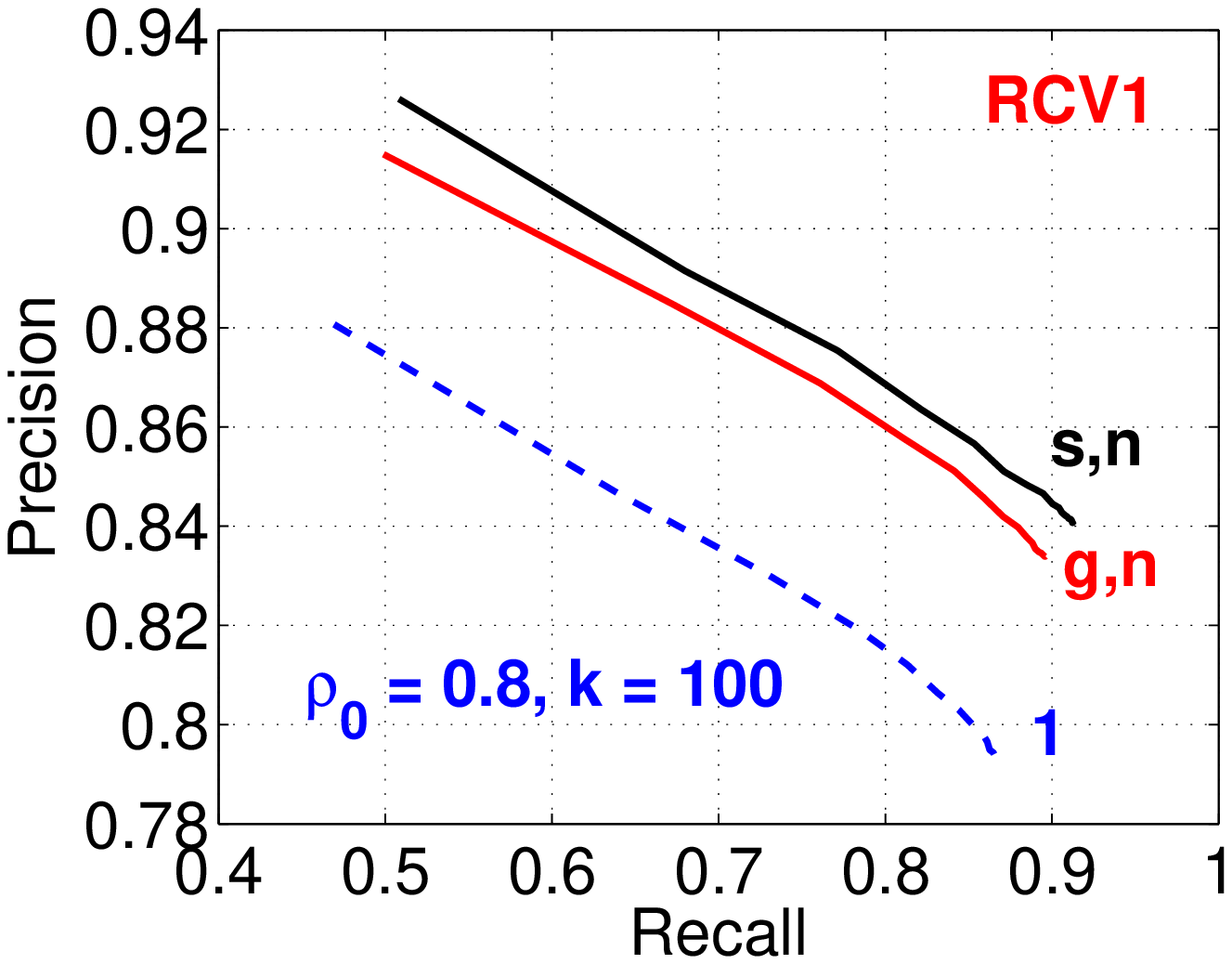}\hspace{-0.15in}
\includegraphics[width=2.2in]{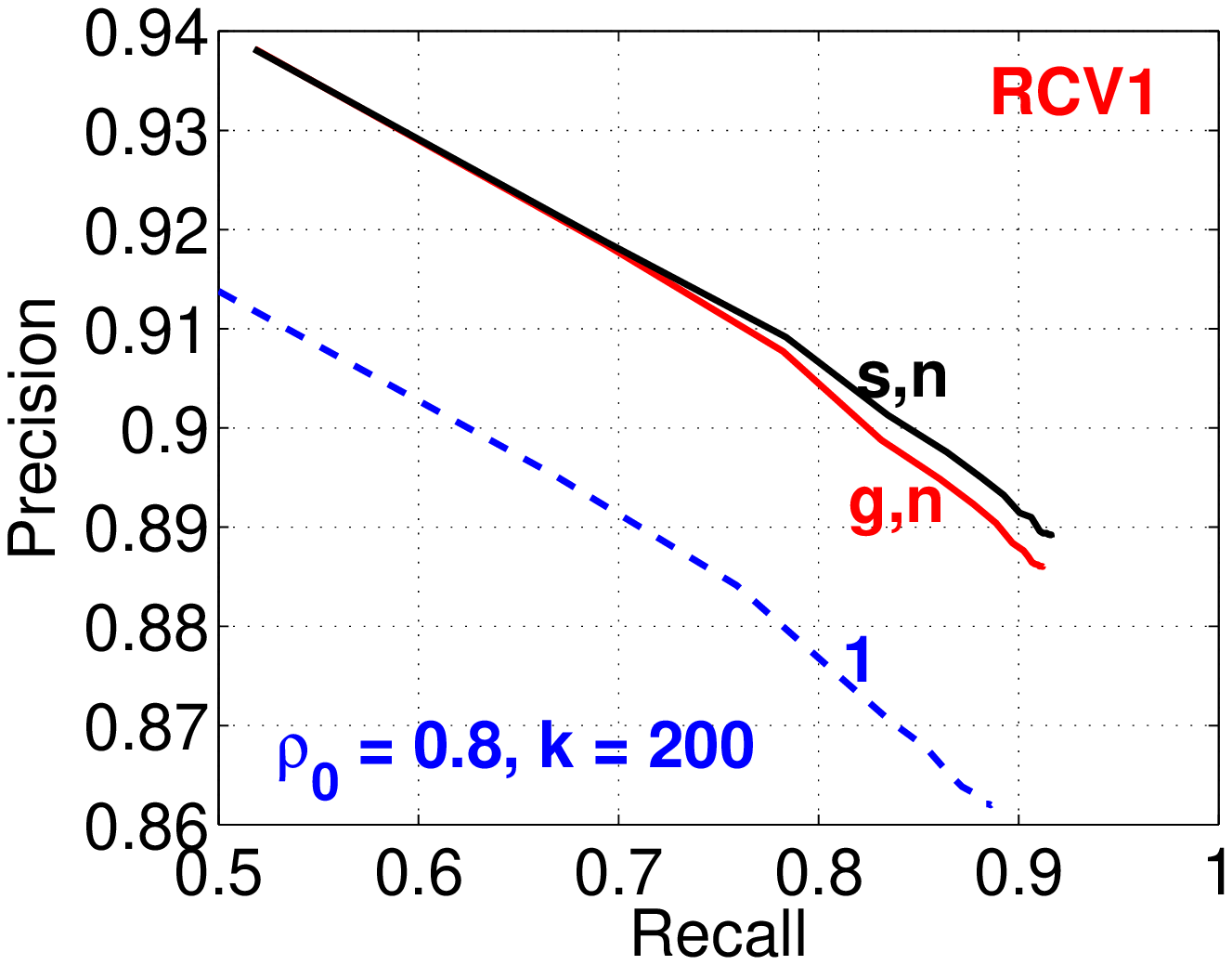}\hspace{-0.15in}
}

\mbox{
\includegraphics[width=2.2in]{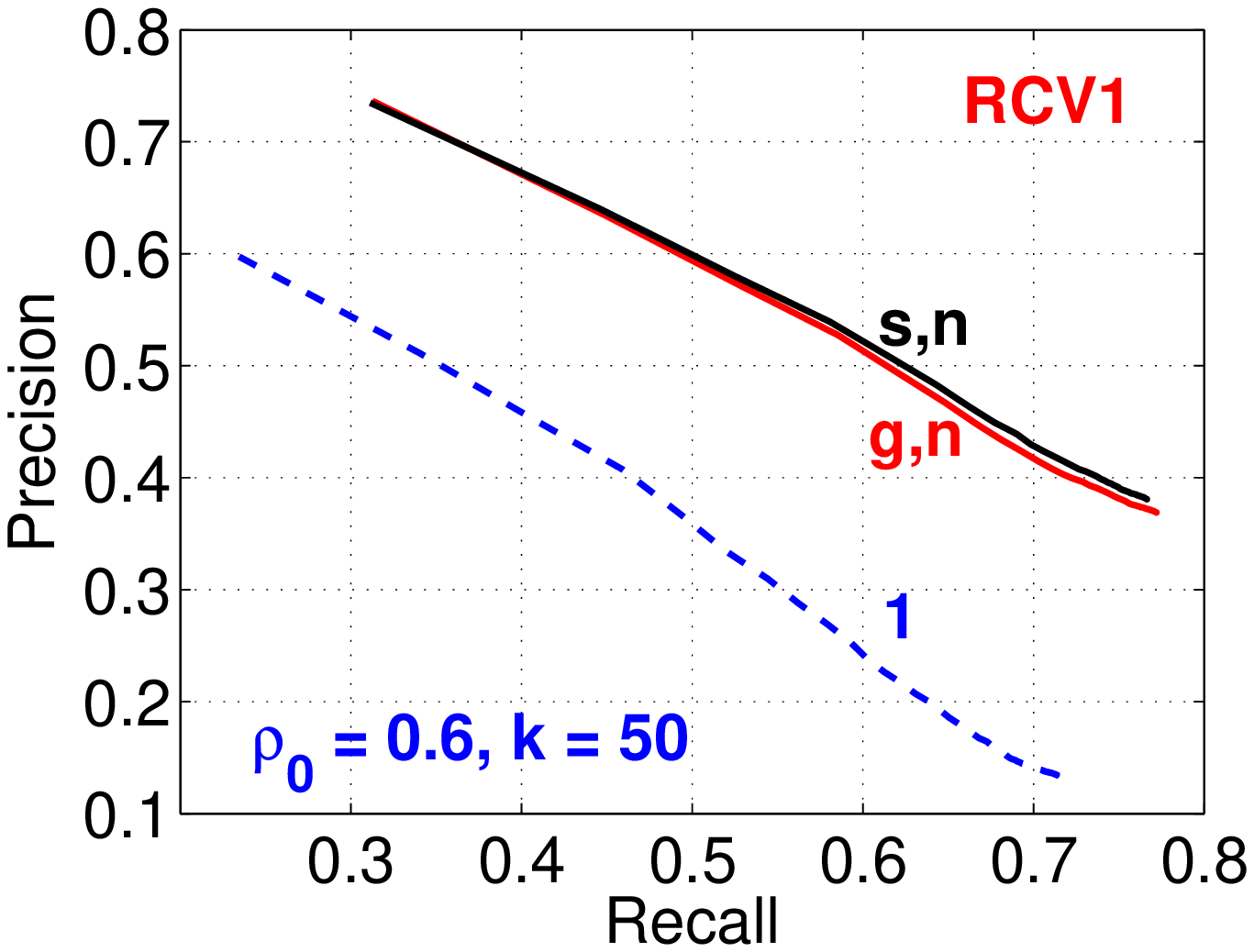}\hspace{-0.15in}
\includegraphics[width=2.2in]{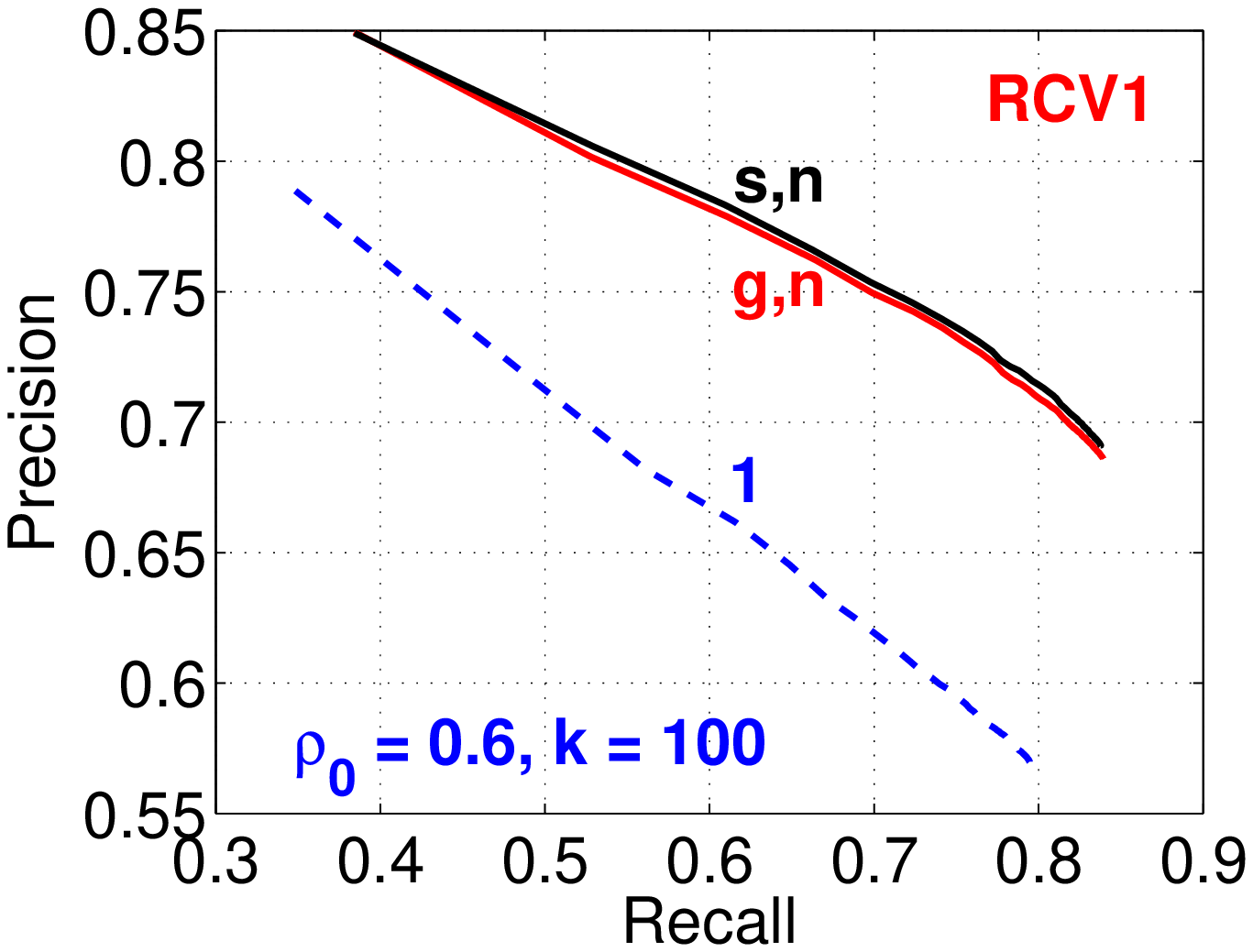}\hspace{-0.15in}
\includegraphics[width=2.2in]{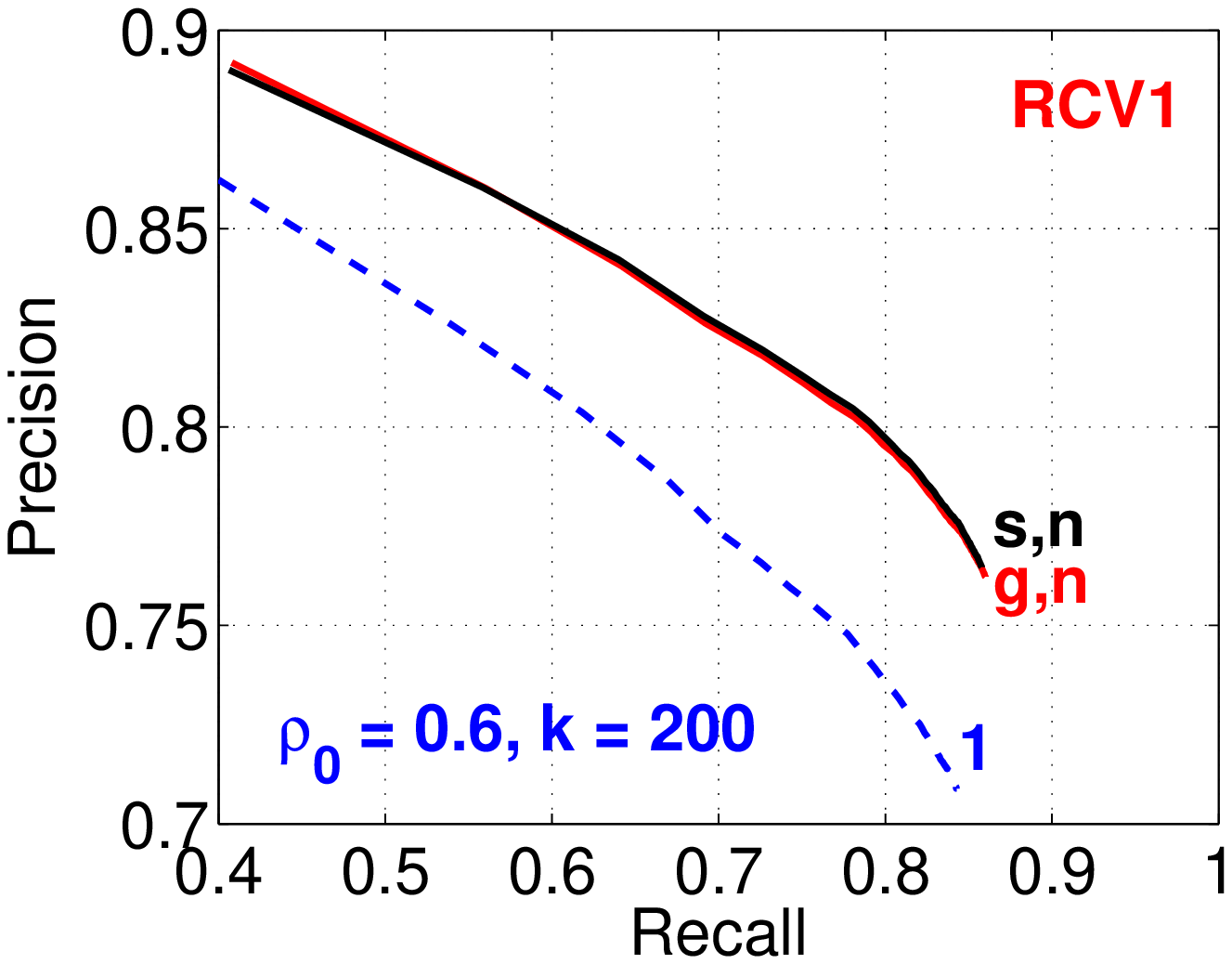}\hspace{-0.15in}
}

\mbox{
\includegraphics[width=2.2in]{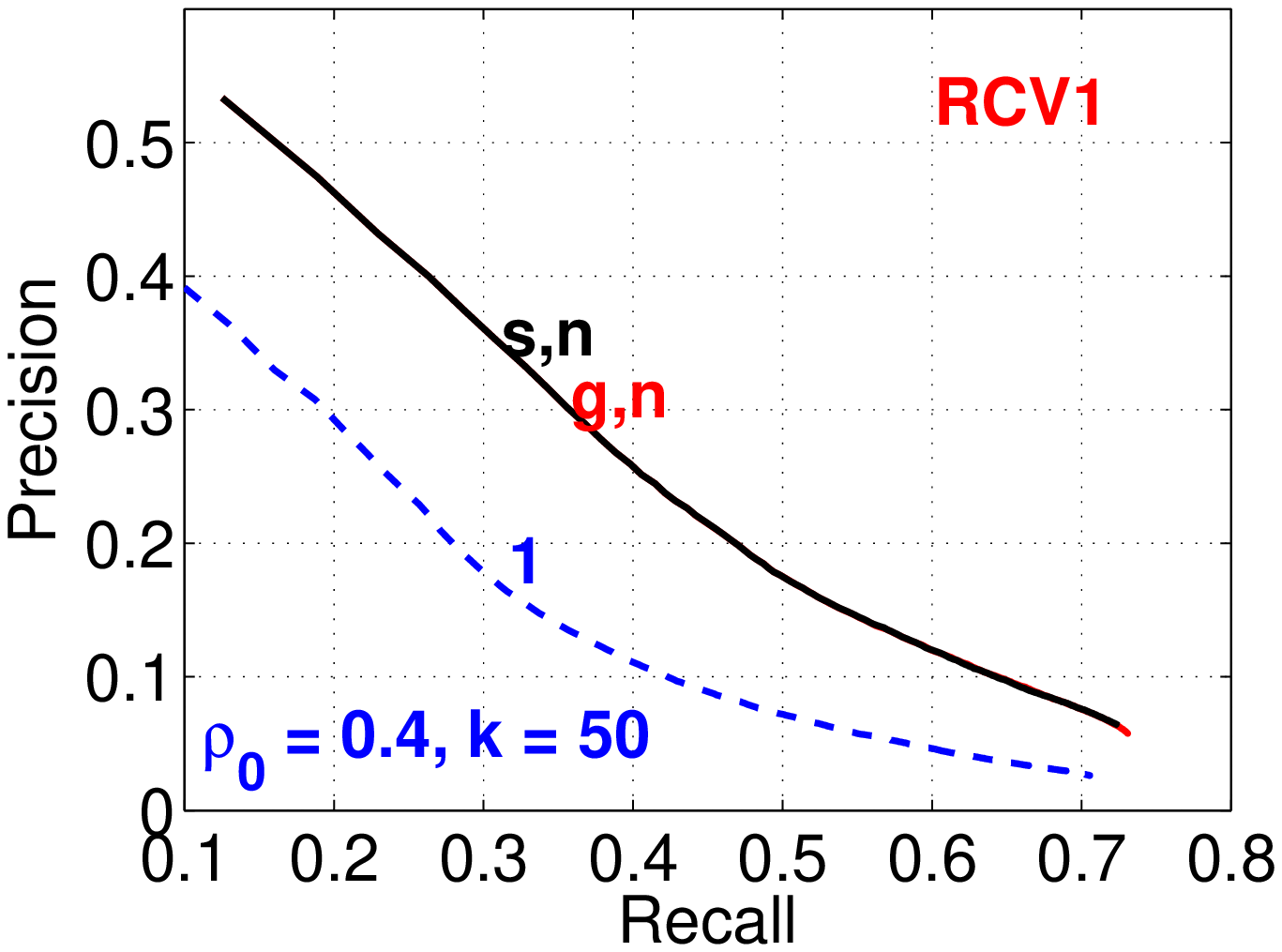}\hspace{-0.15in}
\includegraphics[width=2.2in]{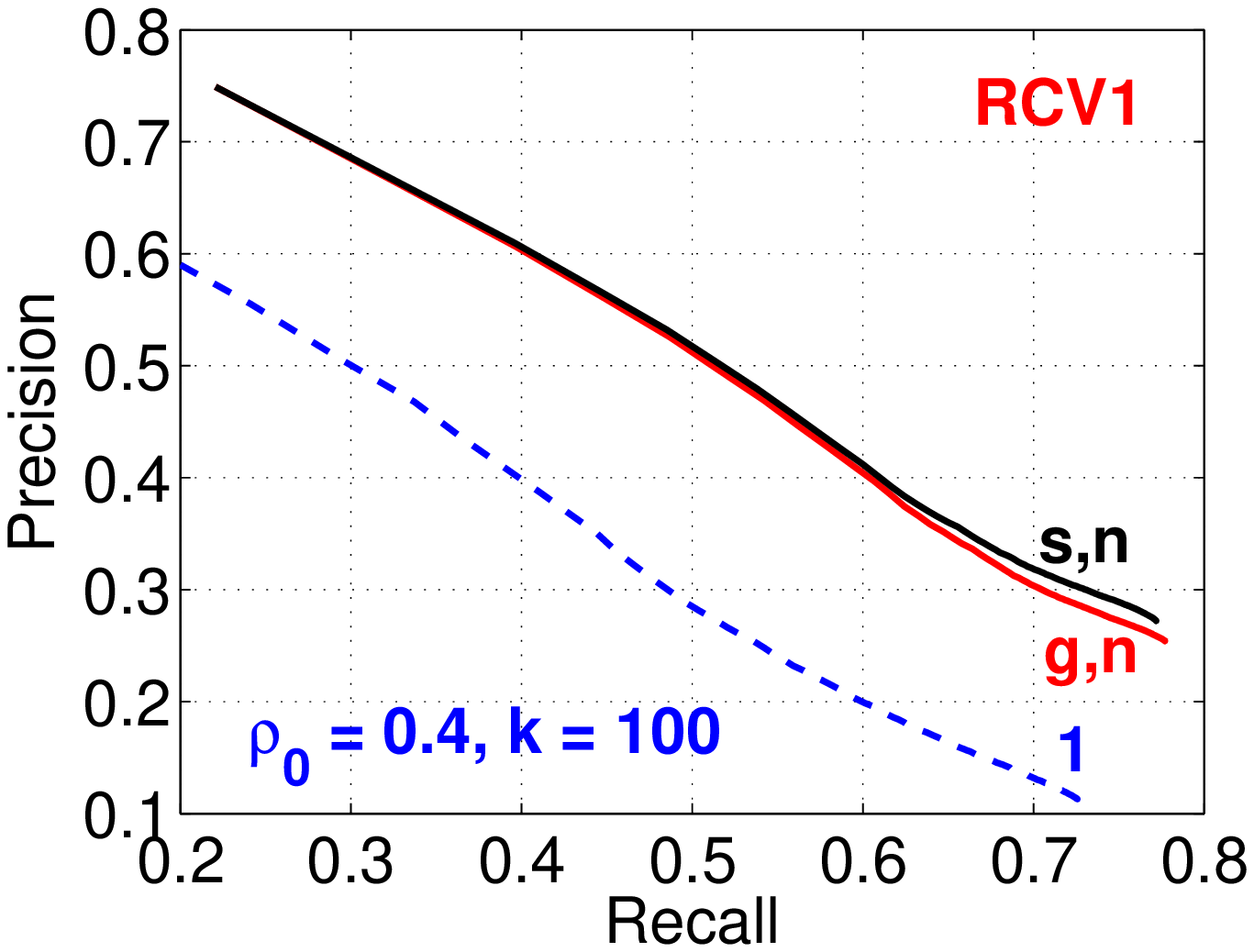}\hspace{-0.15in}
\includegraphics[width=2.2in]{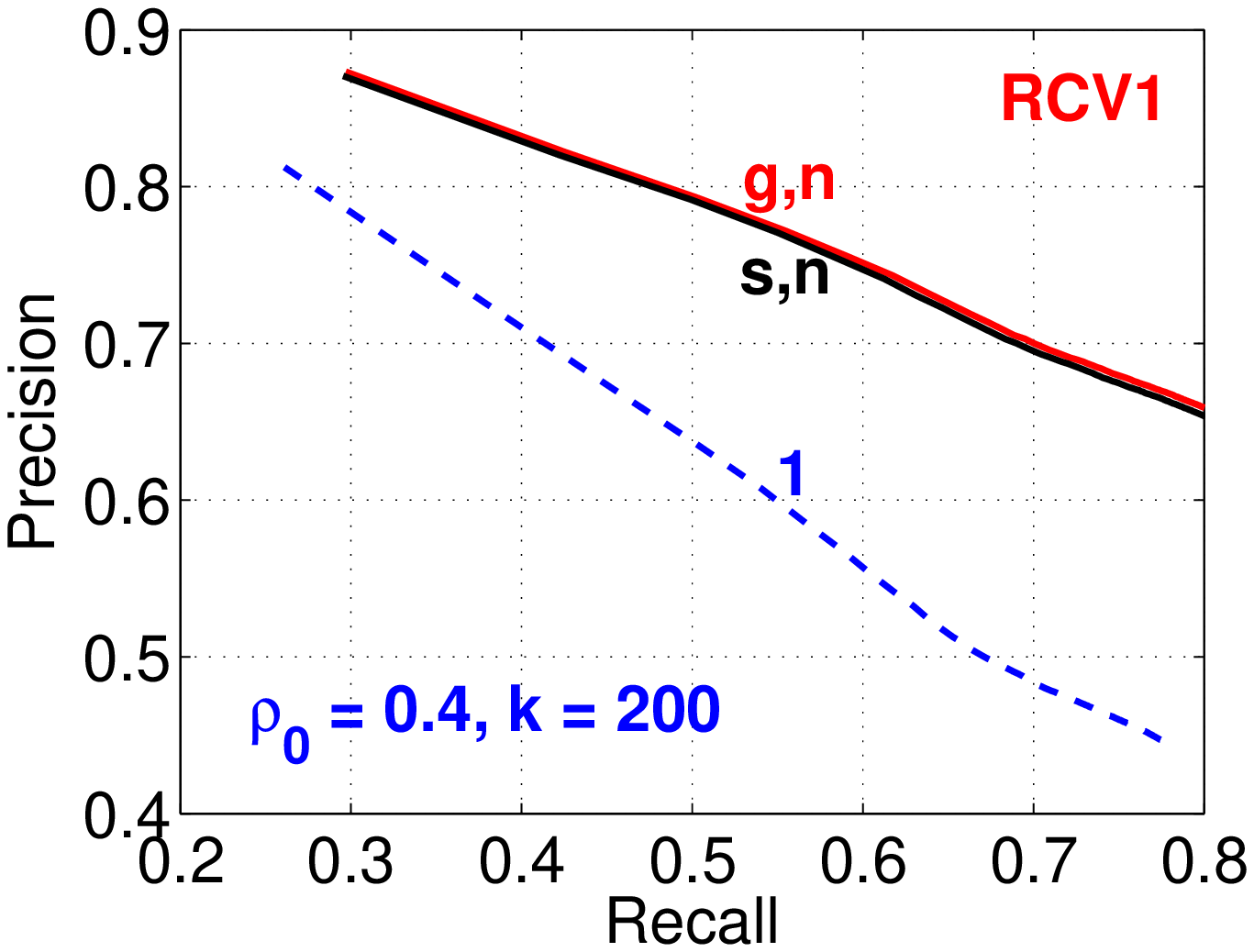}\hspace{-0.15in}
}

\end{center}
\vspace{-0.2in}
\caption{\textbf{RCV1}: precision-recall curves for selected $\rho_0$ and $k$ values, and  for three estimators： $\hat{\rho}_{s,n}$ (recommended), $\hat{\rho}_{g,n}$, $\hat{\rho}_1$.}\label{fig_pr_RCV1}
\end{figure*}

Figure~\ref{fig_pr_RCV1}  presents the results for the RCV1 datasets, for $\rho_0 = 0.95, 0.9, 0.8, 0.6, 0.4$ (top to bottom, one $\rho_0$ per row), and for $k = 50, 100, 200$ (left to right, one $k$ per column). In the first row (i.e., $\rho_0 = 0.95$), we can see that $\hat{\rho}_{s,n}$ is substantially more accurate than both $\hat{\rho}_{1}$ and $\hat{\rho}_{g,n}$. Since this case represents the high-similarity region, as expected, $\hat{\rho}_{g,n}$ performs poorly. When $\rho_0\leq 0.6$, $\hat{\rho}_{s,n}$ and $\hat{\rho}_{g,n}$ are essentially identical and substantially better than $\hat{\rho}_1$, also as expected.\\

\begin{figure*}[h!]
\begin{center}
\mbox{
\includegraphics[width=2.2in]{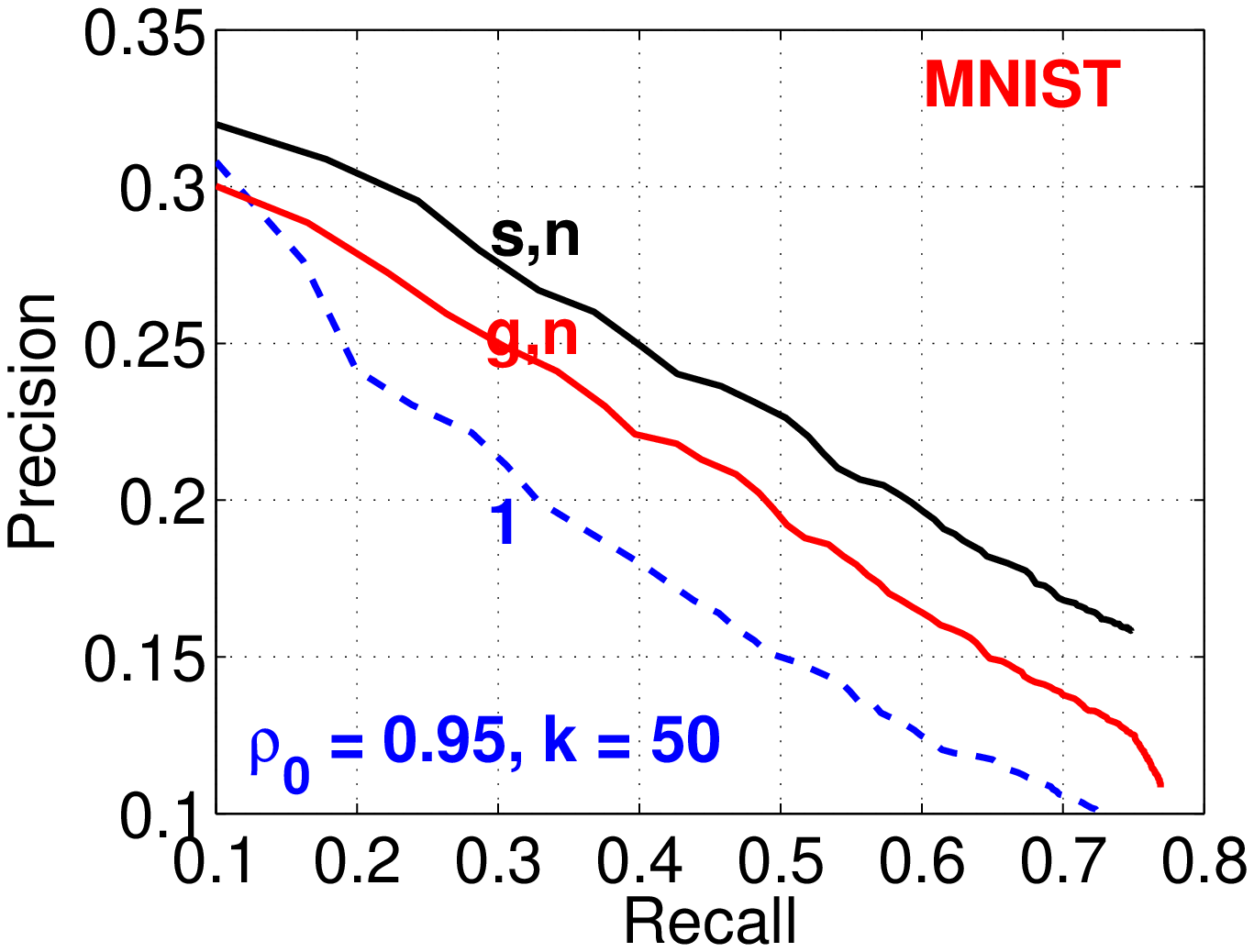}\hspace{-0.15in}
\includegraphics[width=2.2in]{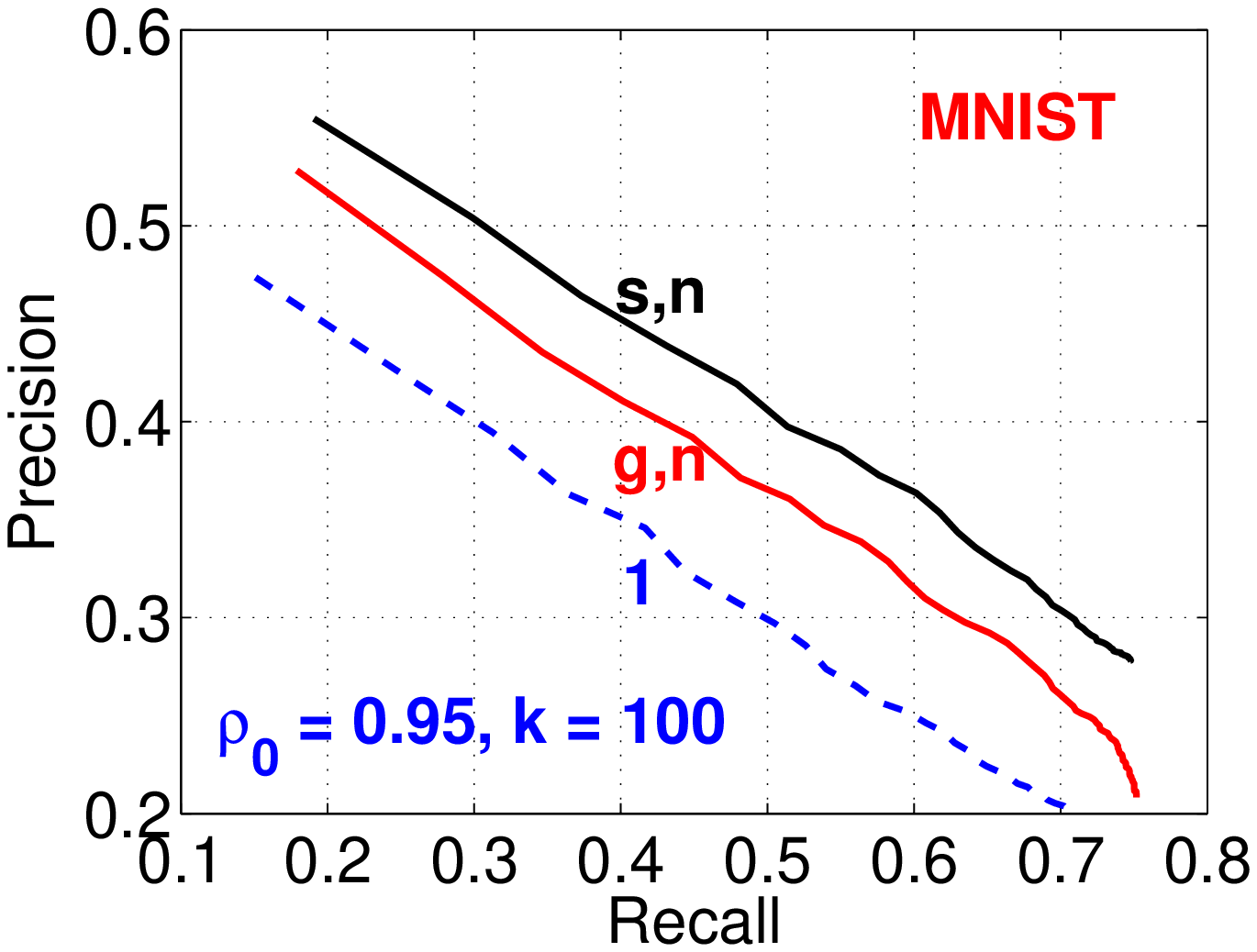}\hspace{-0.15in}
\includegraphics[width=2.2in]{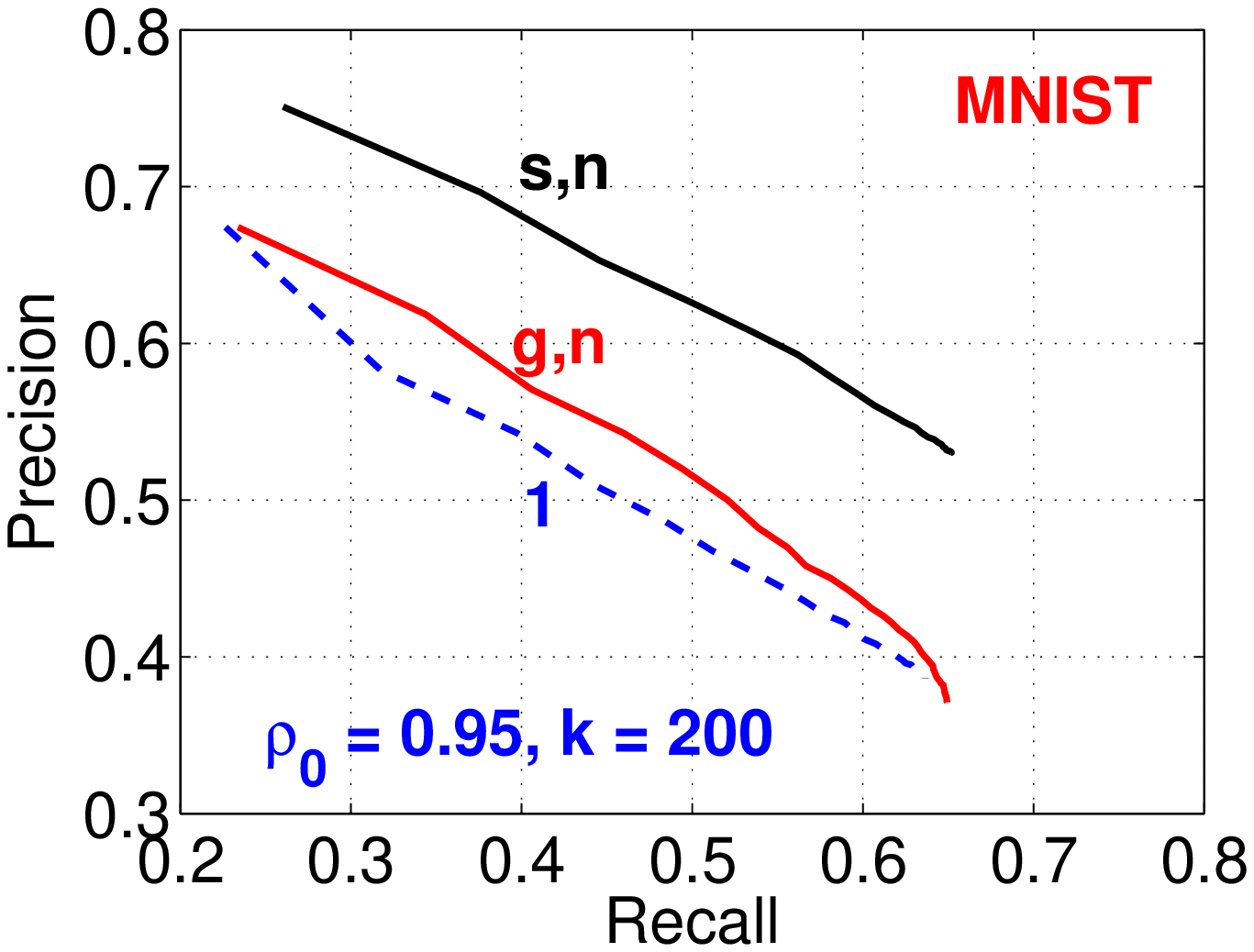}\hspace{-0.15in}
}

\mbox{
\includegraphics[width=2.2in]{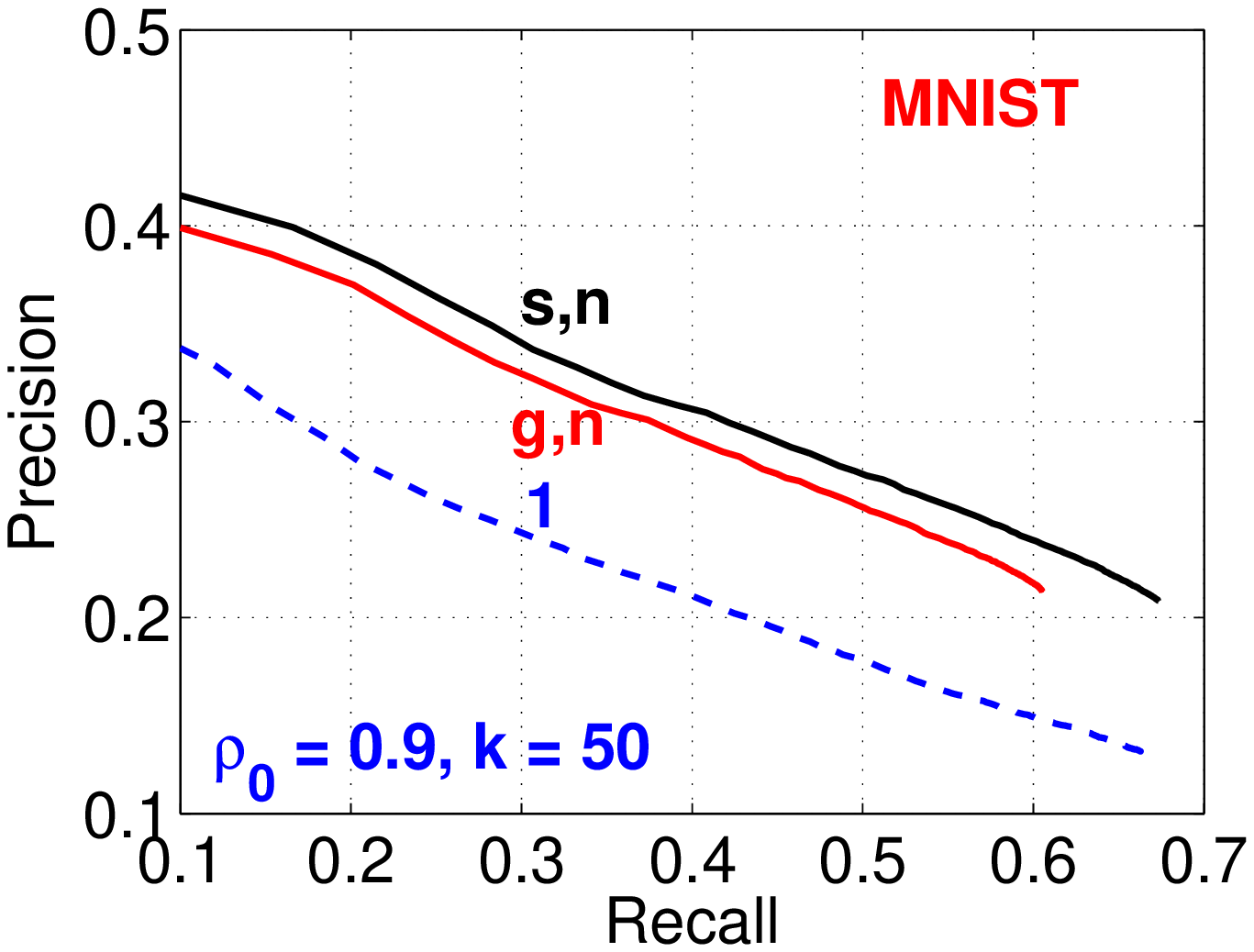}\hspace{-0.15in}
\includegraphics[width=2.2in]{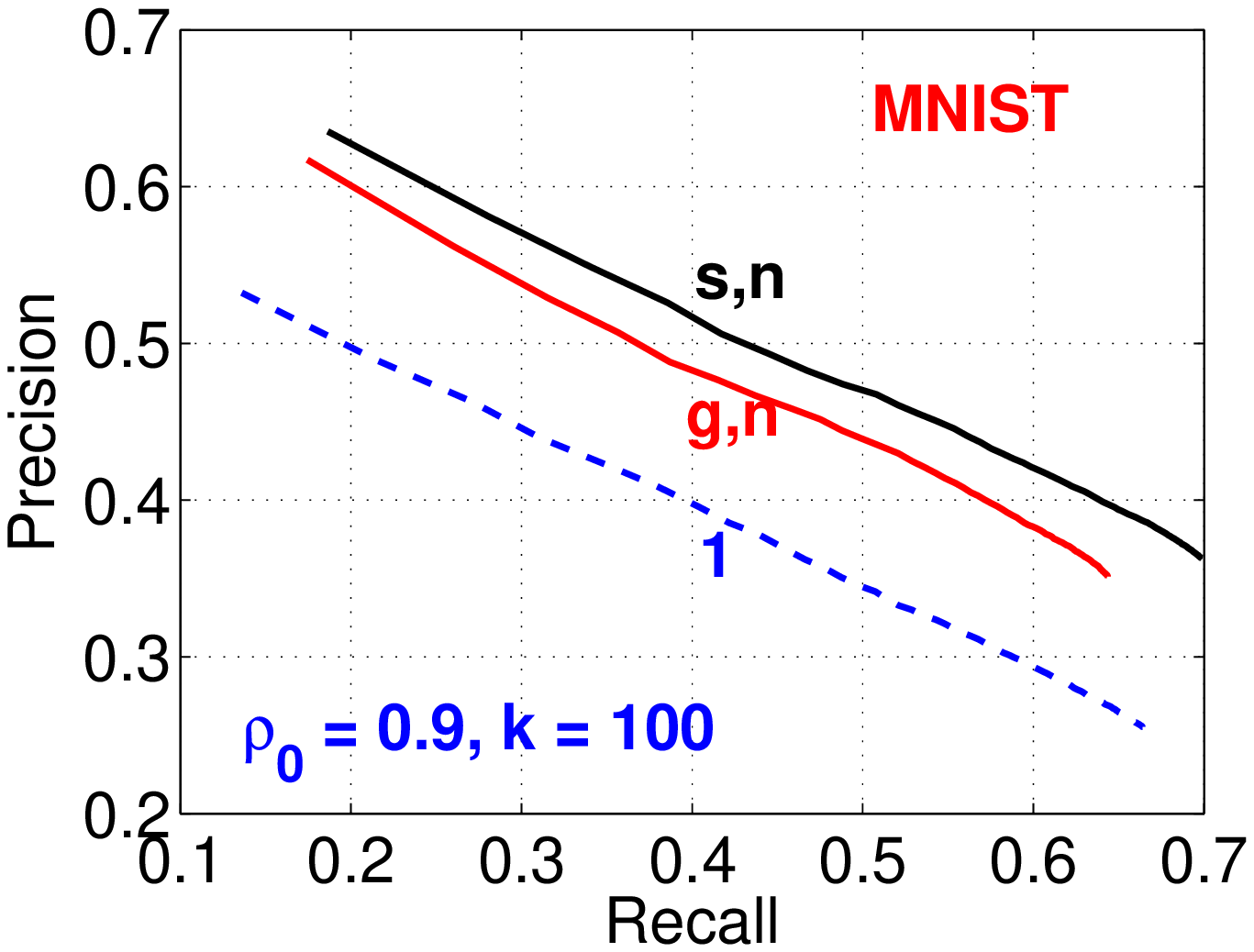}\hspace{-0.15in}
\includegraphics[width=2.2in]{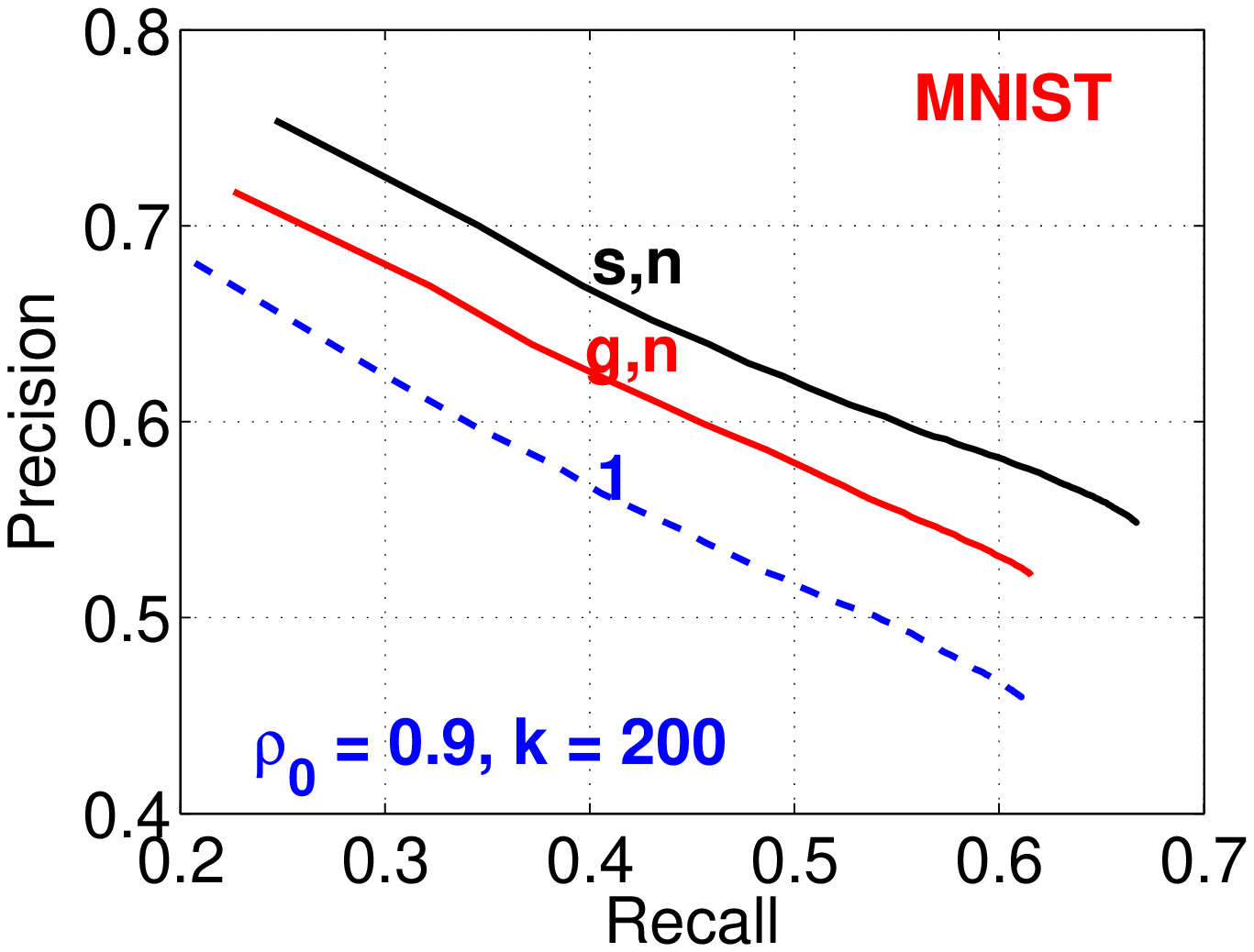}\hspace{-0.15in}
}

\mbox{
\includegraphics[width=2.2in]{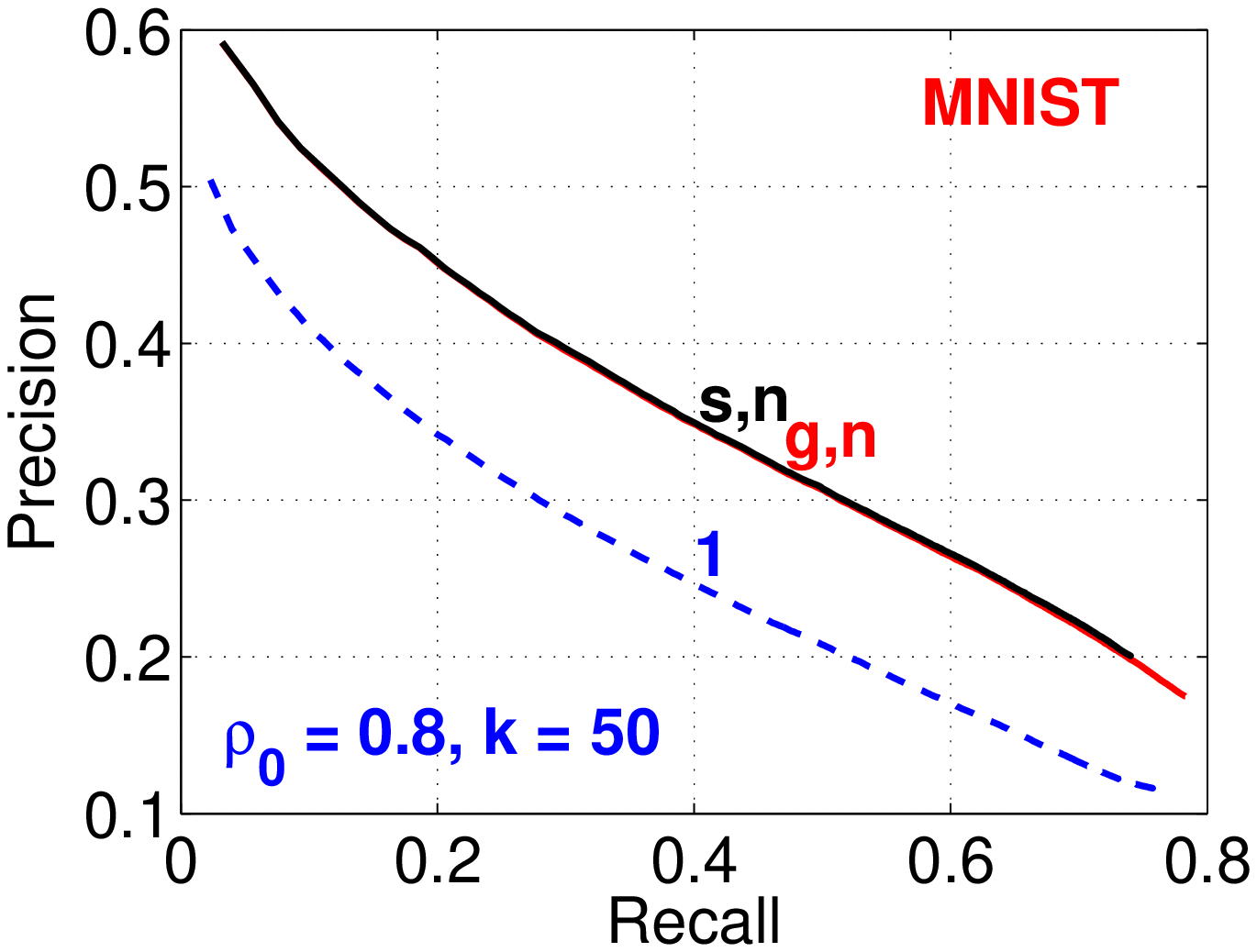}\hspace{-0.15in}
\includegraphics[width=2.2in]{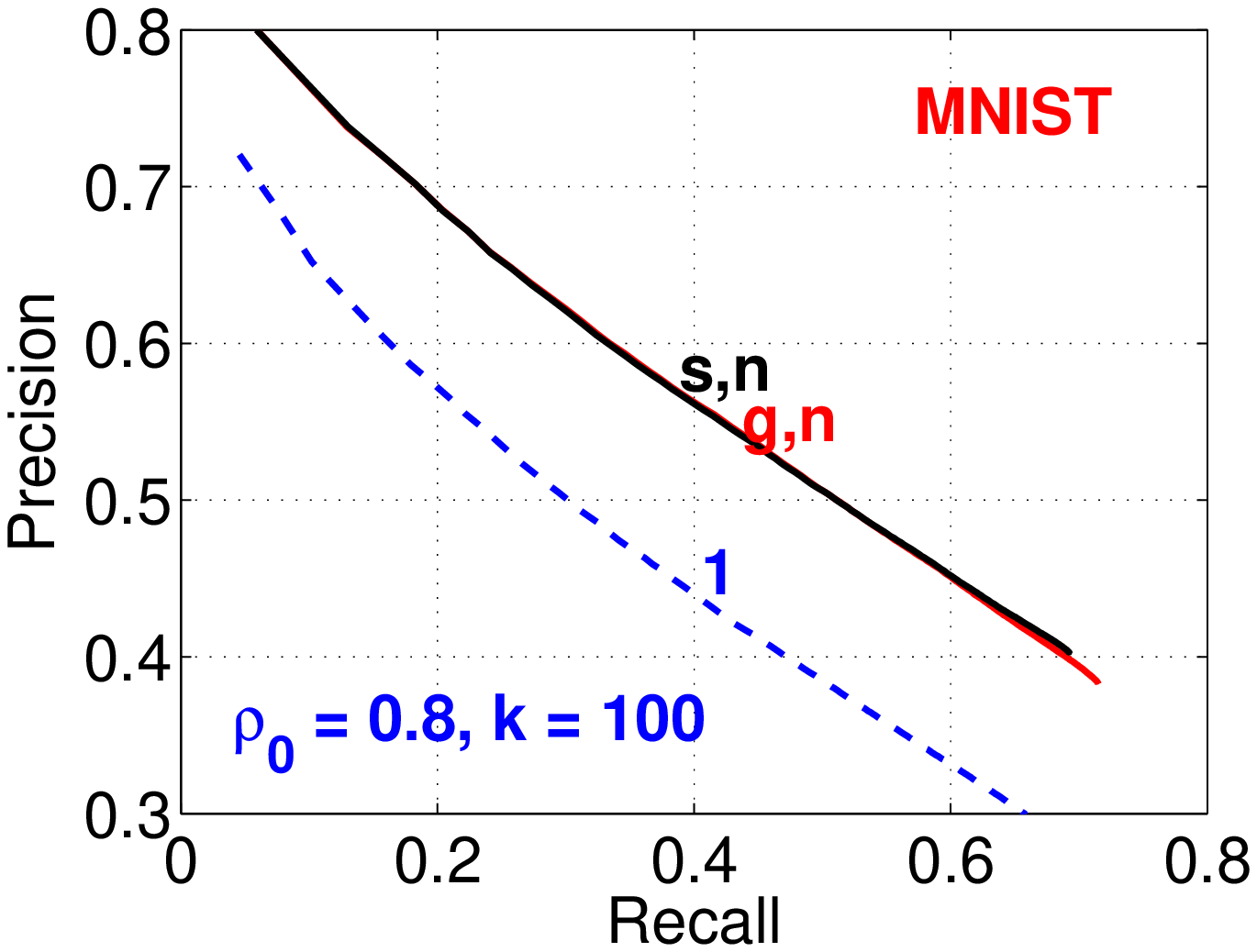}\hspace{-0.15in}
\includegraphics[width=2.2in]{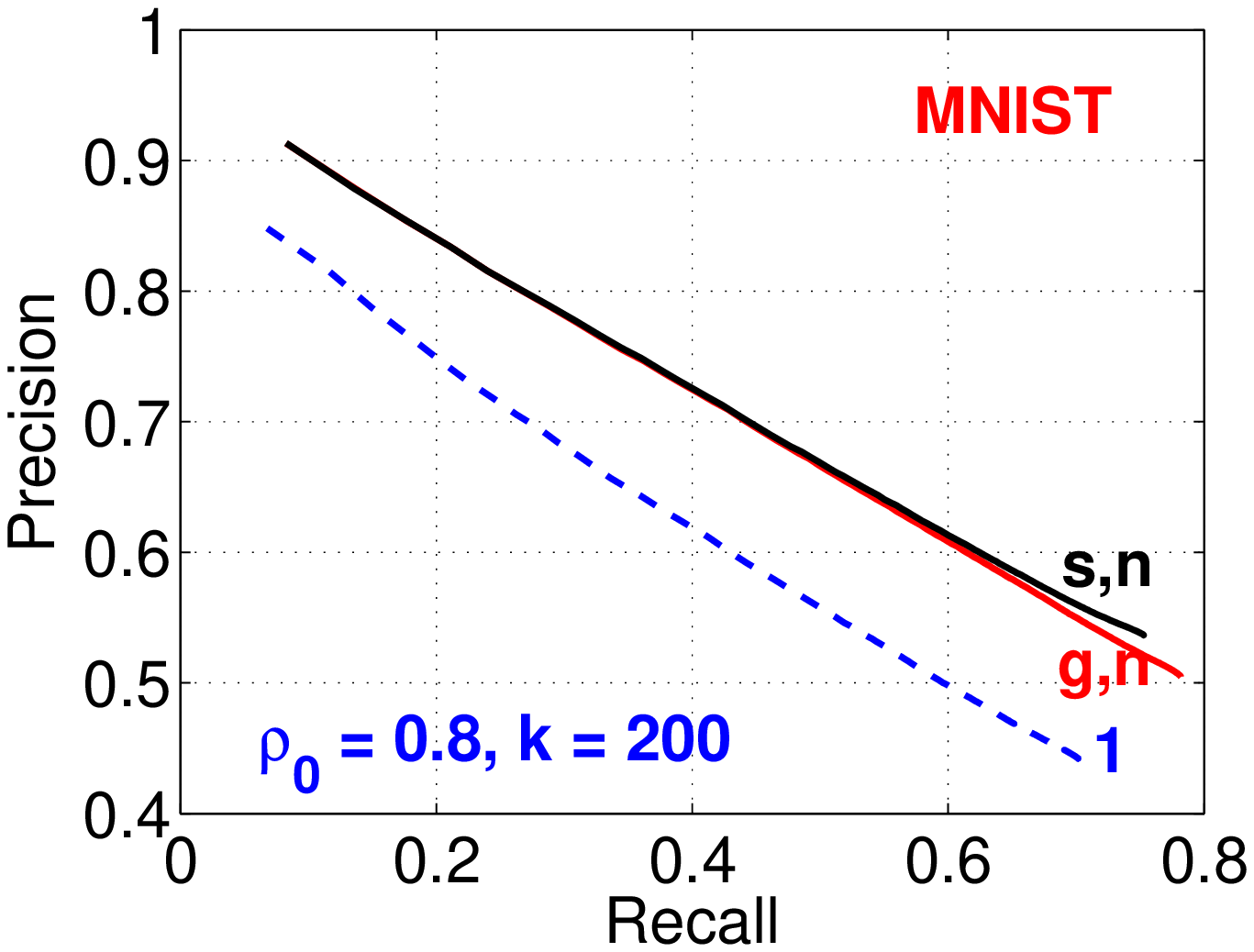}\hspace{-0.15in}
}

\mbox{
\includegraphics[width=2.2in]{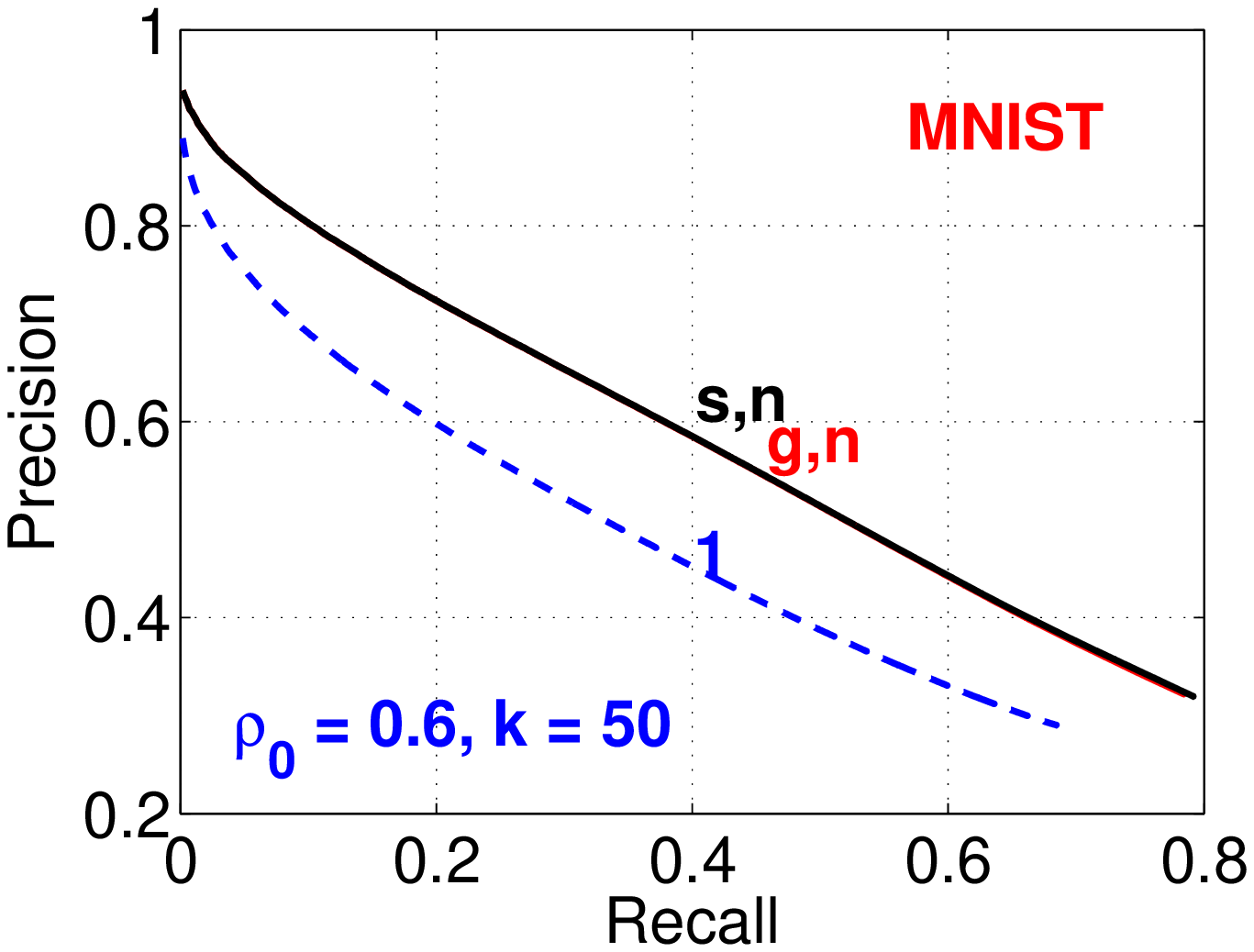}\hspace{-0.15in}
\includegraphics[width=2.2in]{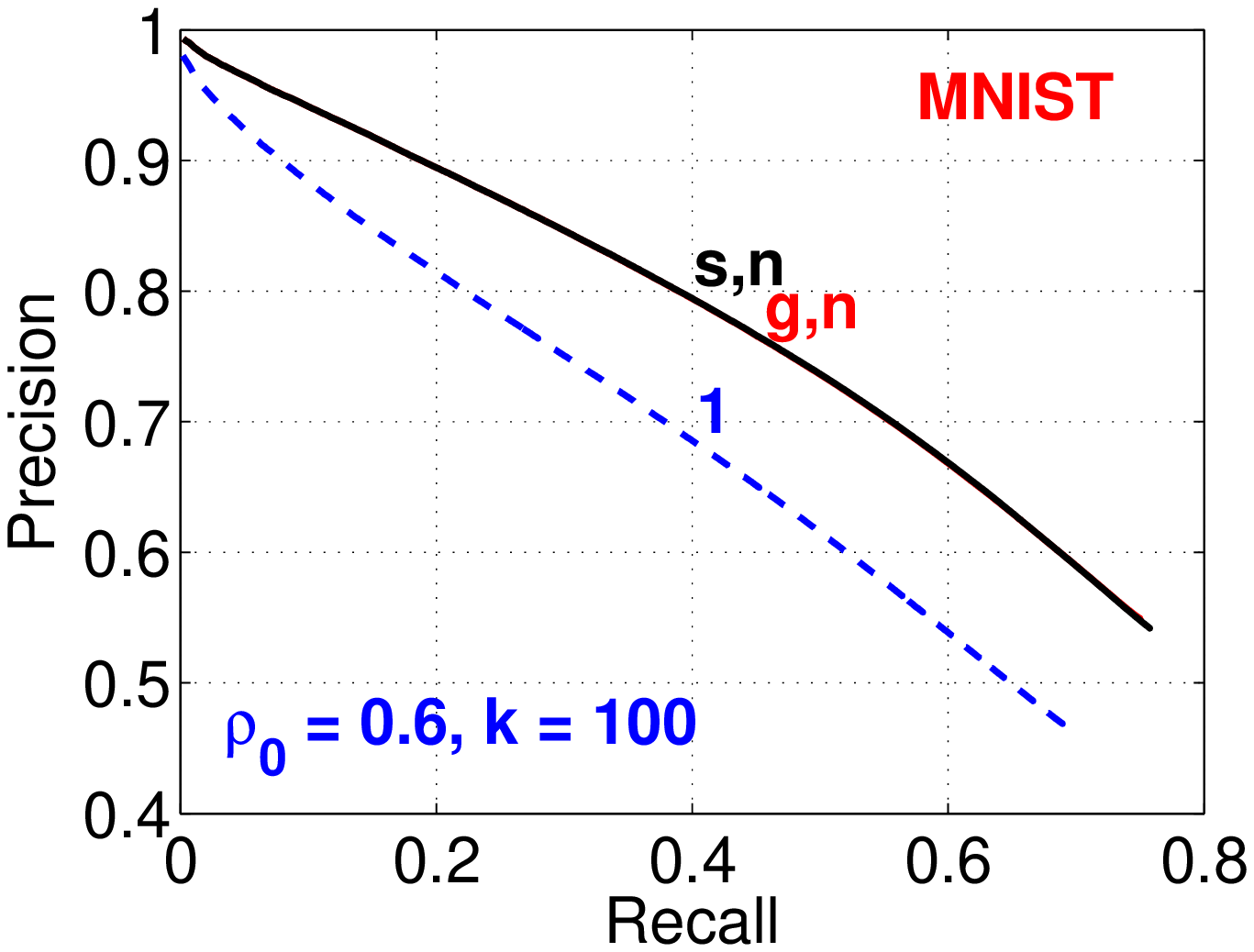}\hspace{-0.15in}
\includegraphics[width=2.2in]{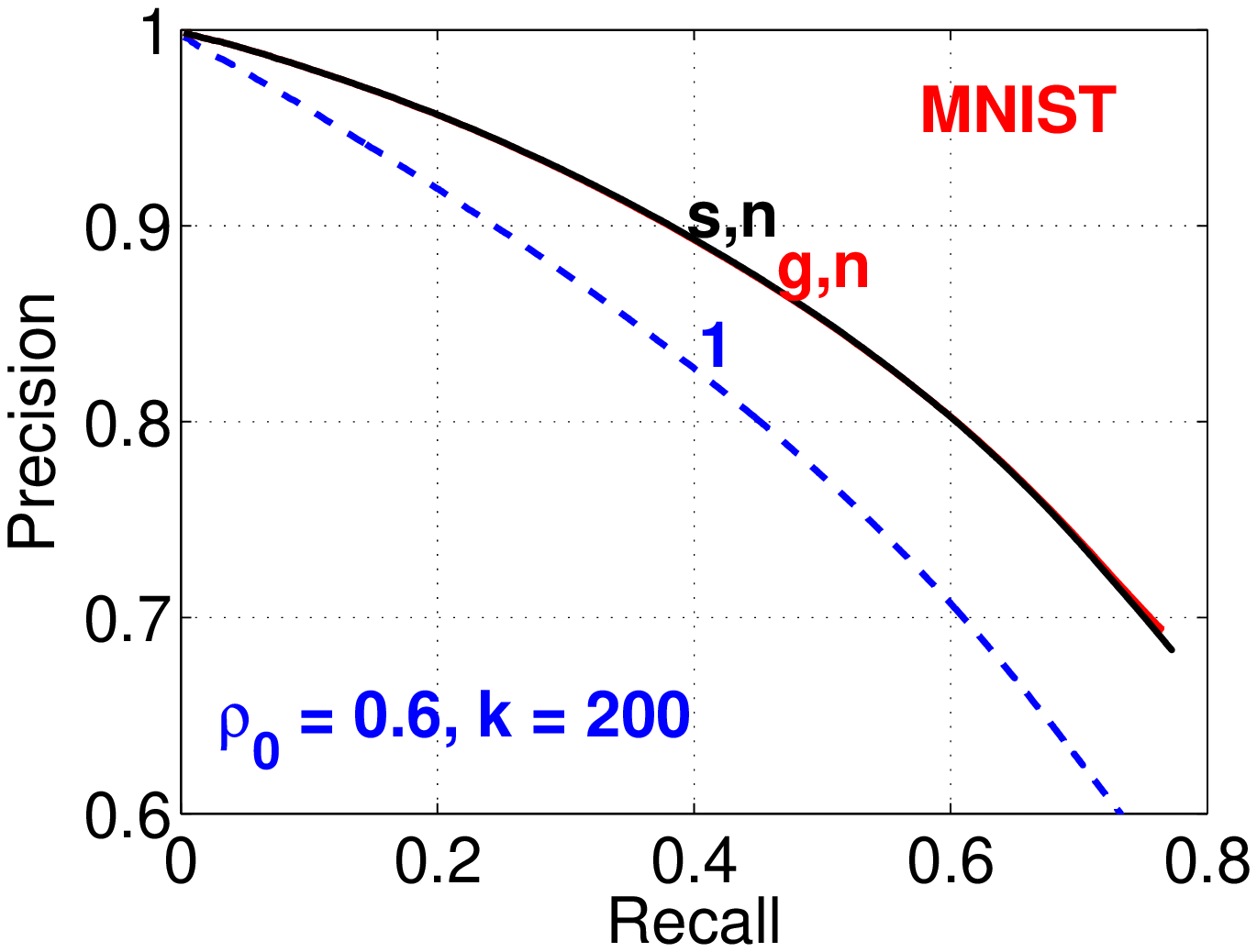}\hspace{-0.15in}
}

\mbox{
\includegraphics[width=2.2in]{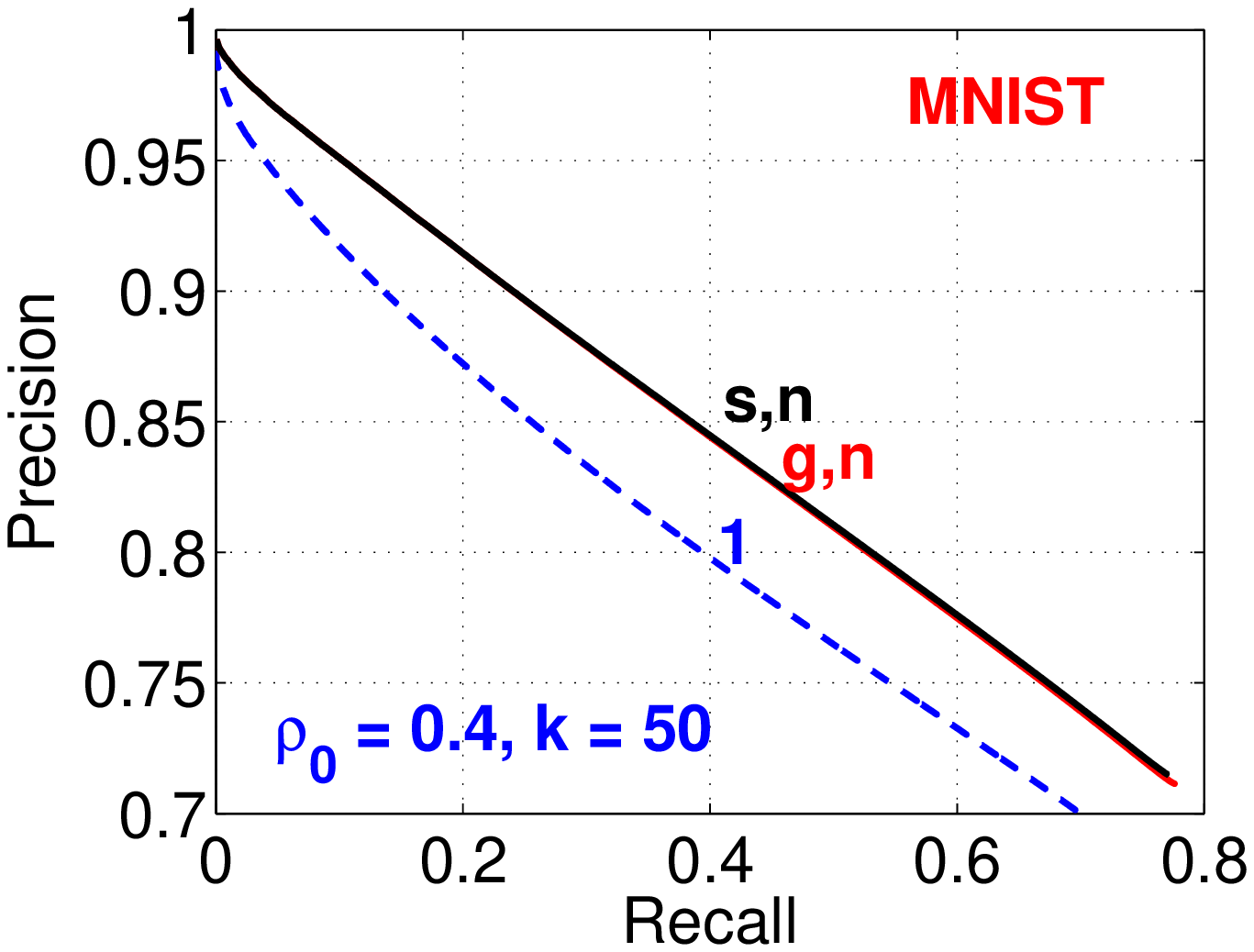}\hspace{-0.15in}
\includegraphics[width=2.2in]{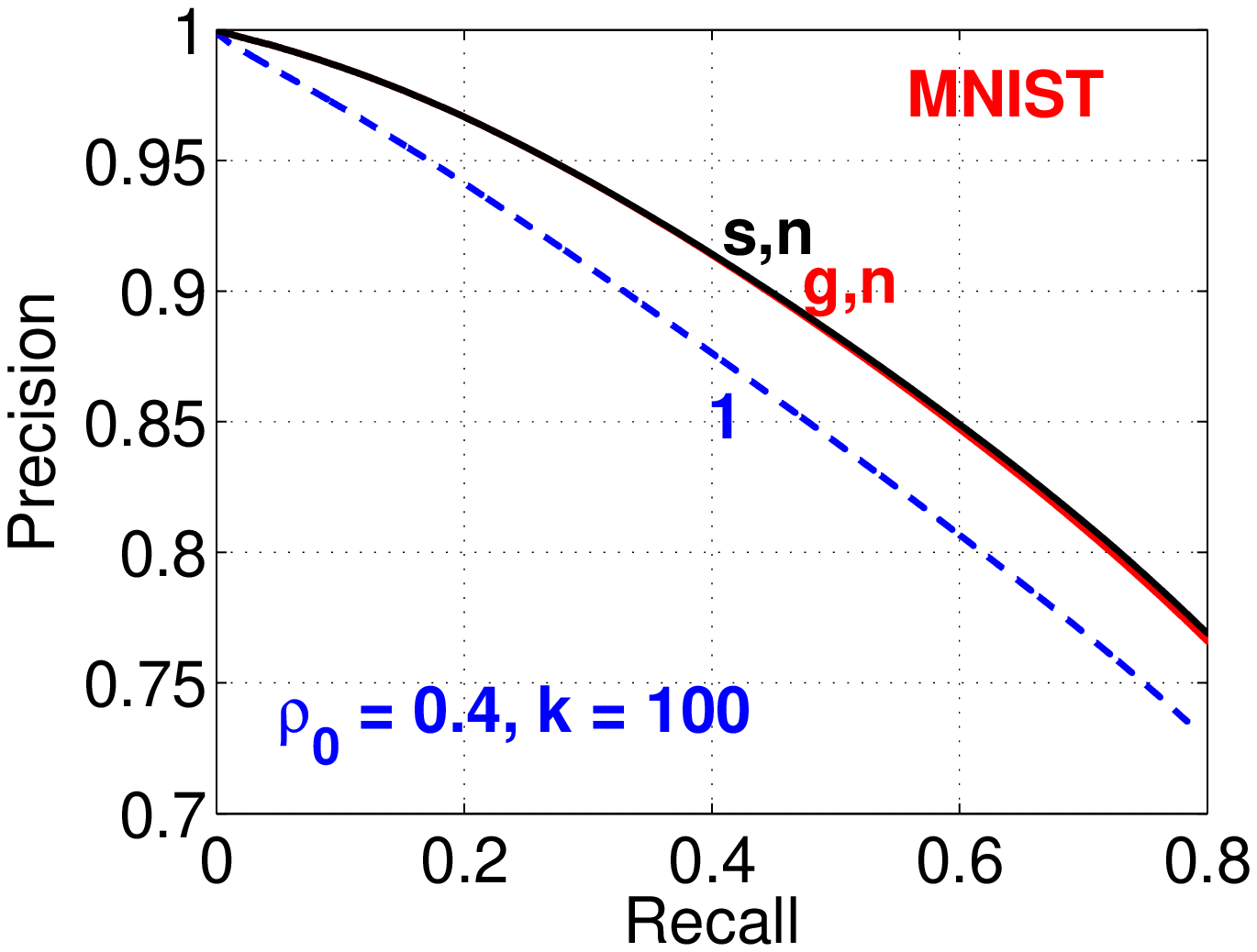}\hspace{-0.15in}
\includegraphics[width=2.2in]{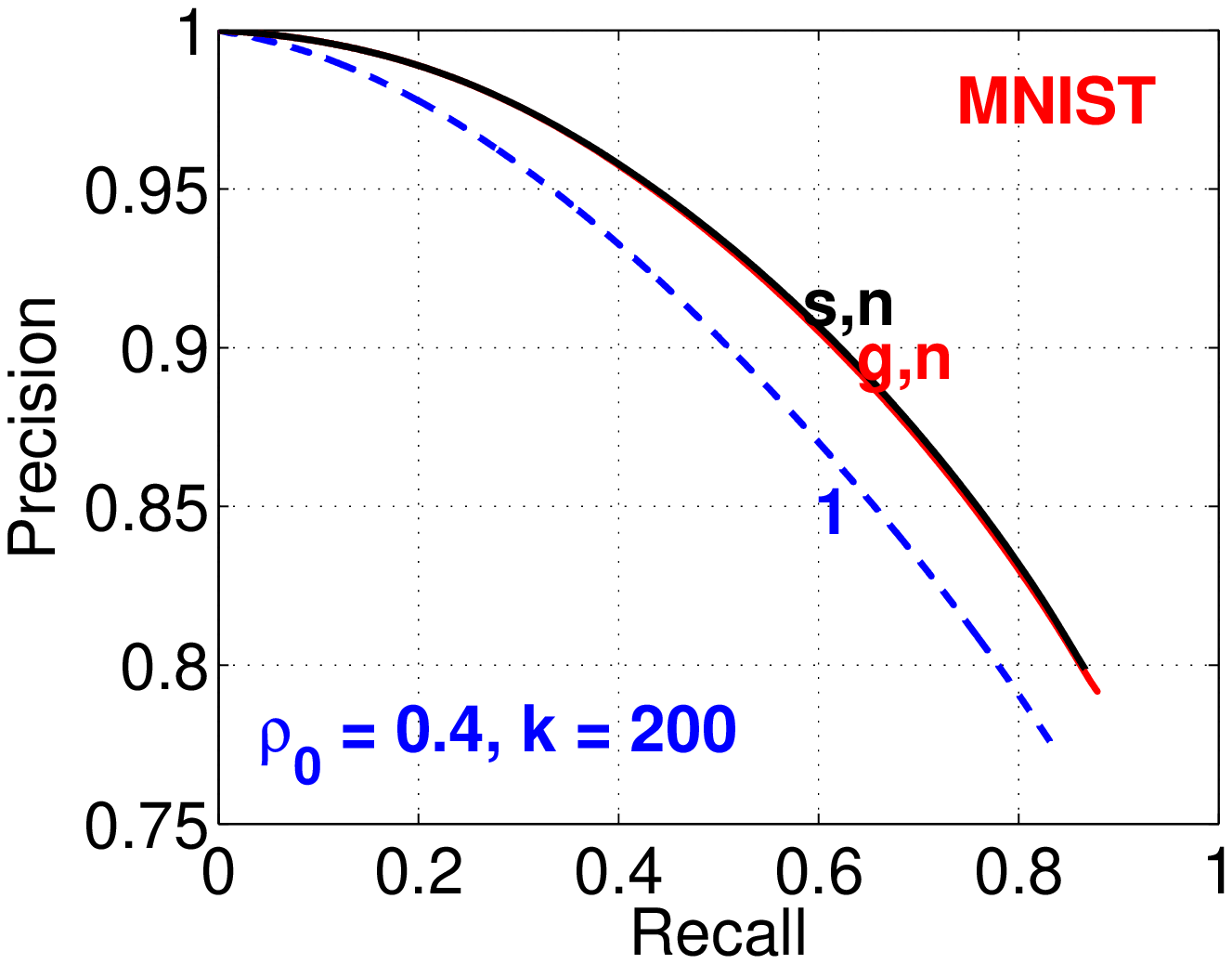}\hspace{-0.15in}
}

\end{center}
\vspace{-0.2in}
\caption{\textbf{MNIST}: precision-recall curves for selected $\rho_0$ and $k$ values, and  for three estimators： $\hat{\rho}_{s,n}$ (recommended), $\hat{\rho}_{g,n}$, $\hat{\rho}_1$.}\label{fig_pr_MNIST}
\end{figure*}

\begin{figure*}[h!]
\begin{center}
\mbox{
\includegraphics[width=2.2in]{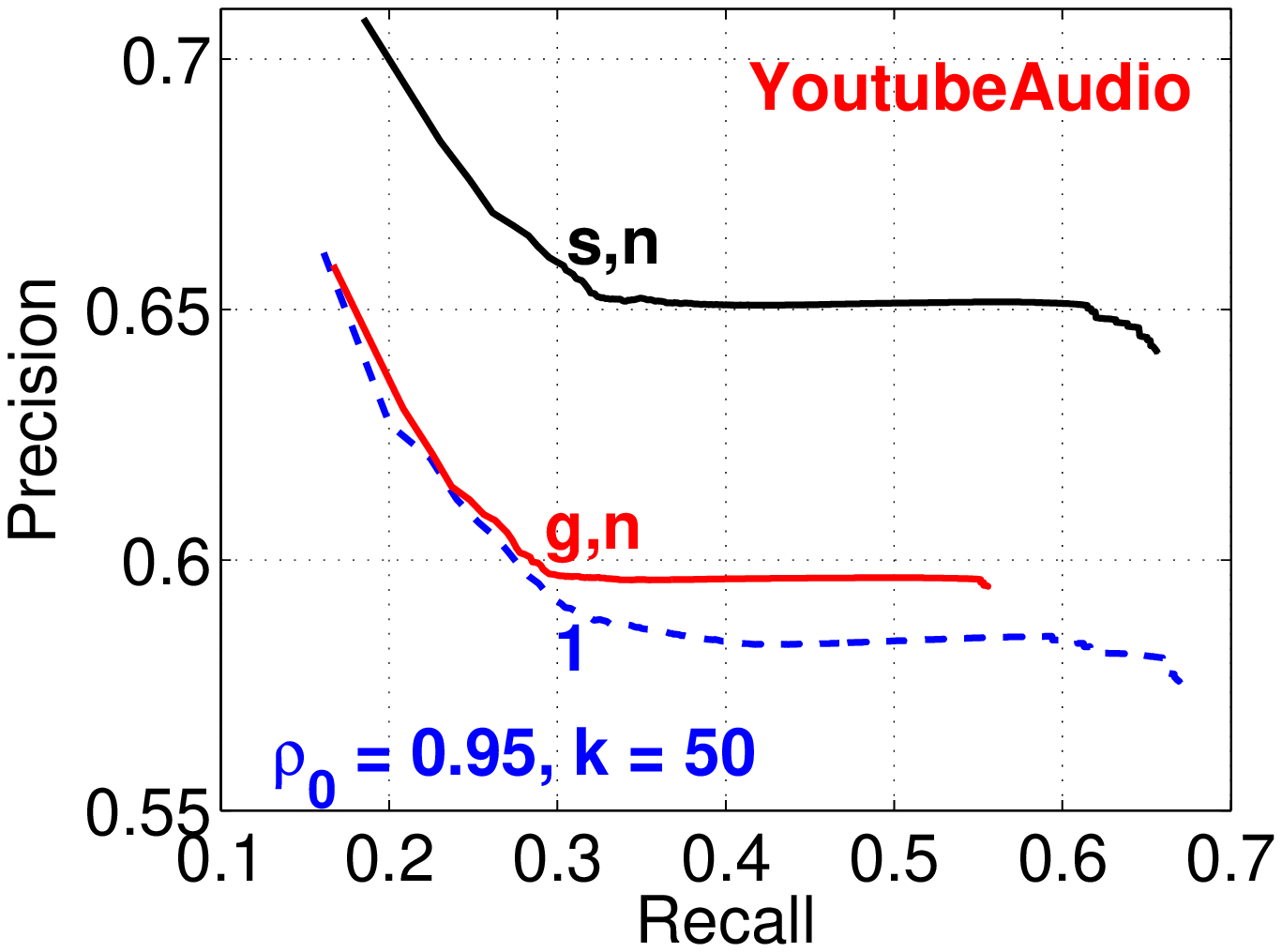}\hspace{-0.15in}
\includegraphics[width=2.2in]{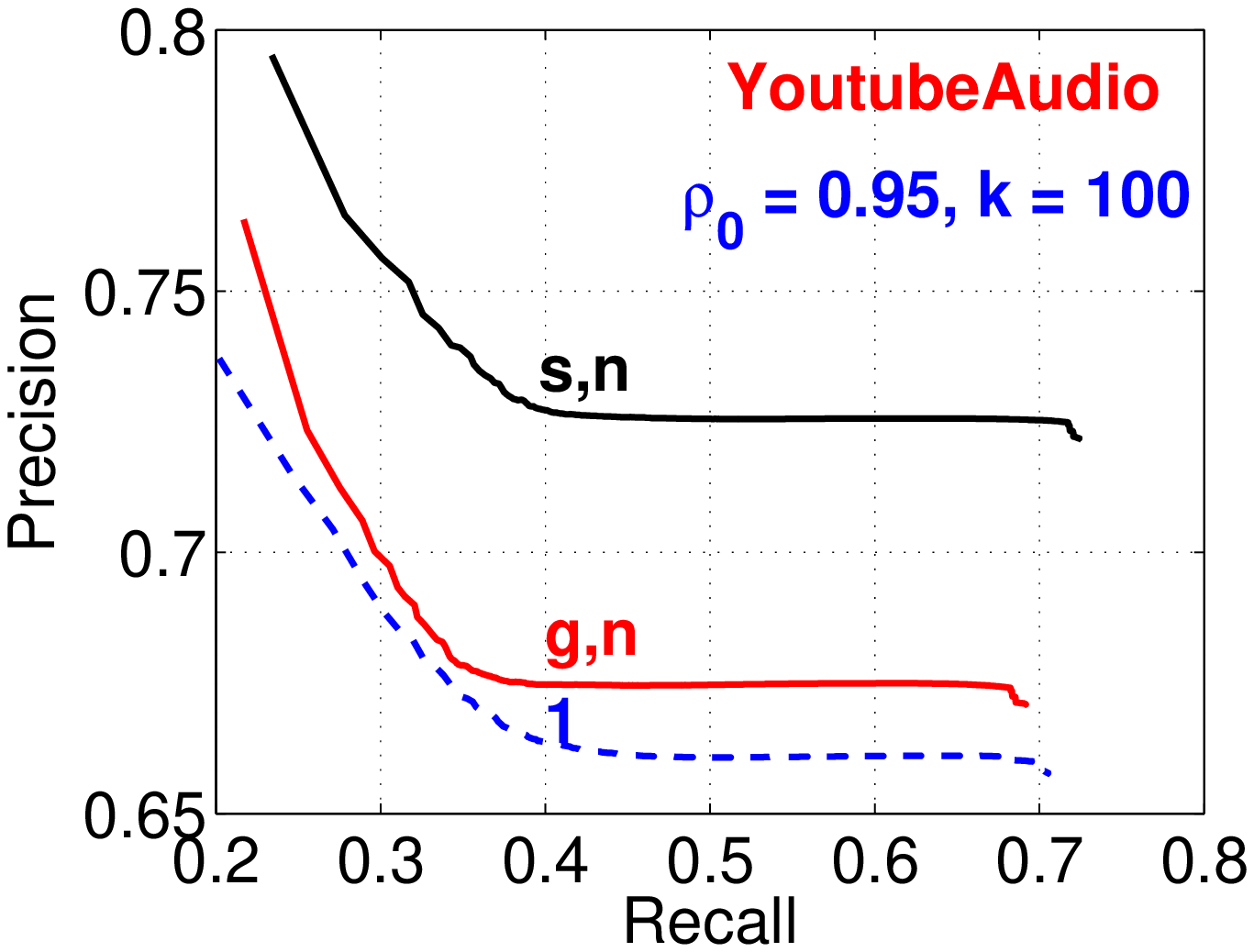}\hspace{-0.15in}
\includegraphics[width=2.2in]{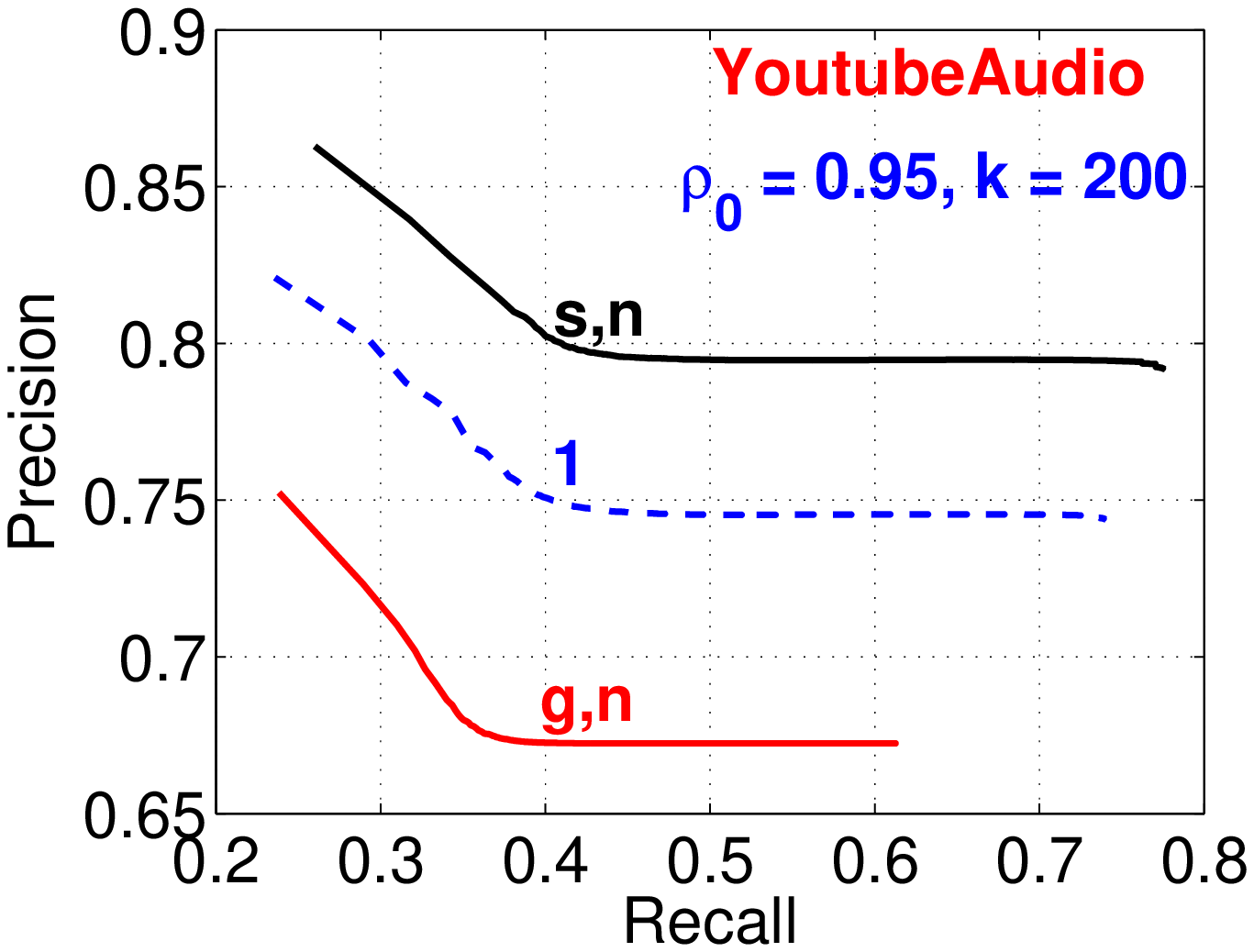}\hspace{-0.15in}
}

\mbox{
\includegraphics[width=2.2in]{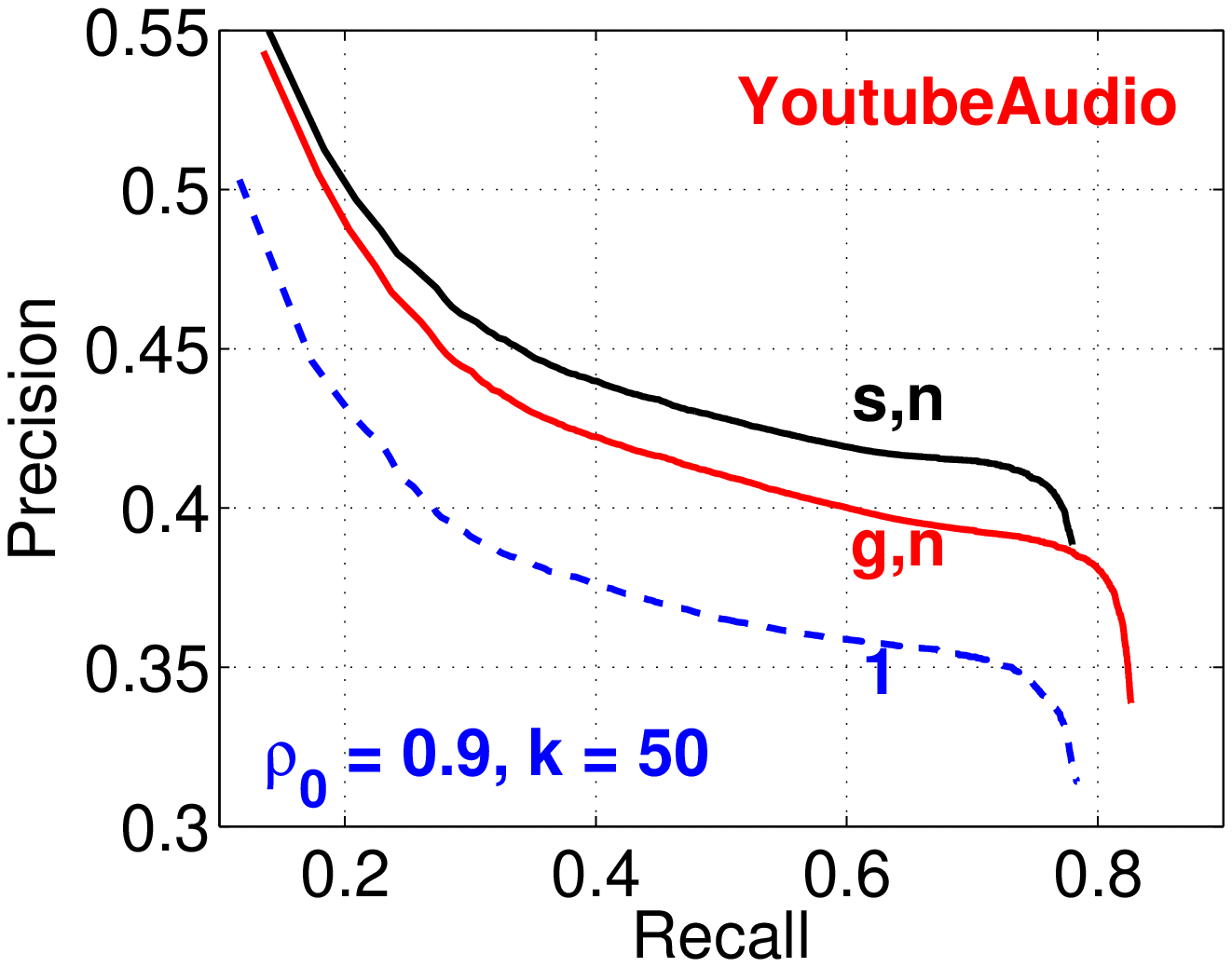}\hspace{-0.15in}
\includegraphics[width=2.2in]{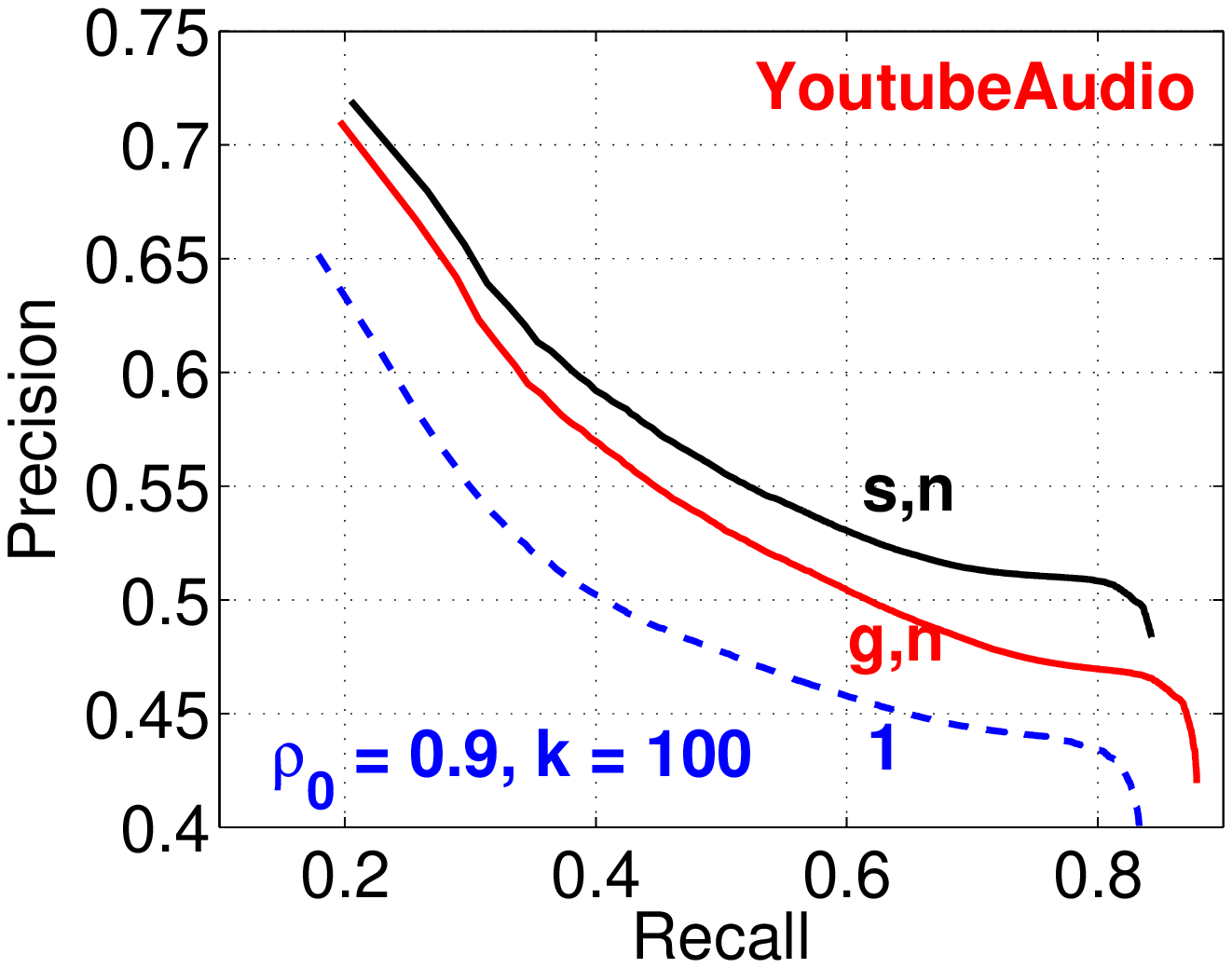}\hspace{-0.15in}
\includegraphics[width=2.2in]{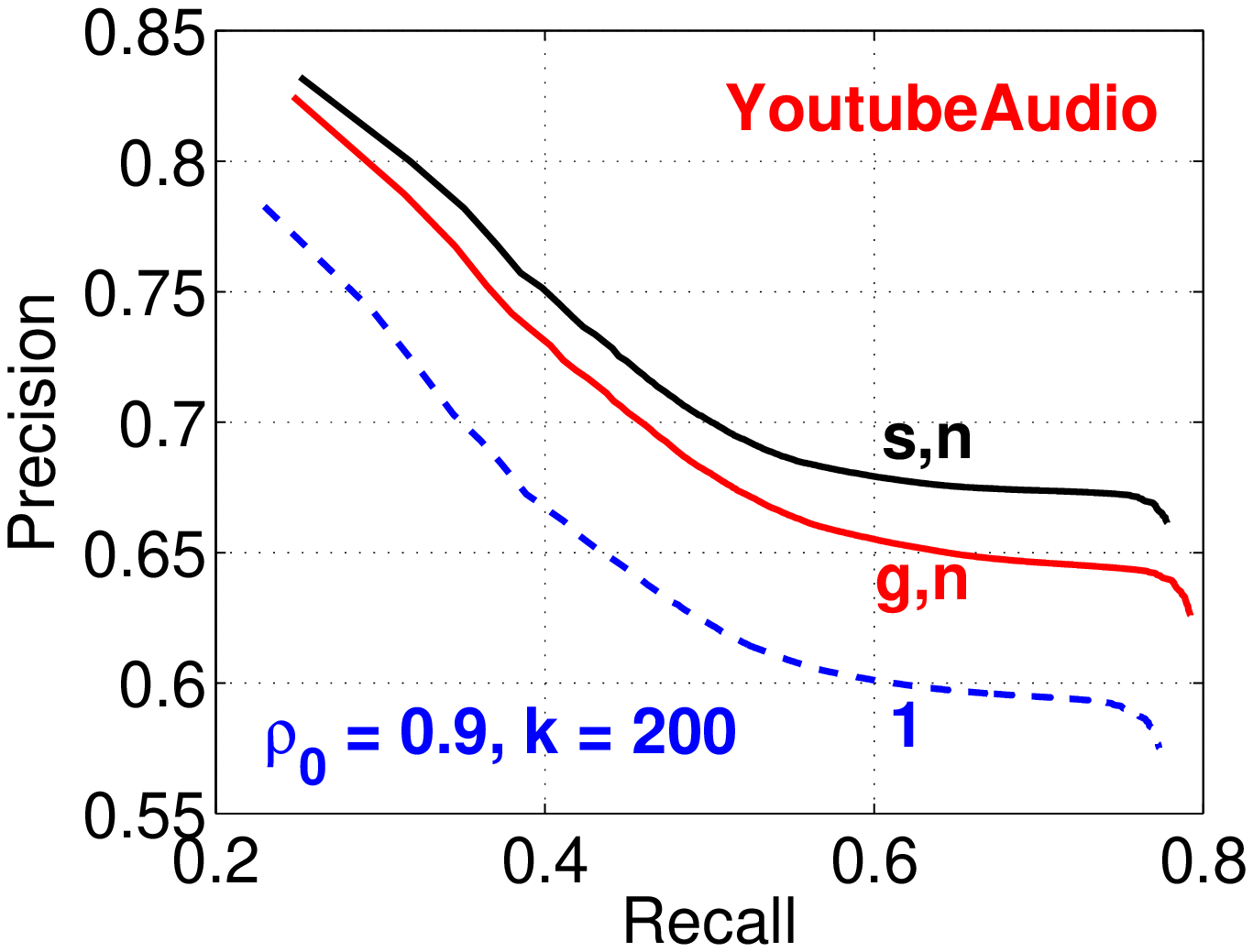}\hspace{-0.15in}
}

\mbox{
\includegraphics[width=2.2in]{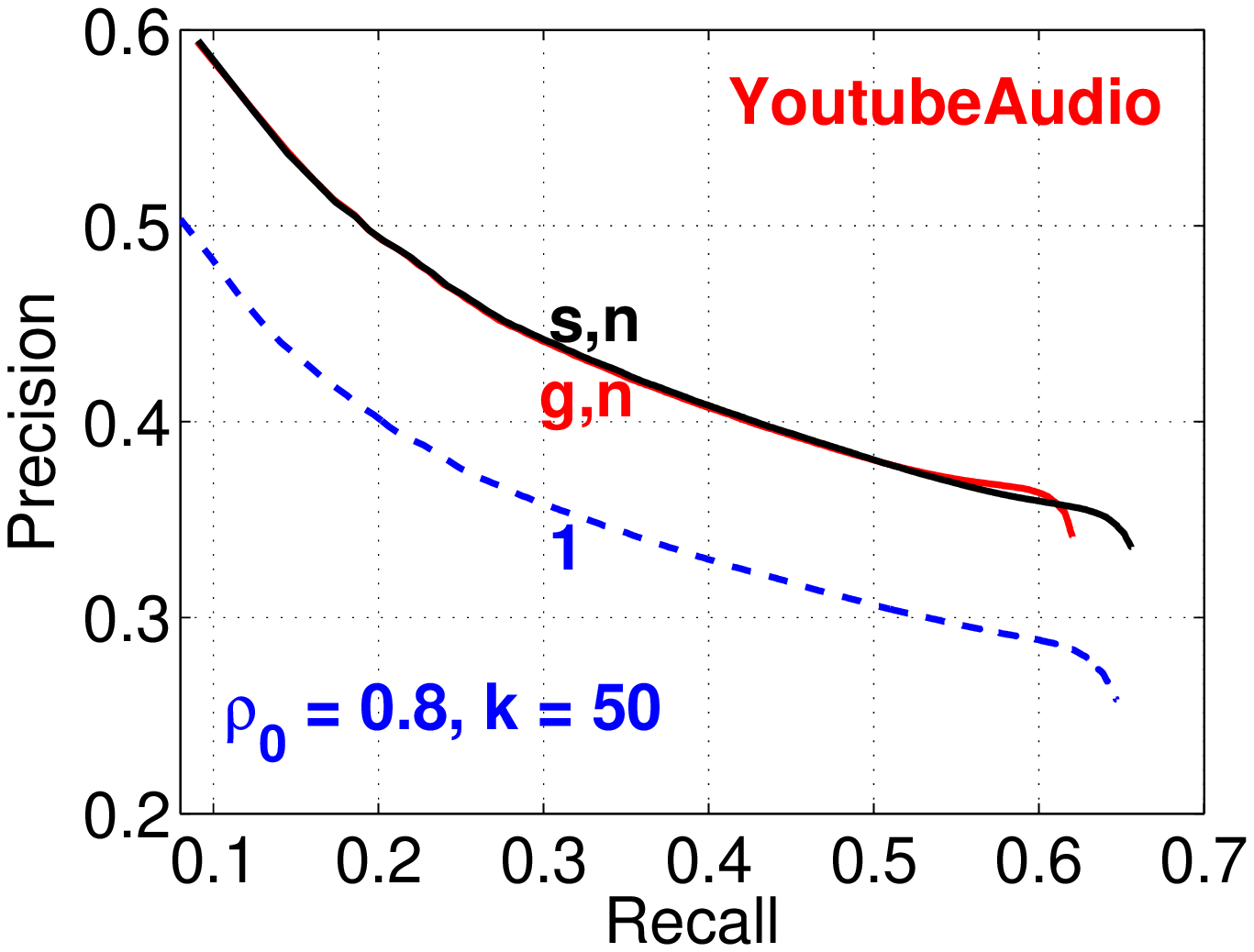}\hspace{-0.15in}
\includegraphics[width=2.2in]{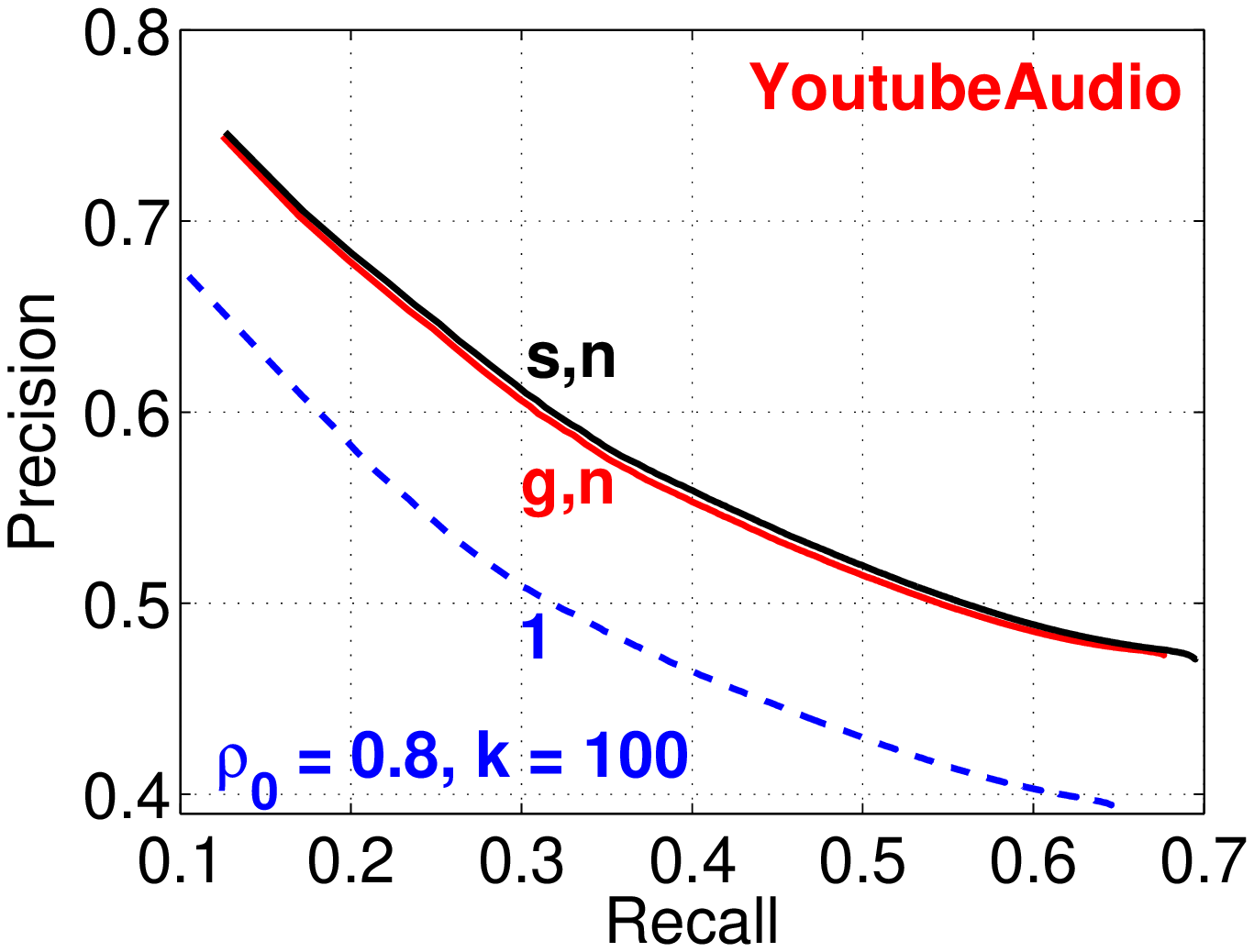}\hspace{-0.15in}
\includegraphics[width=2.2in]{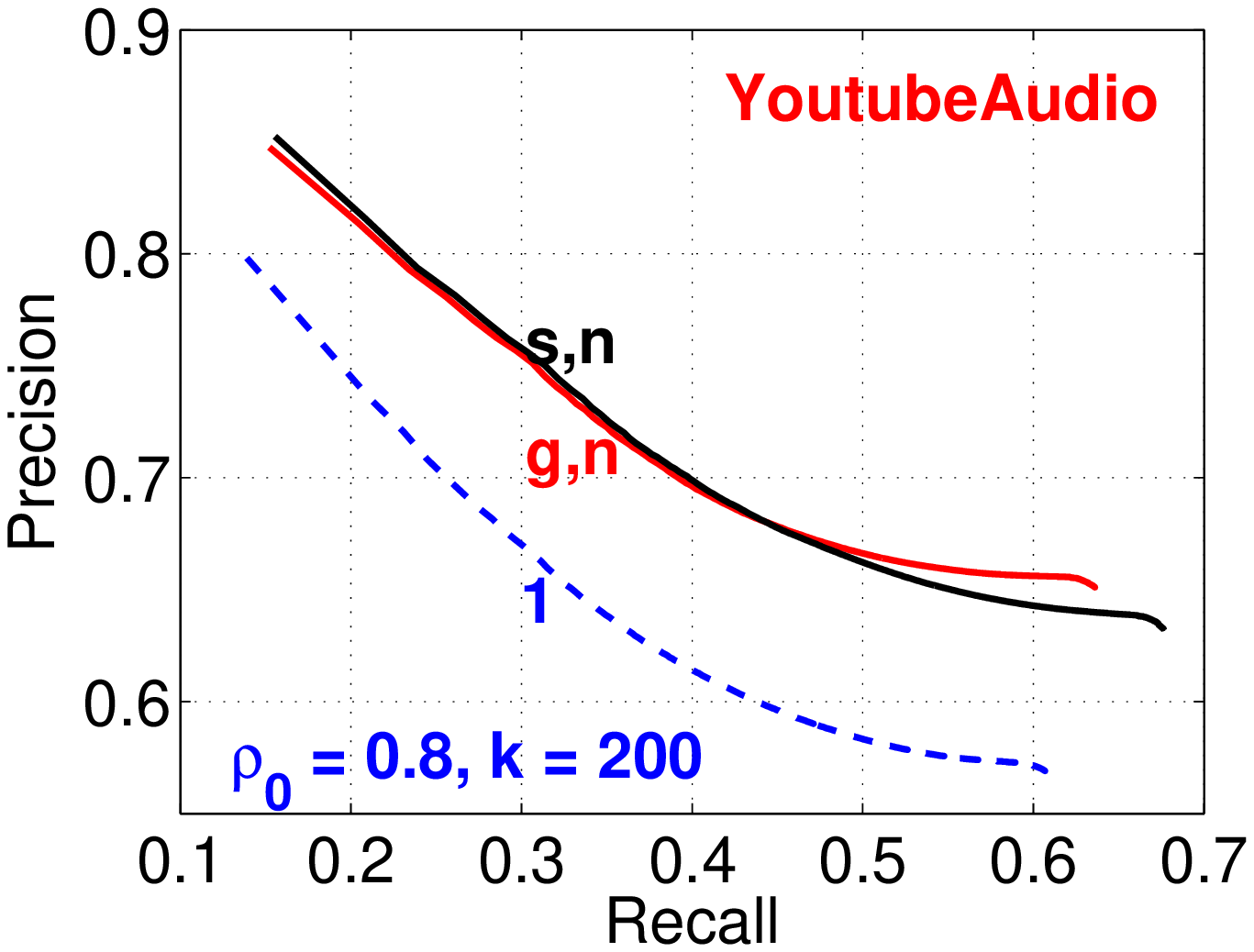}\hspace{-0.15in}
}

\mbox{
\includegraphics[width=2.2in]{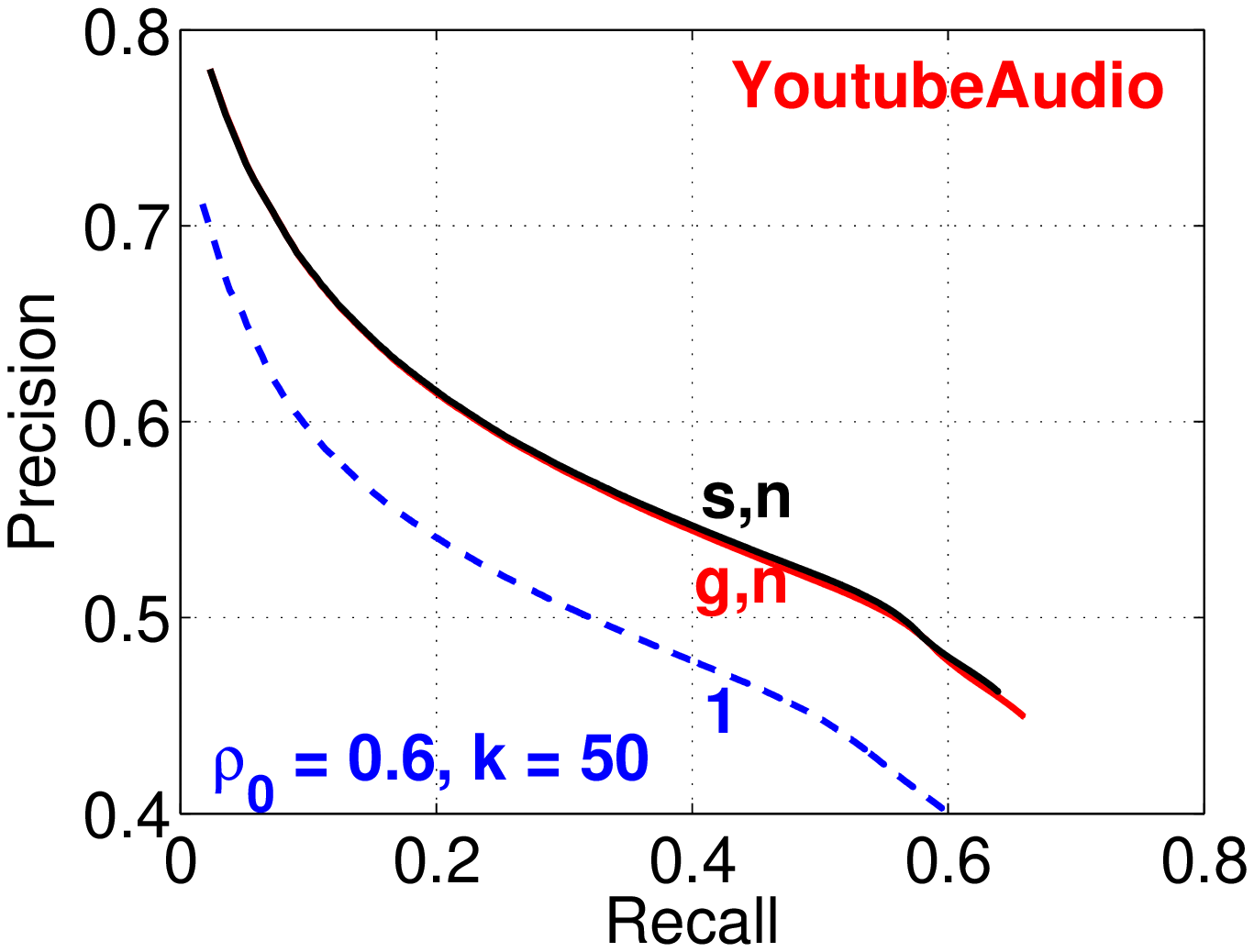}\hspace{-0.15in}
\includegraphics[width=2.2in]{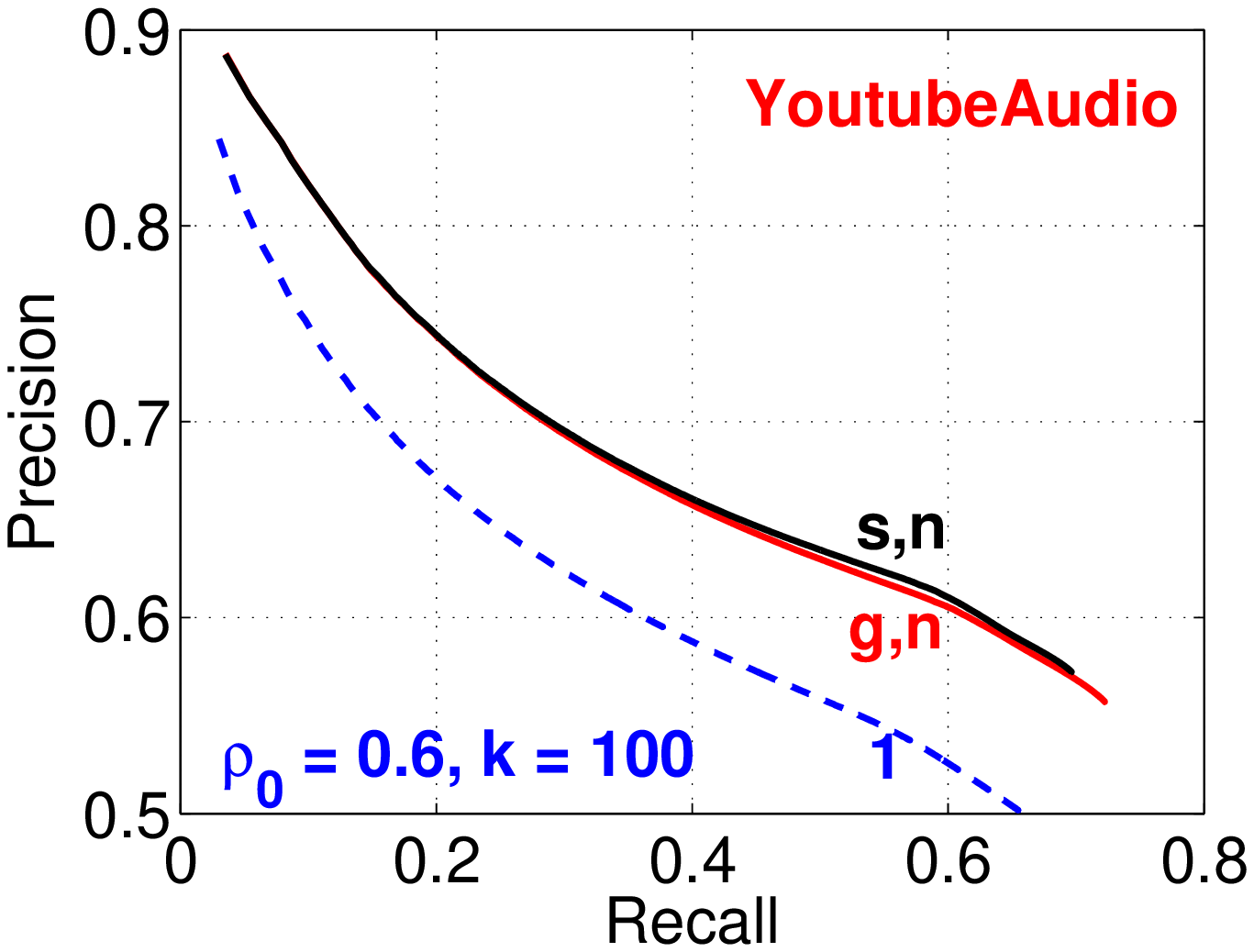}\hspace{-0.15in}
\includegraphics[width=2.2in]{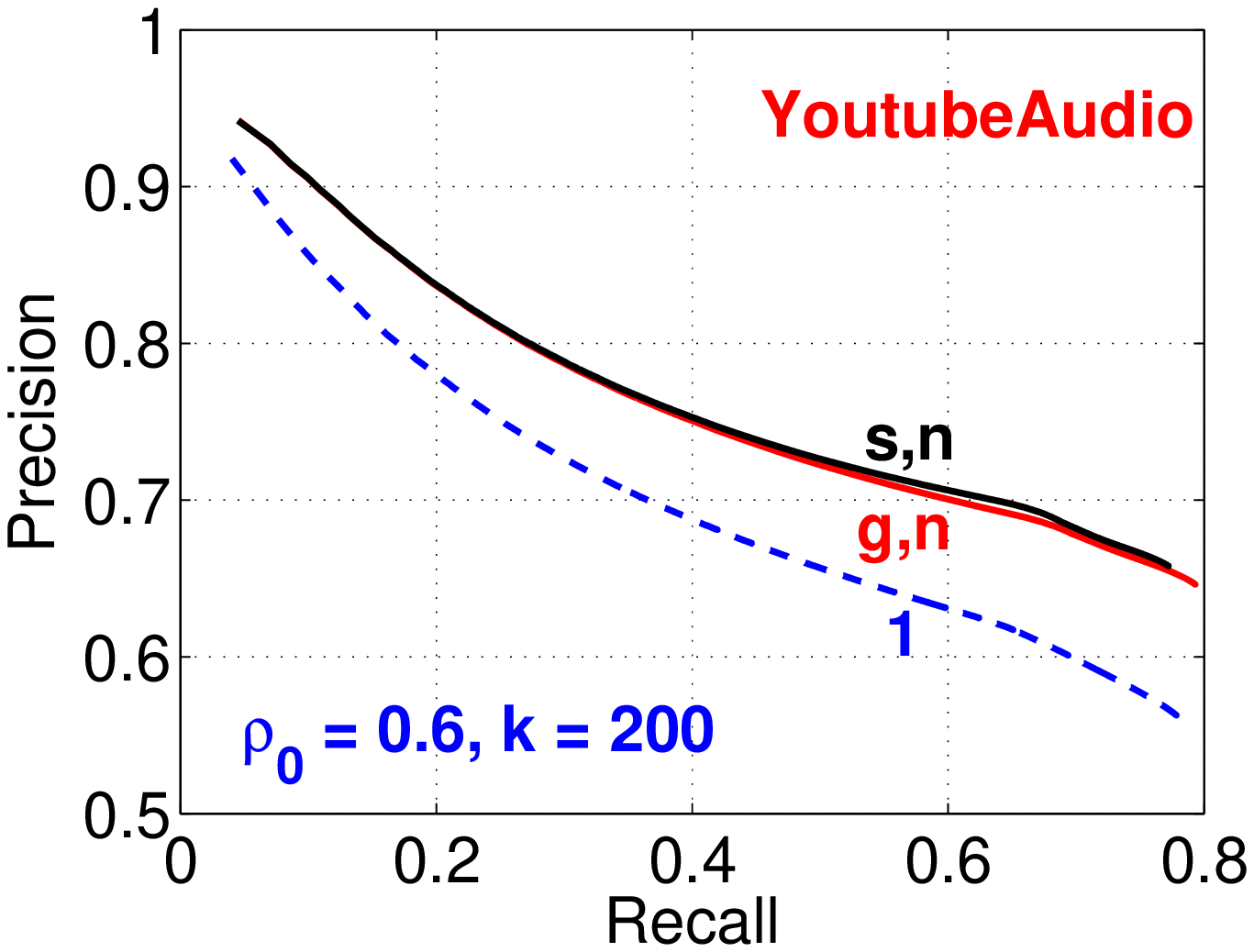}\hspace{-0.15in}
}

\mbox{
\includegraphics[width=2.2in]{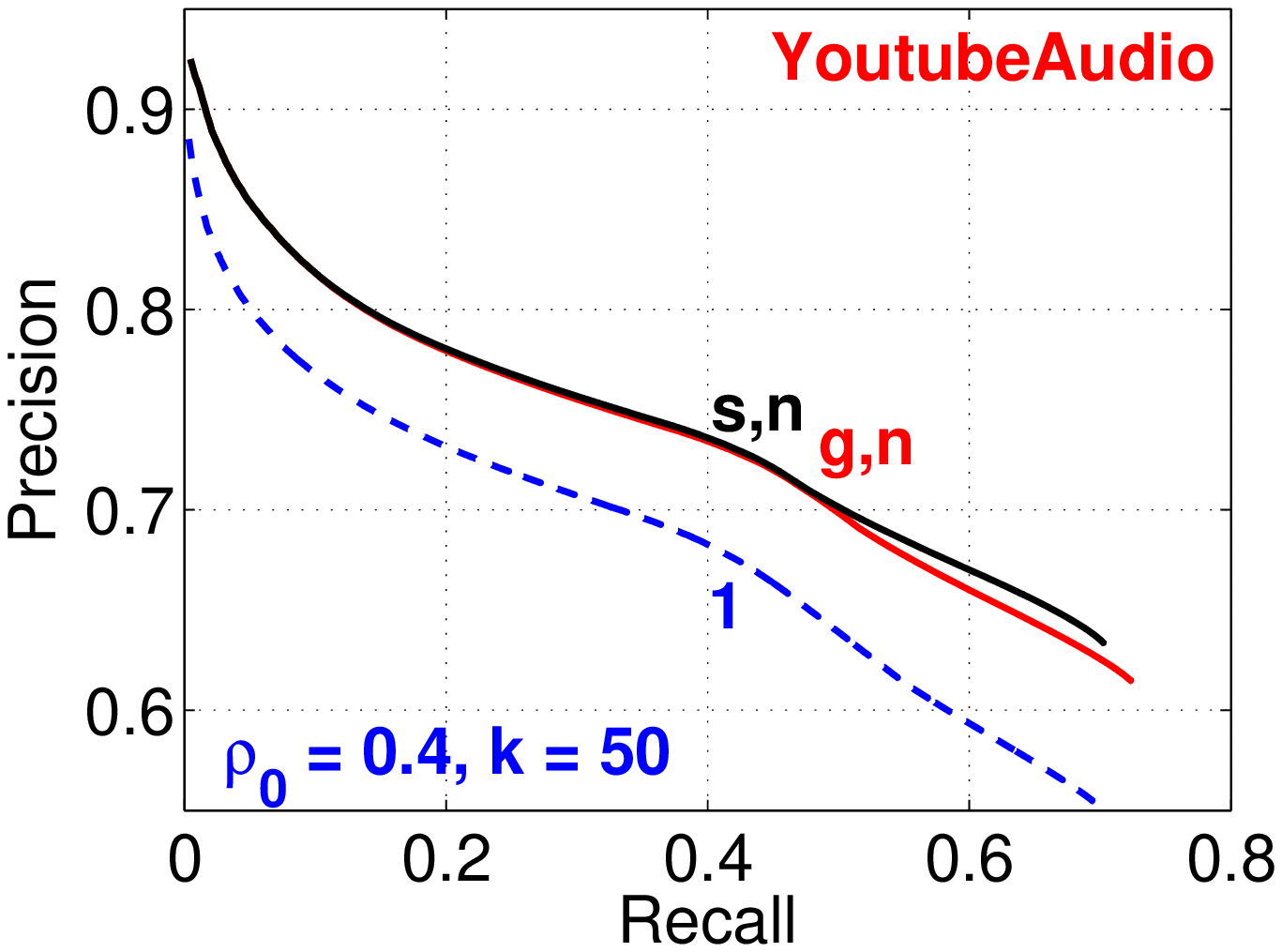}\hspace{-0.15in}
\includegraphics[width=2.2in]{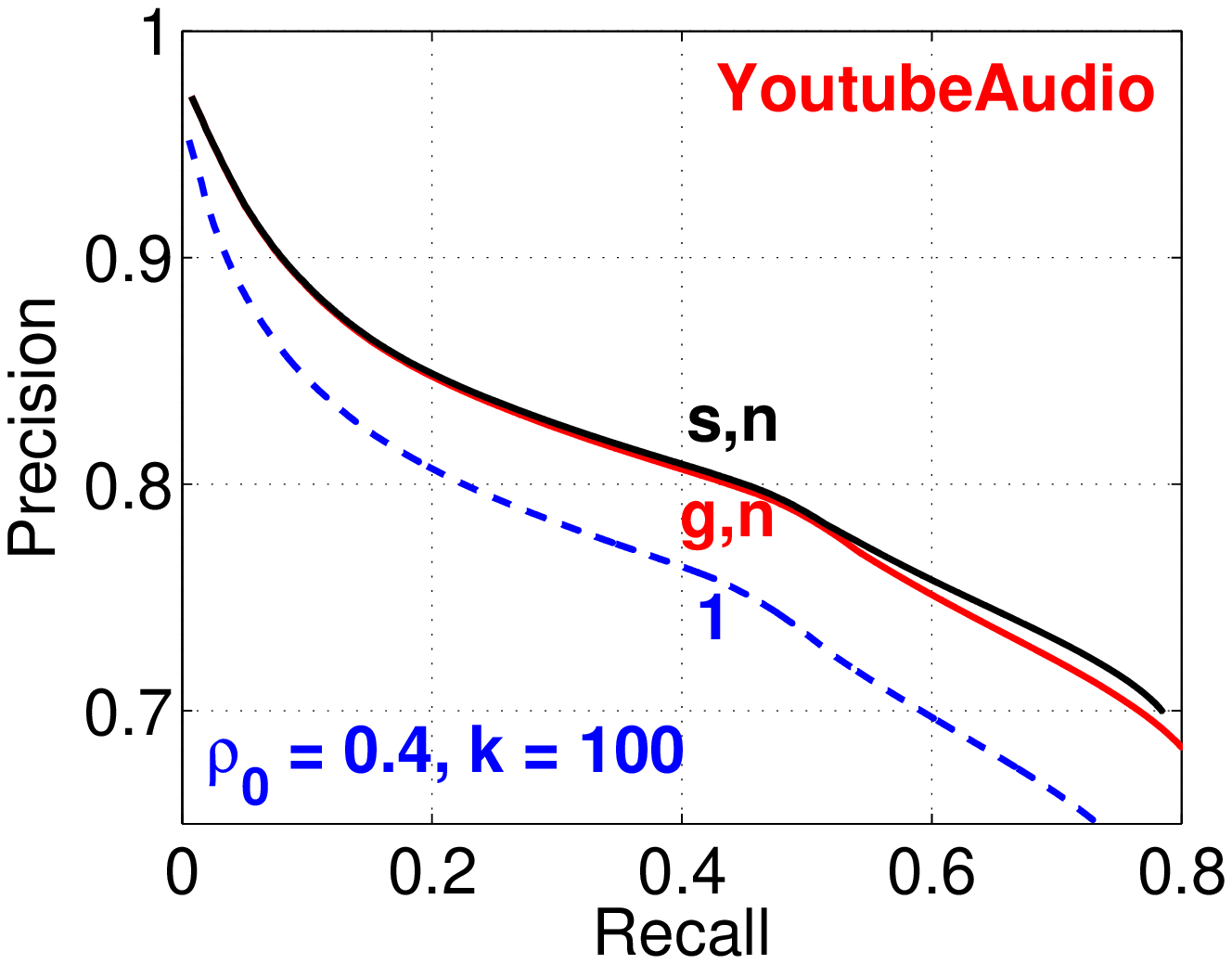}\hspace{-0.15in}
\includegraphics[width=2.2in]{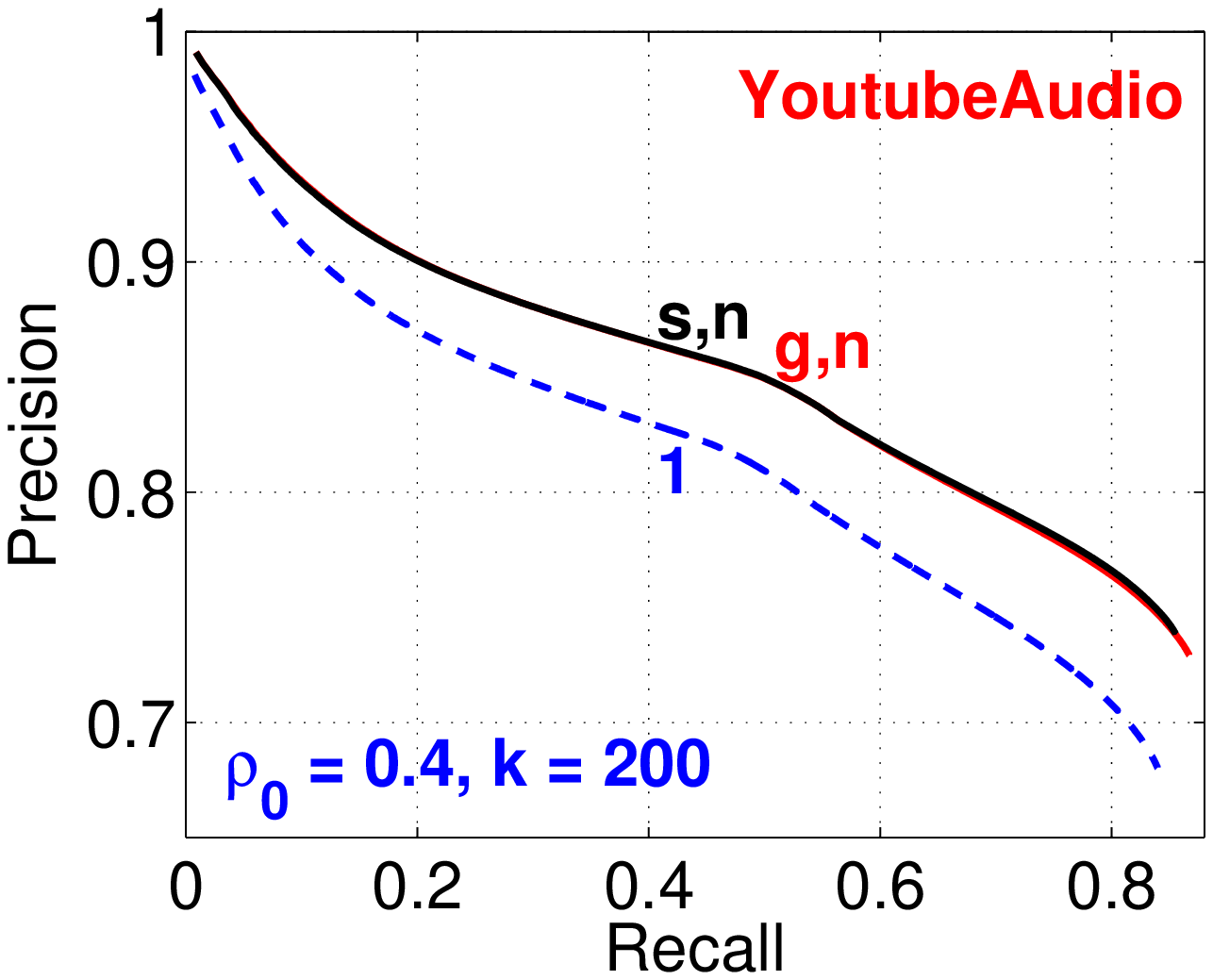}\hspace{-0.15in}
}

\end{center}
\vspace{-0.2in}
\caption{\textbf{YoutubeAudio}: precision-recall curves for selected $\rho_0$ and $k$ values, and  for three estimators： $\hat{\rho}_{s,n}$ (recommended), $\hat{\rho}_{g,n}$, $\hat{\rho}_1$.}\label{fig_pr_YoutubeAudio}
\end{figure*}

\begin{figure*}[h!]
\begin{center}
\mbox{
\includegraphics[width=2.2in]{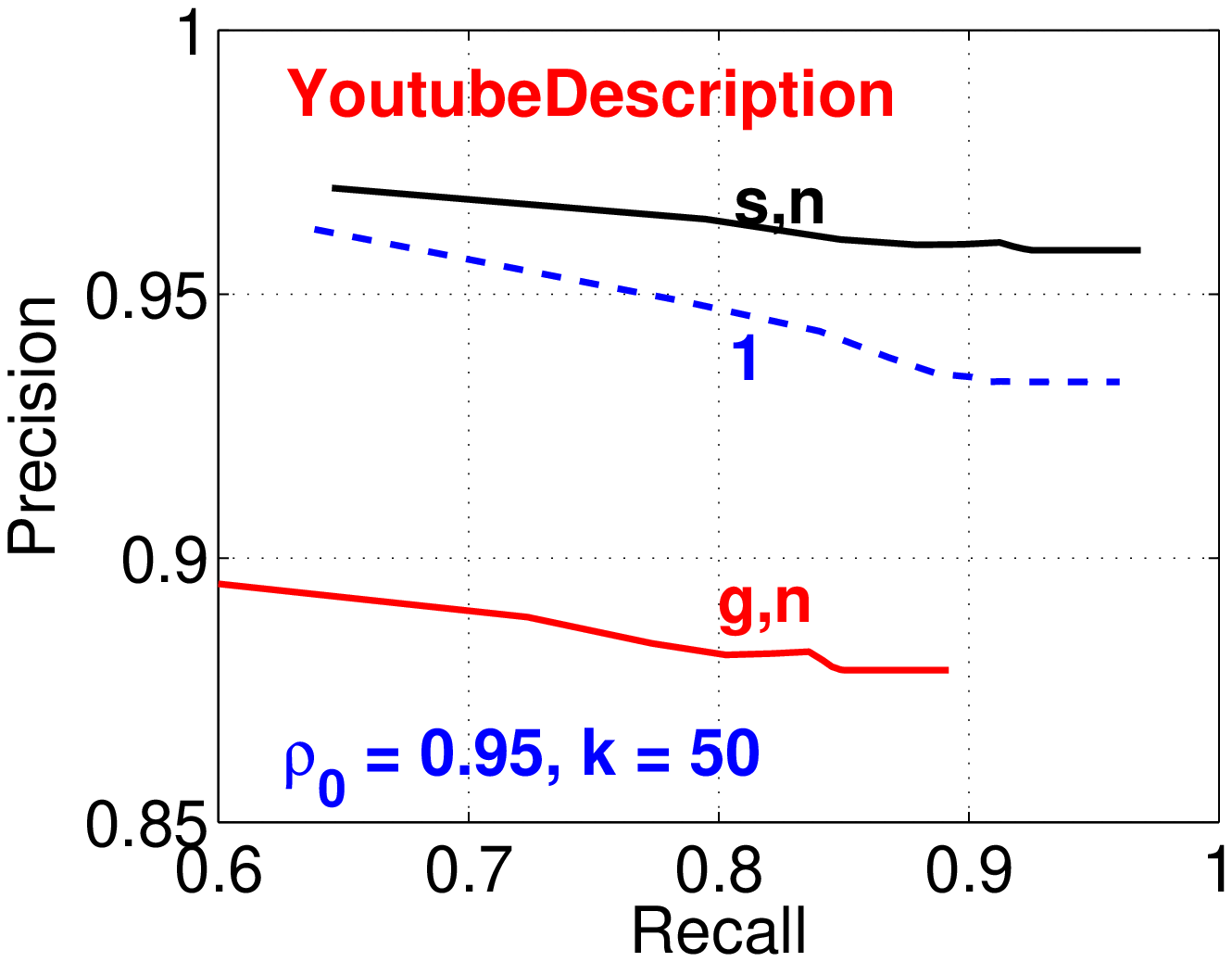}\hspace{-0.15in}
\includegraphics[width=2.2in]{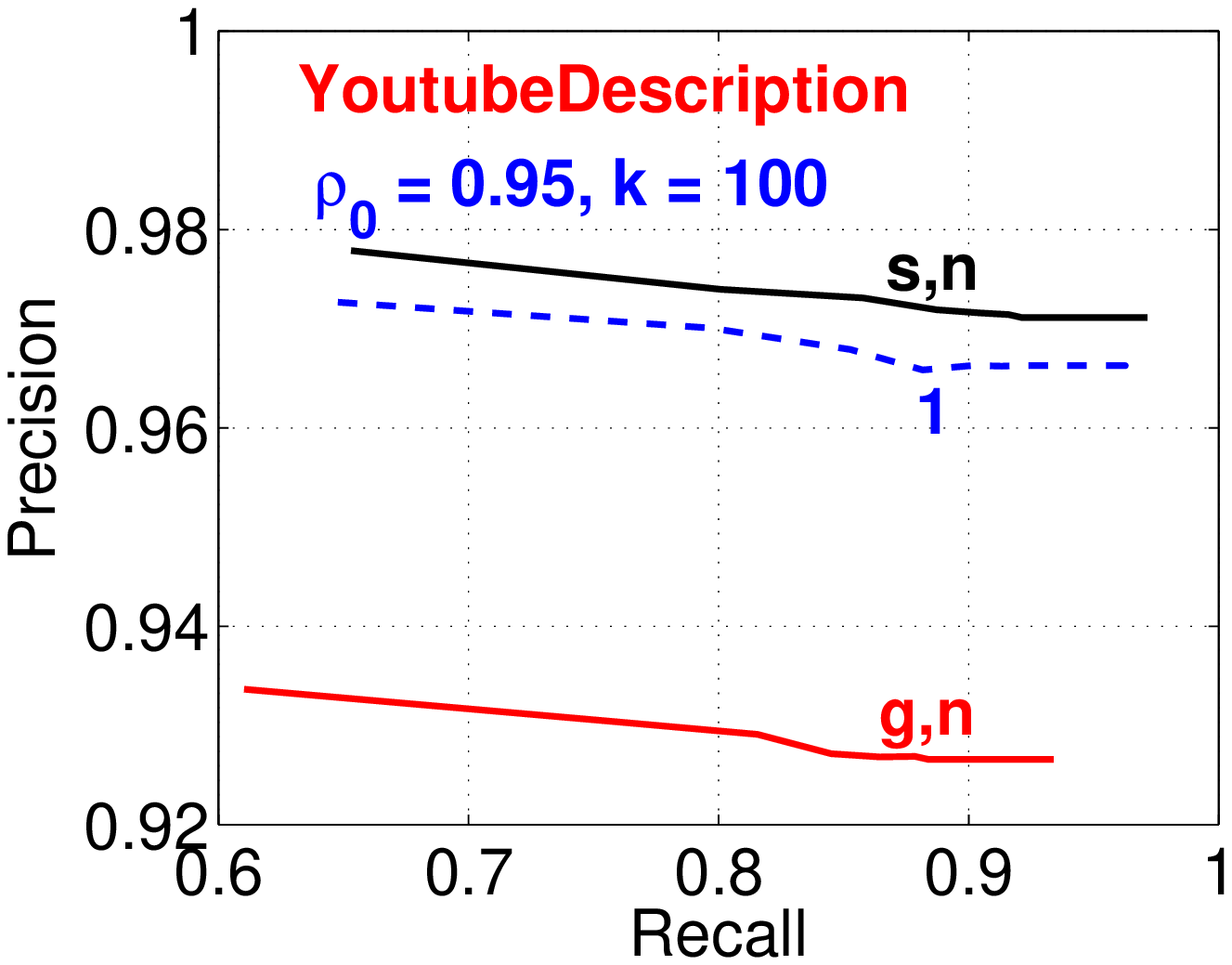}\hspace{-0.15in}
\includegraphics[width=2.2in]{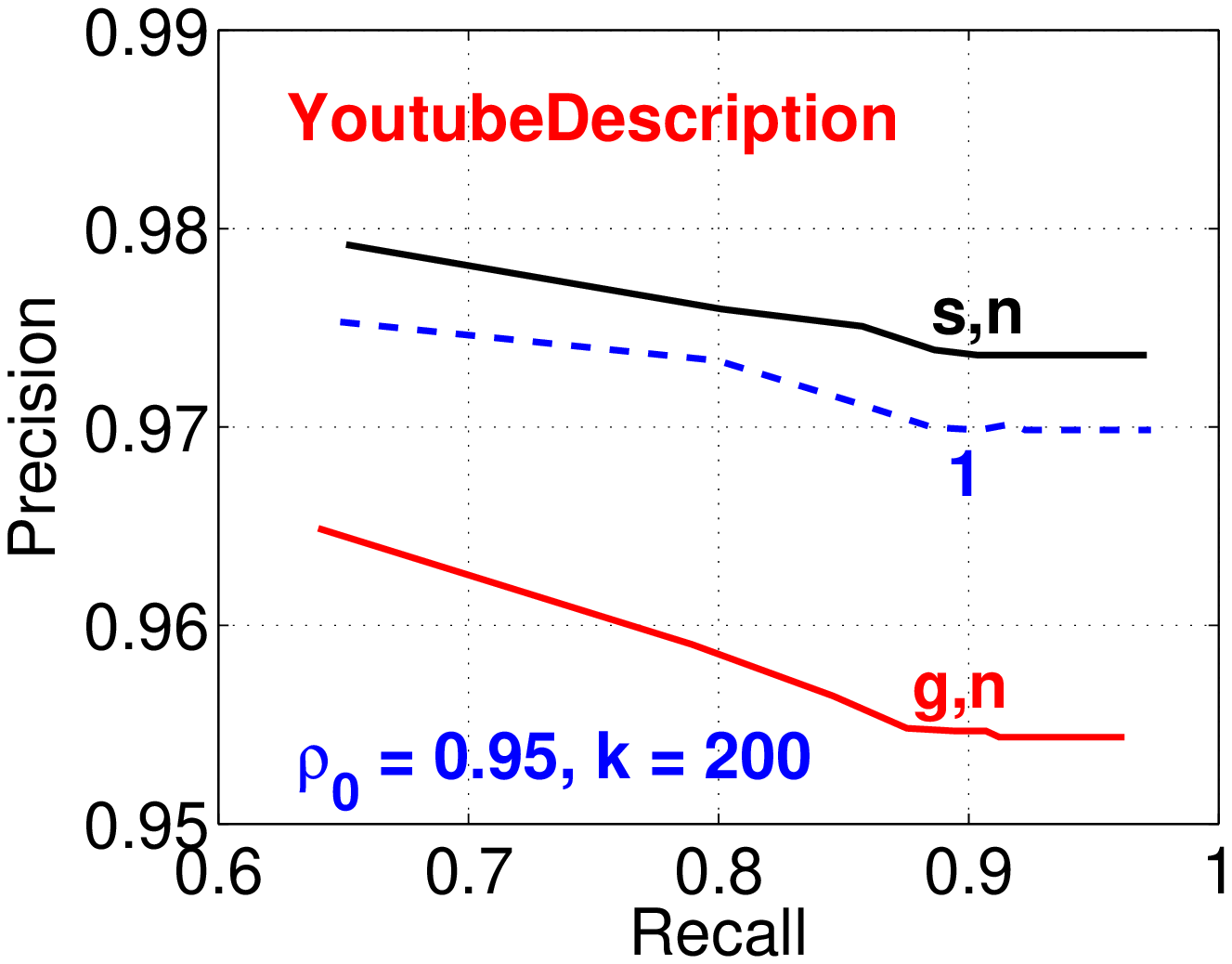}\hspace{-0.15in}
}

\mbox{
\includegraphics[width=2.2in]{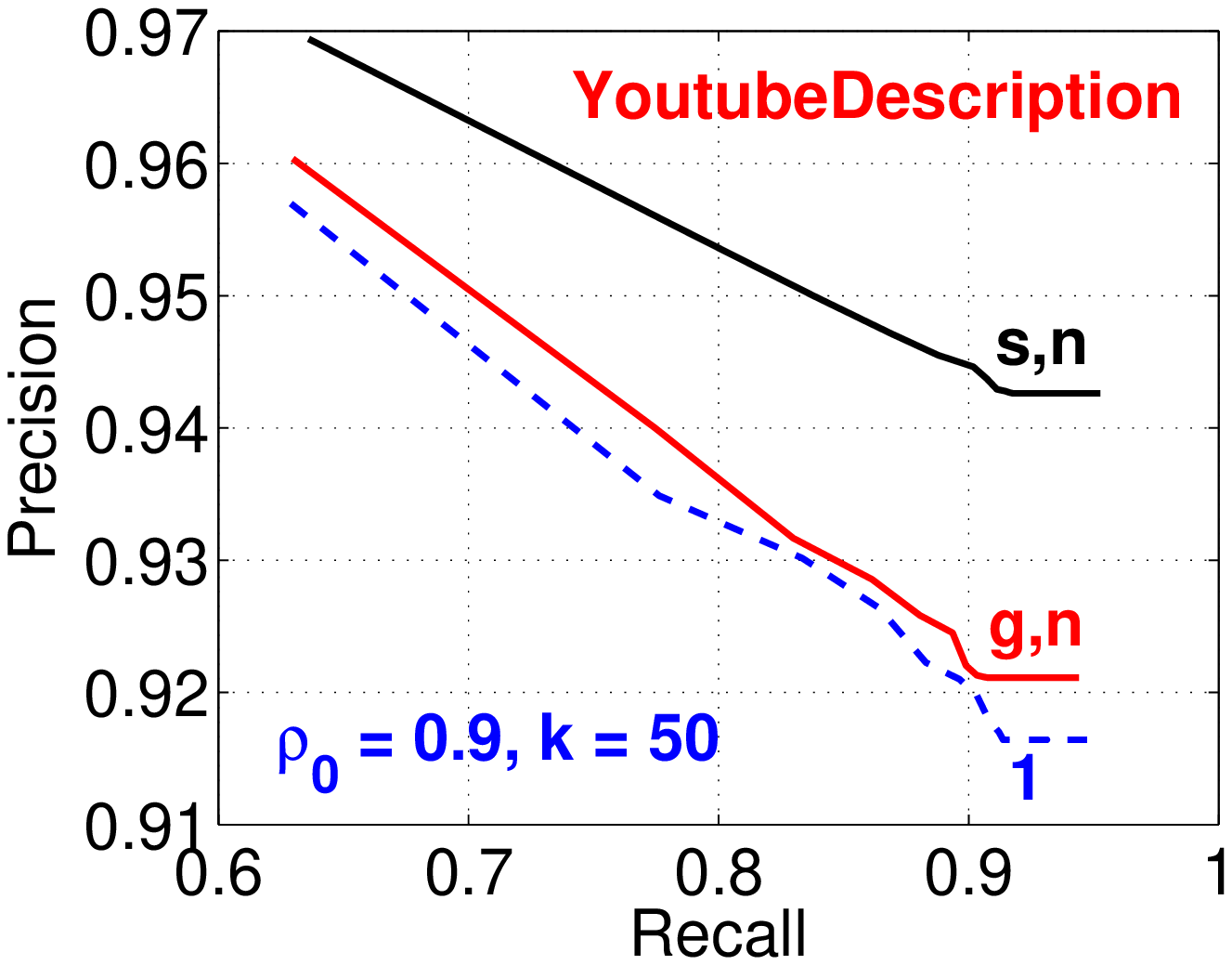}\hspace{-0.15in}
\includegraphics[width=2.2in]{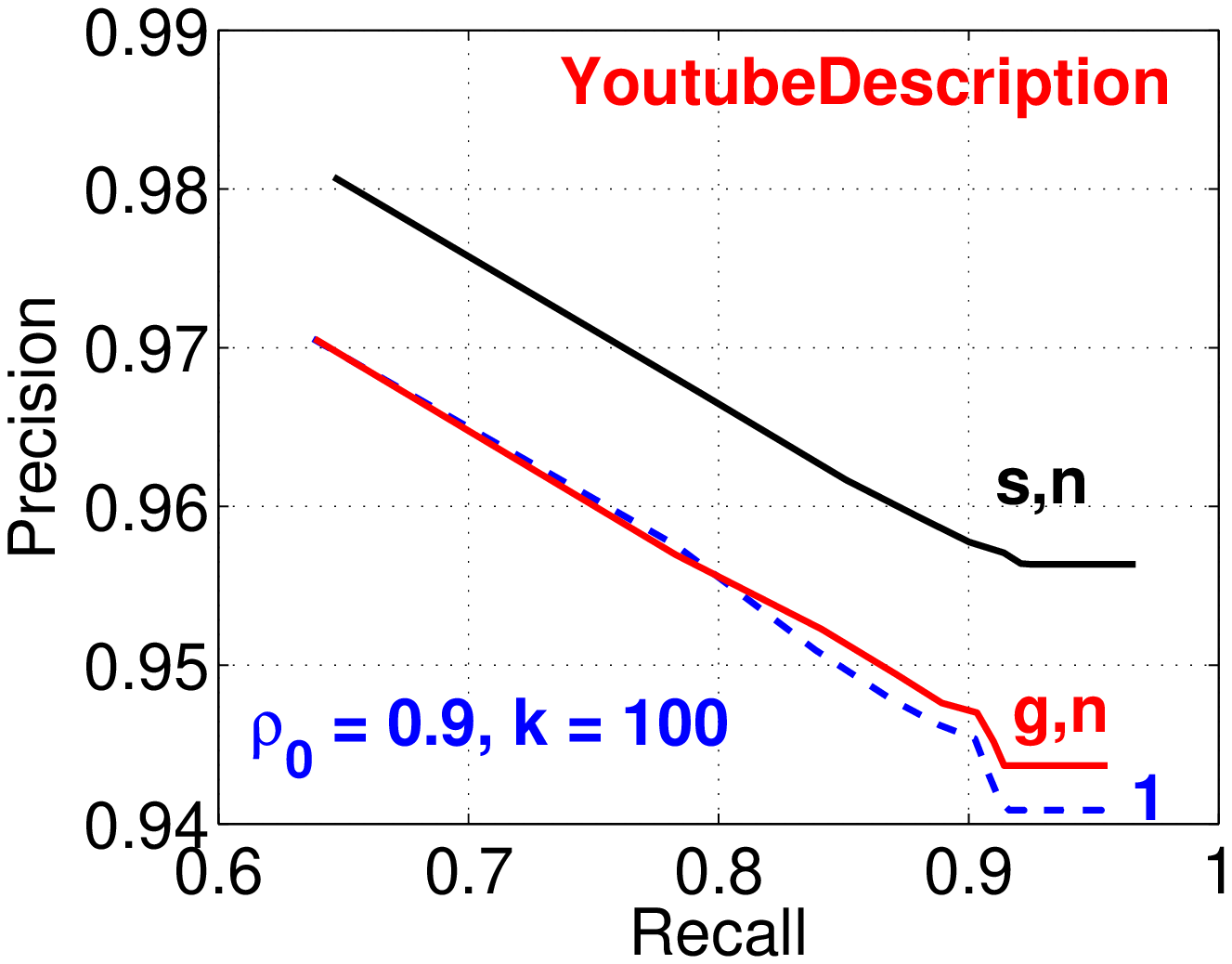}\hspace{-0.15in}
\includegraphics[width=2.2in]{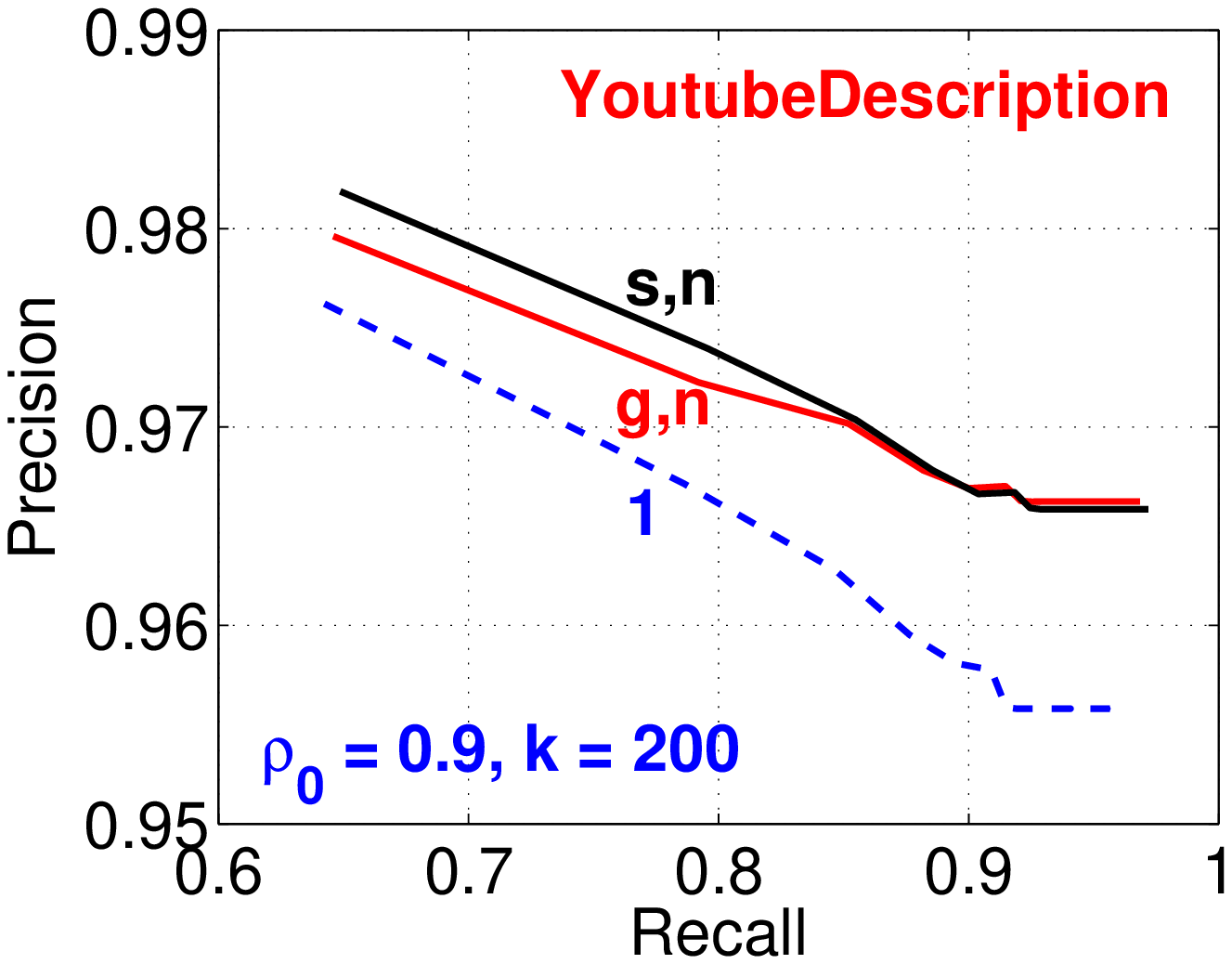}\hspace{-0.15in}
}

\mbox{
\includegraphics[width=2.2in]{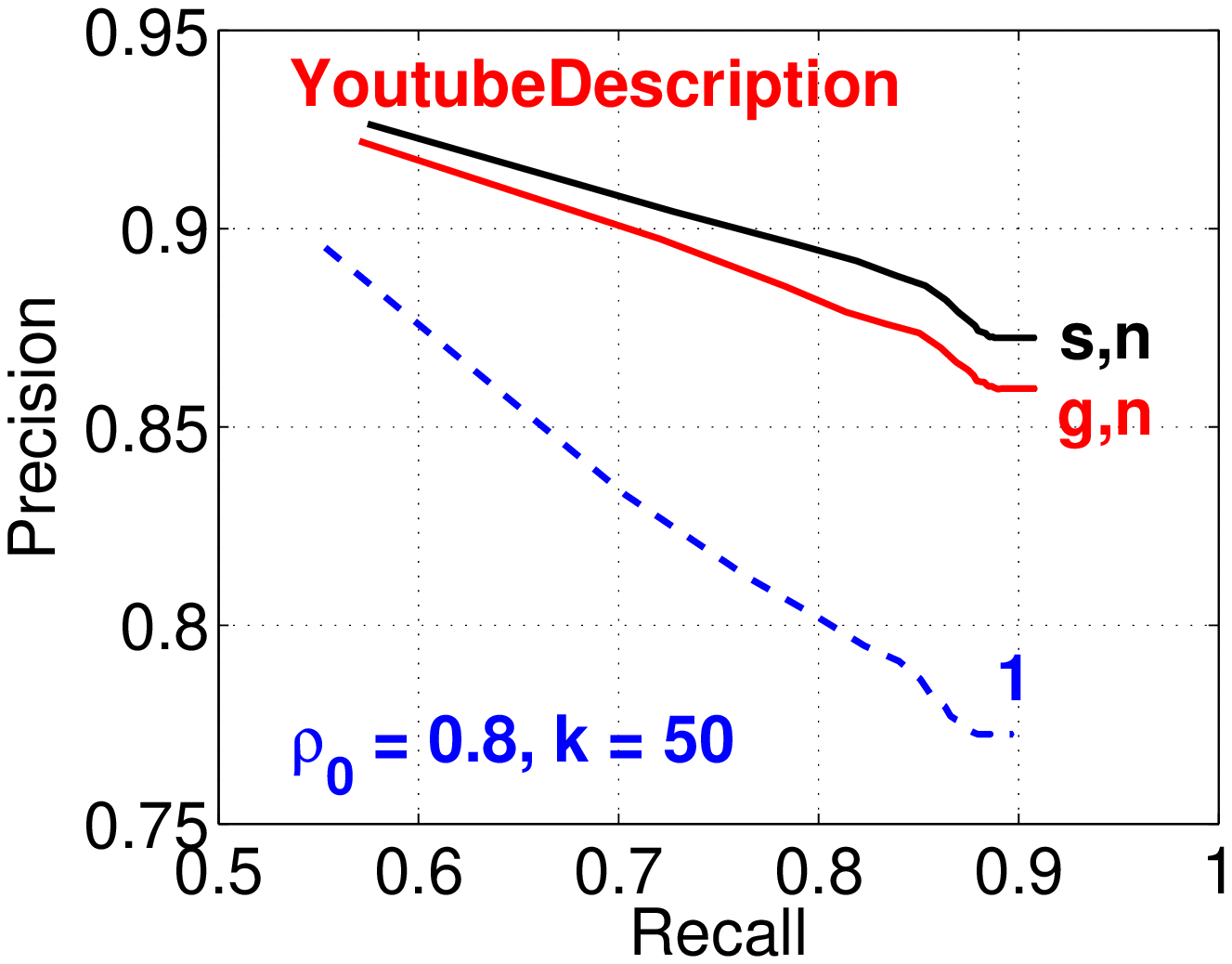}\hspace{-0.15in}
\includegraphics[width=2.2in]{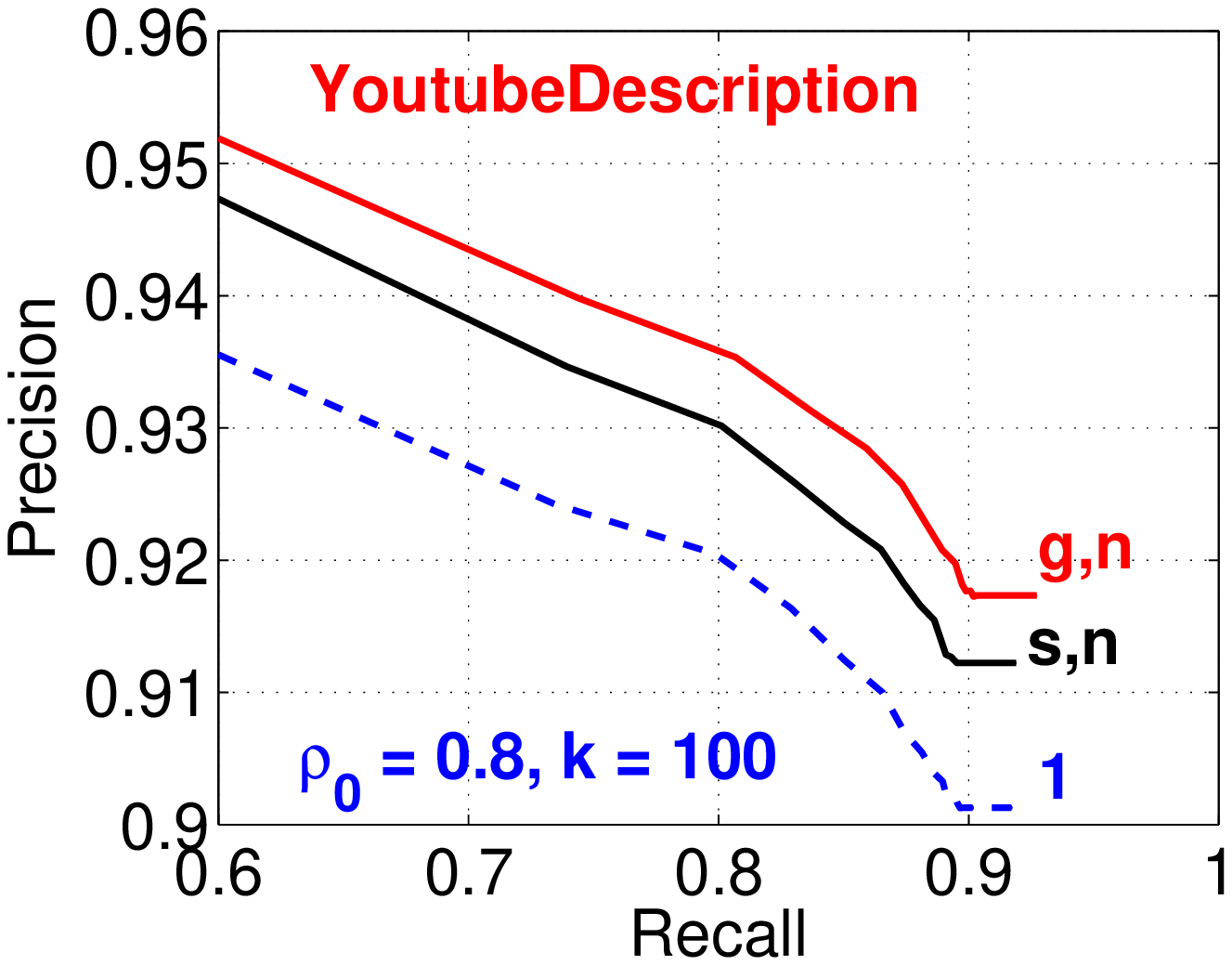}\hspace{-0.15in}
\includegraphics[width=2.2in]{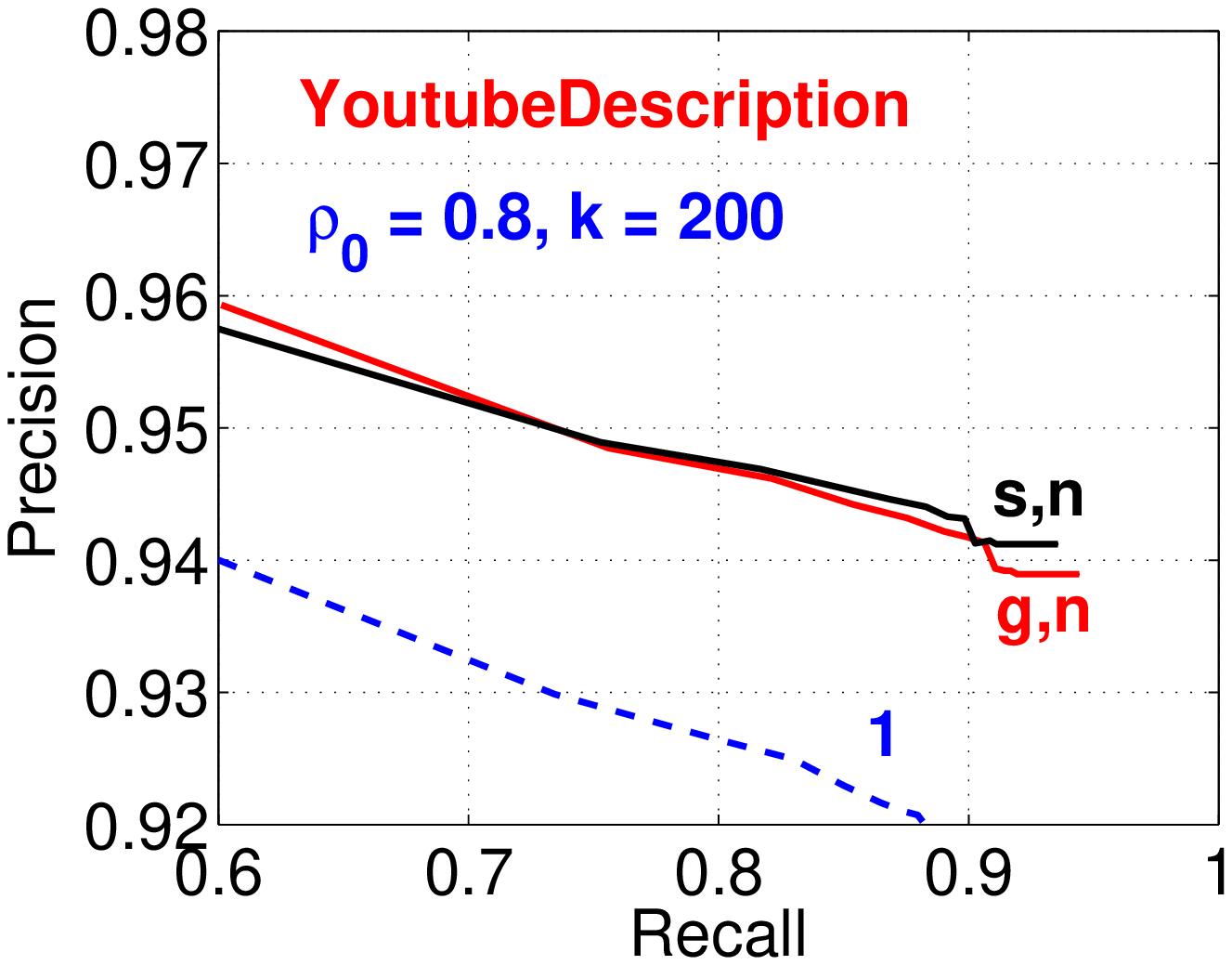}\hspace{-0.15in}
}

\mbox{
\includegraphics[width=2.2in]{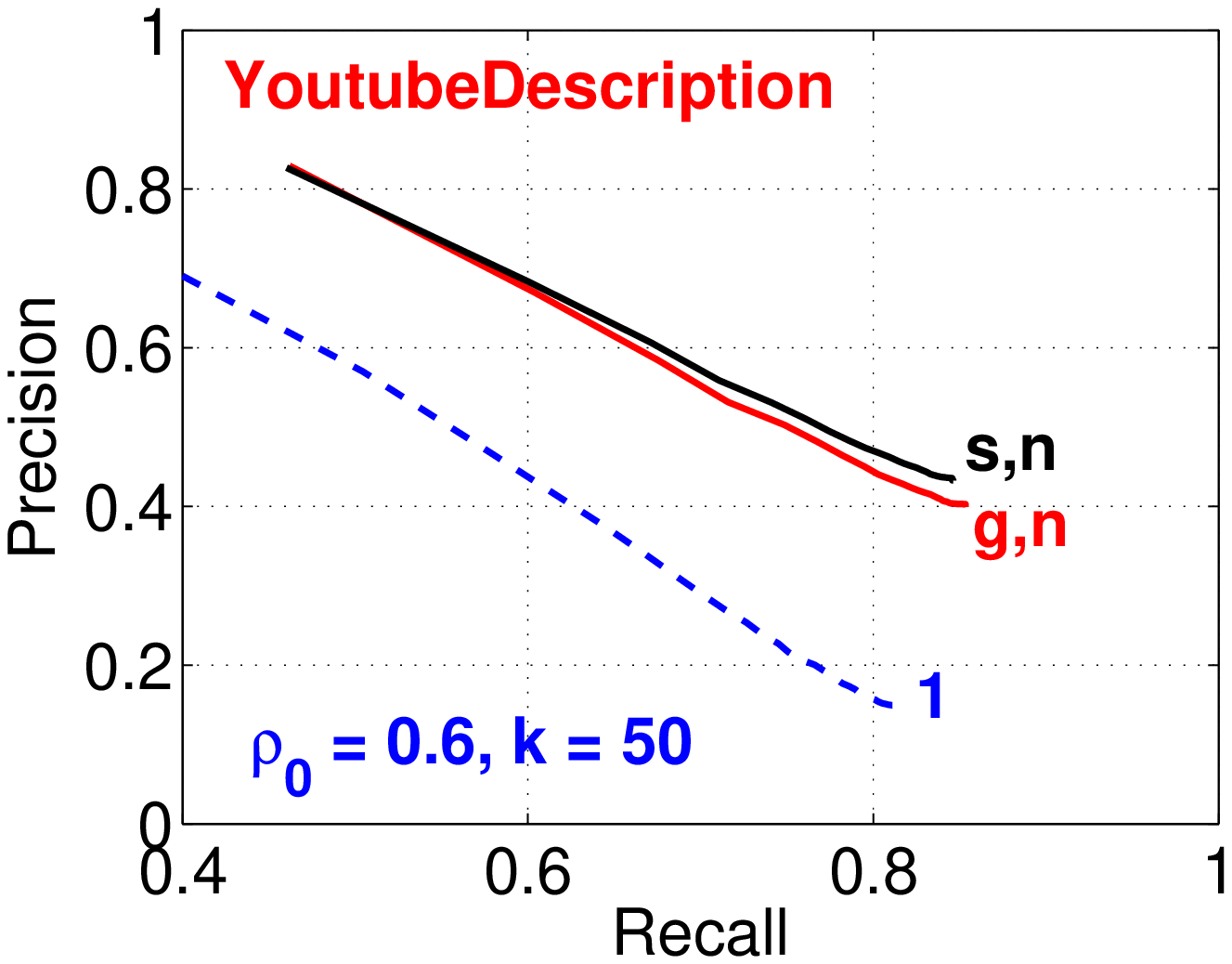}\hspace{-0.15in}
\includegraphics[width=2.2in]{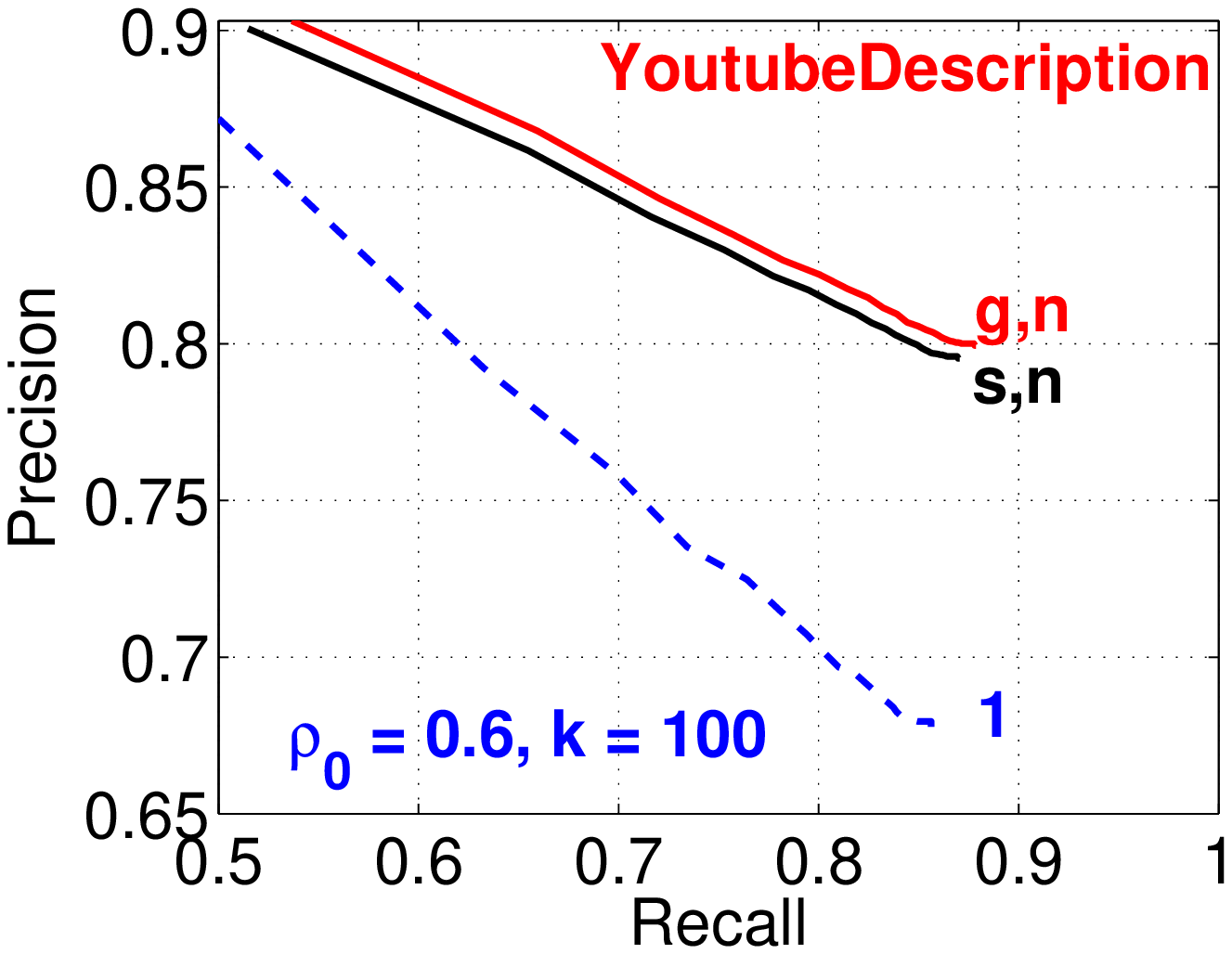}\hspace{-0.15in}
\includegraphics[width=2.2in]{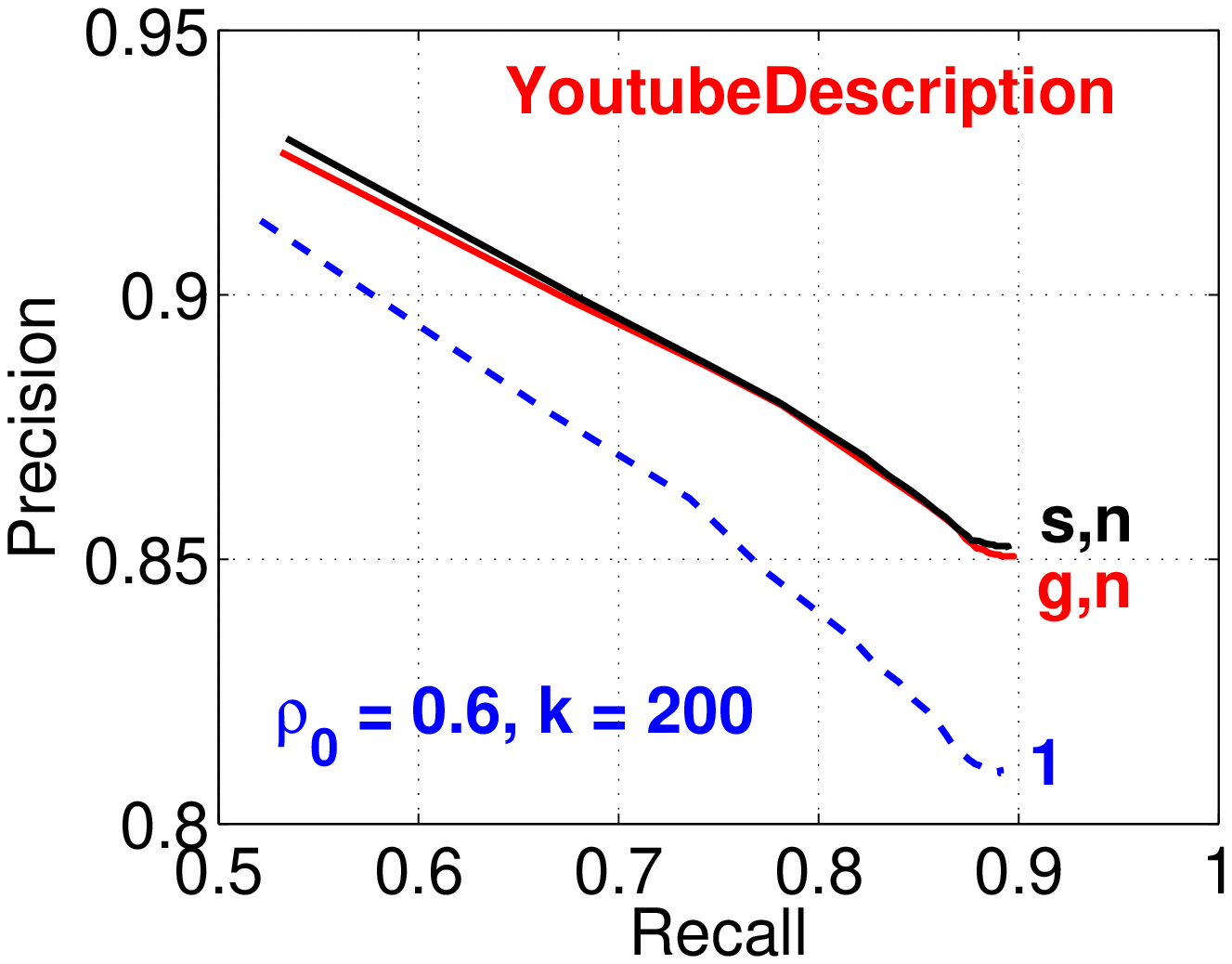}\hspace{-0.15in}
}

\mbox{
\includegraphics[width=2.2in]{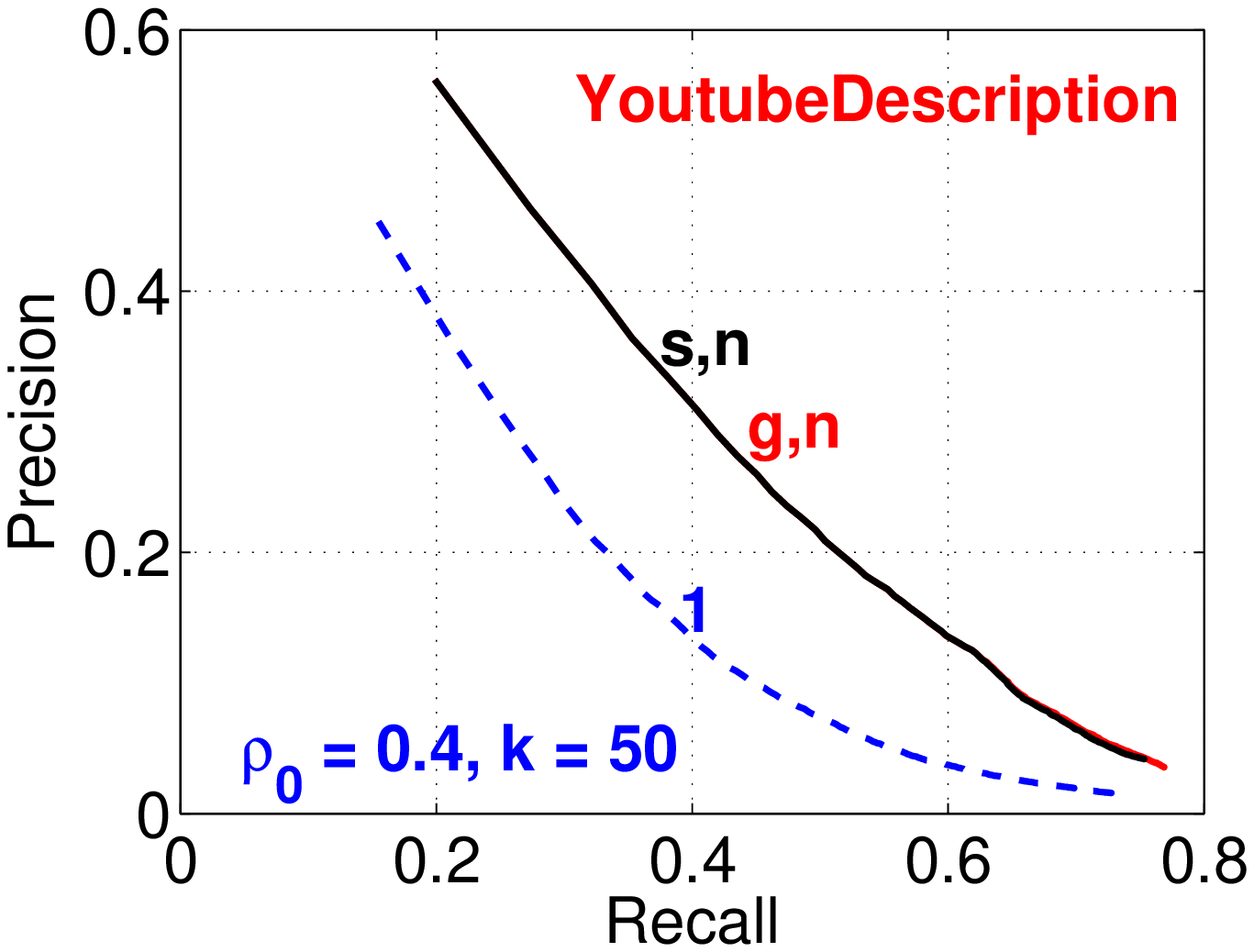}\hspace{-0.15in}
\includegraphics[width=2.2in]{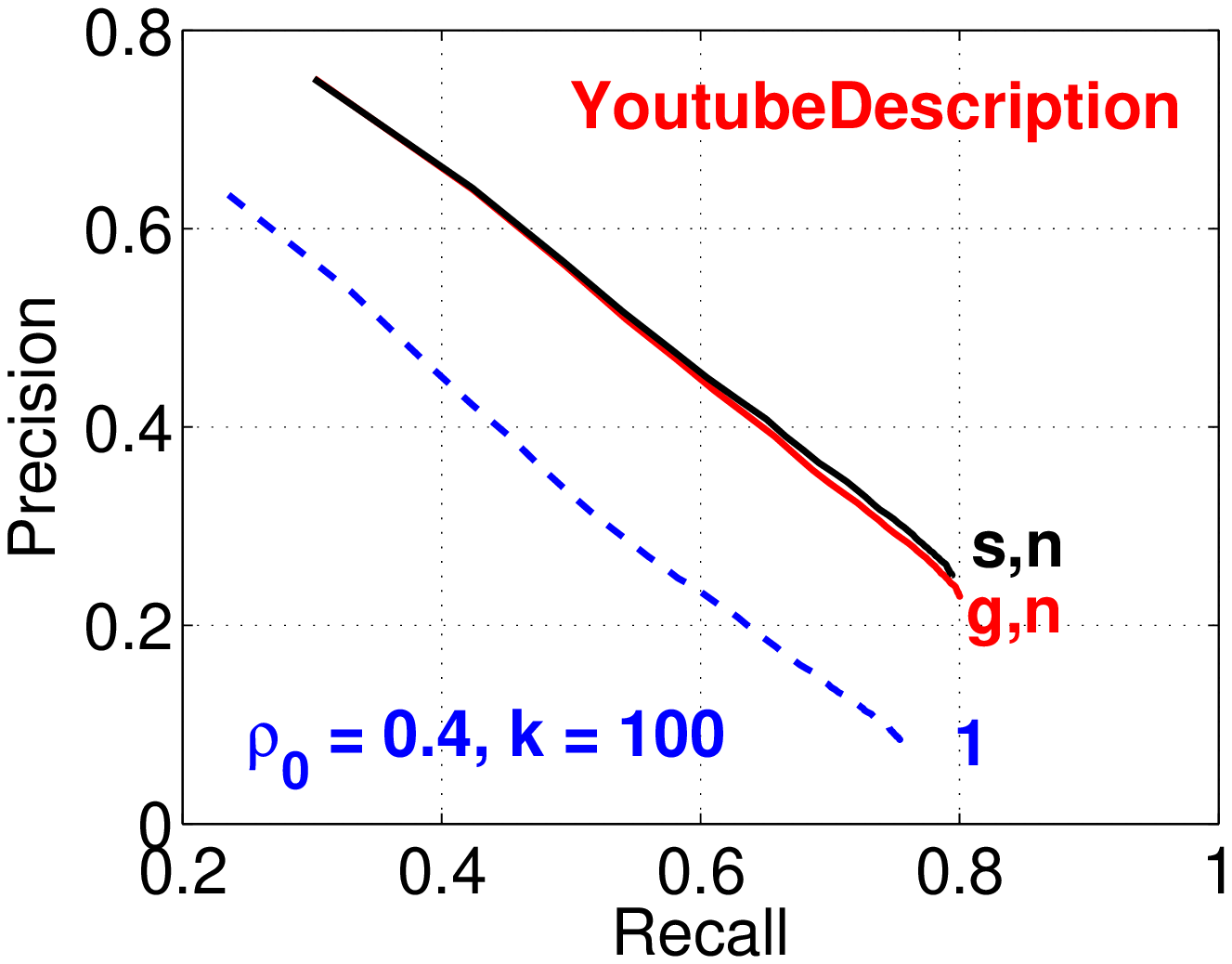}\hspace{-0.15in}
\includegraphics[width=2.2in]{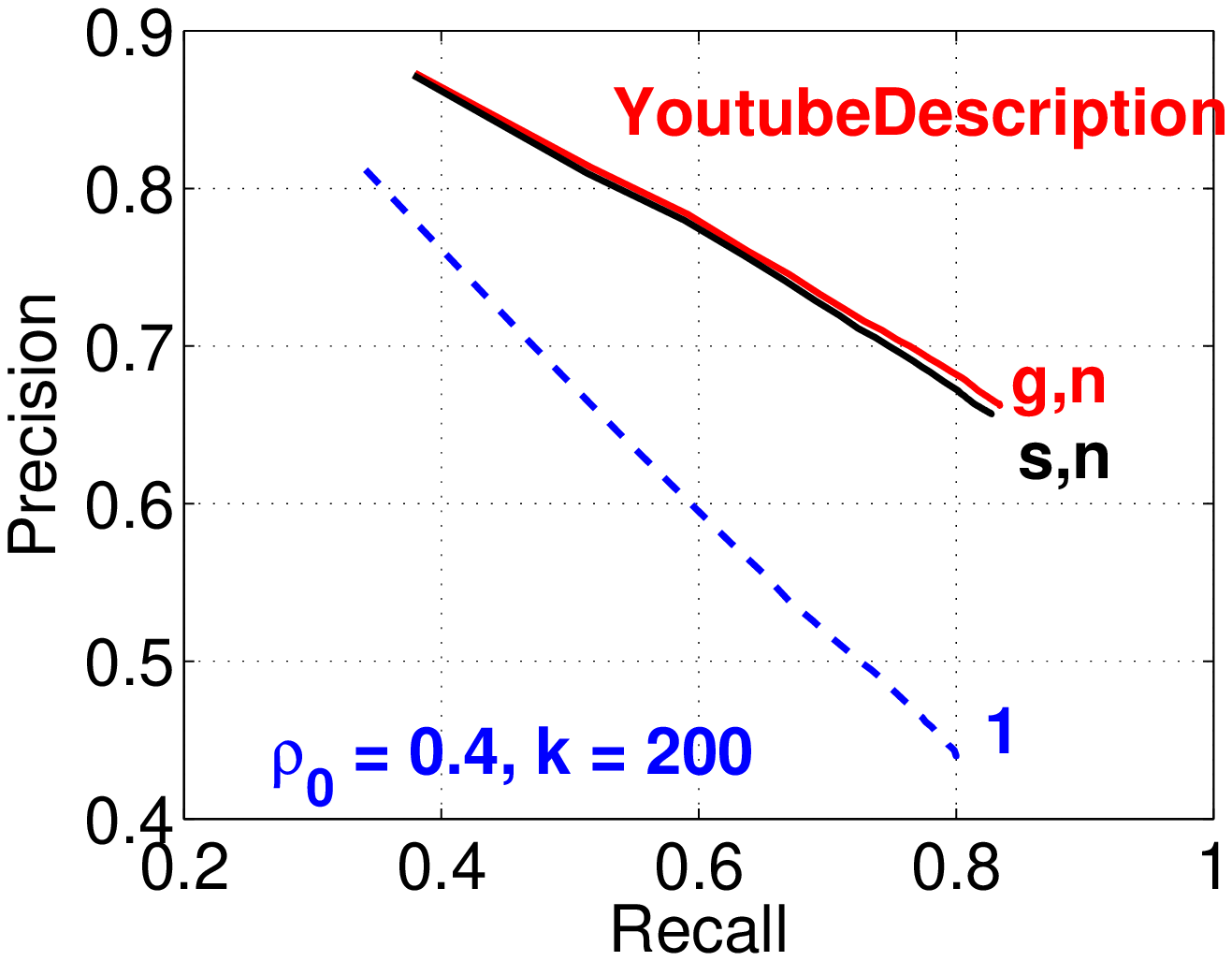}\hspace{-0.15in}
}

\end{center}
\vspace{-0.2in}
\caption{\textbf{YoutubeDescription}: precision-recall curves for selected $\rho_0$ and $k$ values, and  for three estimators： $\hat{\rho}_{s,n}$, $\hat{\rho}_{g,n}$, $\hat{\rho}_1$.}\label{fig_pr_YoutubeDescription}
\end{figure*}

Figure~\ref{fig_pr_MNIST}, Figure~\ref{fig_pr_YoutubeAudio}, and Figure~\ref{fig_pr_YoutubeDescription} present the results for the other three datasets. The trends are pretty much similar to what we observe in Figure~\ref{fig_pr_RCV1}. These results confirm that $\hat{\rho}_{s,n}$ is an overall good estimator, which we recommend for practical use.

\section{Conclusion}

The method of sign-sign (1-bit) random projections has been a standard tool in practice. In many practical scenarios such as near-neighbor search and near-neighbor classification, we can store signs of the projected data  and discard the original high-dimensional data. When a new data point arrives, we generate its projected vector and we can use the full-time to estimate the similarity. We develop four simple estimators for sign-full random projections. In particular, we recommend $\hat{\rho}_{s,n}$ which almost matches the accuracy of the MLE at least for nonnegative data. The improvement over 1-bit  projections is substantial especially for high similarity region. 

\newpage\clearpage

\appendix

\section{Proof of Theorem 1}

Consider two high-dimensional vectors, $u, v\in\mathbb{R}^D$. The idea is to multiply them with a random normal projection matrix $\mathbf{R}\in\mathbb{R}^{D\times k}$, to generate two (much) shorter vectors $x, y$:
\begin{align}\notag
x = u\times \mathbf{R} \in\mathbb{R}^k,\hspace{0.2in} y = v\times \mathbf{R} \in\mathbb{R}^k, \hspace{0.2in} \mathbf{R} = \{r_{ij}\}{_{i=1}^D}{_{j=1}^k}, \hspace{0.2in} r_{ij} \sim N(0,1) \text{ i.i.d. }
\end{align}

In this context, without loss of generality, we assume $\|u\|=\|v\|=1$ in this paper.  The joint distribution of $(x_j, y_j)$ is hence a bi-variant normal:
\begin{align}\notag
\left[\begin{array}{c}x_j\\ y_j\end{array} \right] \sim N
\left(
\left[\begin{array}{c}0\\ 0\end{array} \right],\
\left[\begin{array}{cc}1 &\rho\\ \rho &1
\end{array} \right]
\right), \text{ i.i.d.}\hspace{0.25in} j = 1, 2, ..., k.
\end{align}
where $\rho = \sum_{i=1}^D u_iv_i$ (assuming $\|u\|=\|v\|=1$).   The joint likelihood is
\begin{align}\notag
L({sgn}(x_j), y_j) = \prod_{{sgn}(x_j)=-1 } \int_{-\infty}^0\phi(x,y)dx \prod_{{sgn}(x_j)=1 } \int_0^{\infty}\phi(x,y)dx
\end{align}
In this paper, we  denote
\begin{align}\notag
&\phi(x,y;\rho) = \frac{1}{2\pi\sqrt{1-\rho^2}}e^{-\frac{x^2-2\rho xy +y^2}{2(1-\rho^2)}}, \ \ -1\leq \rho\leq 1\\\notag
&\phi(x) = \frac{1}{\sqrt{2\pi}}e^{-x^2/2},\hspace{0.2in} \Phi(x) = \int_{-\infty}^x \phi(x)dx
\end{align}
 The joint log-likelihood is
\begin{align}\notag
l&=l({sgn}(x_j), y_j) = \log L({sgn}(x_j), y_j) \\\notag
&=  \sum_{{sgn}(x_j)=-1 }\log \int_{-\infty}^0\phi(x,y;\rho)dx+ \sum_{{sgn}(x_j)=1 }\log \int_0^{\infty}\phi(x,y;\rho)dx
\end{align}
where
\begin{align}\notag
&\int_0^{\infty}\phi(x,y;\rho)dx= \int_{0}^{\infty}\frac{1}{2\pi\sqrt{1-\rho^2}}e^{-\frac{x^2-2\rho xy+y^2}{2(1-\rho^2)}}dx \\\notag
=&\frac{1}{2\pi\sqrt{1-\rho^2}}e^{-\frac{y^2}{2}}\int_{0}^{\infty}
e^{-\frac{(x-\rho y)^2}{2(1-\rho^2)}}dx
=\frac{1}{2\pi\sqrt{1-\rho^2}}e^{-\frac{y^2}{2}}\int_{\frac{-\rho y}{\sqrt{1-\rho^2}}}^{\infty}
e^{-\frac{u^2}{2}}\sqrt{1-\rho^2}du\\\notag
=&\frac{1}{\sqrt{2\pi}}e^{-\frac{y^2}{2}}\int_{\frac{-\rho x}{\sqrt{1-\rho^2}}}^{\infty}\frac{1}{\sqrt{2\pi}}
e^{-\frac{u^2}{2}}du
=\frac{1}{\sqrt{2\pi}}e^{-\frac{y^2}{2}}\Phi\left(\frac{\rho y}{\sqrt{1-\rho^2}}\right)
\end{align}
and
\begin{align}\notag
\int_{-\infty}^0\phi(x,y;\rho)dx
=&\frac{1}{\sqrt{2\pi}}e^{-\frac{y^2}{2}}\Phi\left(\frac{-\rho y}{\sqrt{1-\rho^2}}\right)
\end{align}
Thus,
\begin{align}\notag
l=&\sum_{j,{sgn}(x_j)=-1 }\log \left[\frac{1}{\sqrt{2\pi}}e^{-\frac{y_j^2}{2}}\Phi\left(\frac{-\rho y_j}{\sqrt{1-\rho^2}}\right)\right]
 +\sum_{j,{sgn}(x_j)=1 }\log \left[\frac{1}{\sqrt{2\pi}}e^{-\frac{y_j^2}{2}}\Phi\left(\frac{\rho y_j}{\sqrt{1-\rho^2}}\right)\right]\\\notag
=& \sum_{j=1}^k\log \Phi\left(\frac{\rho}{\sqrt{1-\rho^2}}{sgn}(x_j) y_j\right)\hspace{0.3in} \text{irrelevant terms are neglected}
\end{align}
Once we have the likelihood function, we  obtain the MLE equation by setting its derivative $l^\prime(\rho)=0$
\begin{align}\notag
l^\prime(\rho) =  \sum_{j=1}^k\frac{\phi\left(\frac{\rho}{\sqrt{1-\rho^2}}{sgn}(x_j) y_j\right)}{\Phi\left(\frac{\rho}{\sqrt{1-\rho^2}}{sgn}(x_j) y_j\right)}
\frac{{sgn}(x_j)y_j}{(1-\rho^2)^{3/2}}=0
\end{align}

We will also need to the second derivative $l^{\prime\prime}$ in order to assess the asymptotic variance of the MLE by classification theory of statistics. After some algebra, we obtain
\begin{align}\notag
l^{\prime\prime}(\rho)
=&\frac{-\rho}{(1-\rho^2)^{7/2}}\sum_{j=1}^k\frac{\phi\left(\frac{\rho}{\sqrt{1-\rho^2}}{sgn}(x_j) y_j\right)
}{\Phi\left(\frac{\rho}{\sqrt{1-\rho^2}}{sgn}(x_j) y_j\right)} {sgn}(x_j)y_j^3
 -
\frac{1}{(1-\rho^2)^{3}}\sum_{j=1}^k\frac{\phi^2\left(\frac{\rho}{\sqrt{1-\rho^2}}{sgn}(x_j) y_j\right)
}{\Phi^2\left(\frac{\rho}{\sqrt{1-\rho^2}}{sgn}(x_j) y_j\right)} y_j^2\\\notag
&+
\frac{3\rho}{(1-\rho^2)^{5/2}}\sum_{j=1}^k\frac{\phi\left(\frac{\rho}{\sqrt{1-\rho^2}}{sgn}(x_j) y_j\right)}{\Phi\left(\frac{\rho}{\sqrt{1-\rho^2}}{sgn}(x_j) y_j\right)}
{sgn}(x_j)y_j
\end{align}

We can evaluate the Fisher Information $-E(l^{\prime\prime}(\rho)$ numerically or by simulations. The asymptotic variance is the reciprocal of $-E(l^{\prime\prime}(\rho)$.

\section{Proof of Lemma 1}\label{app_lem_int}

Let $c = \frac{\rho}{\sqrt{1-\rho^2}}$, we have
\begin{align}\notag
&\int_0^\infty te^{-\frac{t^2}{2}} \Phi\left(ct\right)dy
=\int_0^\infty - \Phi\left(ct\right)de^{-\frac{y^2}{2}}\\\notag
=&\int_0^\infty e^{-\frac{y^2}{2}}\phi\left(cy\right)cdy +\frac{1}{2}\\\notag
=&c\sqrt{\frac{1}{1+c^2}}\int_0^\infty\frac{1}{\sqrt{2\pi}}\sqrt{1+c^2}e^{-(1+c^2)y/2} dy+\frac{1}{2}\\\notag
=&c\sqrt{\frac{1}{1+c^2}}\frac{1}{2}+\frac{1}{2}
=\frac{1+\rho}{2}
\end{align}

\begin{align}\notag
&\int_0^\infty t^3 e^{-t^2/2}\Phi(ct)dt
=-\int_0^\infty t^2\Phi(ct)d e^{-t^2/2}\\\notag
=&\int_0^\infty e^{-t^2/2} \left[2t\Phi(ct)+ct^2\phi(ct)\right]dt - 0 \\\notag
=&-\int_0^\infty 2 \Phi(ct) de^{-t^2/2}  + \int_0^\infty ct^2 e^{-t^2/2}\frac{1}{\sqrt{2\pi}} e^{-\frac{c^2t^2}{2}}dt \\\notag
=&\int_0^\infty 2ce^{-t^2/2} \phi(ct)dt +1  + \int_0^\infty ct^2 \frac{1}{\sqrt{2\pi}} e^{-\frac{(1+c^2)t^2}{2}}dt \\\notag
=&\int_0^\infty \frac{1}{\sqrt{1+c^2}}2c \frac{\sqrt{1+c^2}}{\sqrt{2\pi}}e^{-\frac{(1+c^2)t^2}{2}}dt +1\\\notag
&  + \int_0^\infty \frac{1}{\sqrt{1+c^2}}ct^2 \frac{\sqrt{1+c^2}}{\sqrt{2\pi}} e^{-\frac{(1+c^2)t^2}{2}}dt \\\notag
=&\frac{c}{\sqrt{1+c^2}}+1 + \frac{c}{2(1+c^2)^{3/2}}\\\notag
=&\rho+1 + \rho(1-\rho^2)/2=1+3/2\rho-\rho^3/2
\end{align}

Consider $c>0$, we have
\begin{align}\notag
&\int_0^\infty y^2 e^{-y^2/2}\Phi(cy)dy\\\notag
=&\int_0^\infty \frac{2}{c^2}ue^{-u/c^2}\Phi(\sqrt{2u}) \frac{\sqrt{2}}{c}\frac{1}{2} \frac{1}{u^{1/2}} du \\\notag
=&\frac{\sqrt{2}}{c^3}\int_0^\infty \sqrt{u}e^{-u/c^2}\Phi(\sqrt{2u})du\\\notag
=&\frac{\sqrt{2}}{2c^3}\int_0^\infty \sqrt{u}e^{-u/c^2}2\left(\Phi(\sqrt{2u})-1\right)du
+\frac{\sqrt{2}}{c^3}\int_0^\infty \sqrt{u}e^{-u/c^2}du\\\notag
=&-\frac{\sqrt{2}}{2c^3}\frac{1}{\sqrt{\pi}}\left(\frac{\tan^{-1}\frac{1}{c}}{\frac{1}{(c^2)^{3/2}}}-\frac{1}{1/c^2 (1+1/c^2)}\right)+\frac{\sqrt{2\pi}}{2}\\\notag
=&\sqrt{\frac{\pi}{2}}-\sqrt{\frac{1}{2\pi}}\left(\tan^{-1}\frac{1}{c}-\frac{c}{c^2+1}\right)
\end{align}
where we have used the result in \cite[8.258.5]{Book:Gradshteyn_94} which says
\begin{align}\notag
\int_0^\infty \sqrt{x}\text{erfc}(\sqrt{x})e^{-\beta x}dx =& \int_0^\infty \sqrt{x}\left(2-2\Phi(\sqrt{2x})\right)e^{-\beta x}dx\\\notag
 =& \frac{1}{\sqrt{\pi}}\left(\frac{\tan^{-1}\sqrt{\beta}}{\beta^{3/2}}-\frac{1}{\beta(1+\beta)}\right)
\end{align}
Note that \cite[8.258.5]{Book:Gradshteyn_94} incorrectly included a $\frac{1}{2}$ factor.

Now consider $c<0$, we have
\begin{align}\notag
&\int_0^\infty y^2 e^{-y^2/2}\Phi(cy)dy
=\int_0^\infty y^2 e^{-y^2/2}\left(1-\Phi(-cy)\right)dy\\\notag
=&\frac{\sqrt{2\pi}}{2}-\int_0^\infty y^2 e^{-y^2/2}\Phi(-cy)dy\\\notag
=&\frac{\sqrt{2\pi}}{2}-\sqrt{\frac{\pi}{2}}+\sqrt{\frac{1}{2\pi}}\left(\tan^{-1}\frac{1}{-c}-\frac{-c}{c^2+1}\right)\\\notag
=&-\sqrt{\frac{1}{2\pi}}\left(\tan^{-1}\frac{1}{c}-\frac{c}{c^2+1}\right)
\end{align}
Note that when $c=0$, we have
\begin{align}\notag
&\int_0^\infty y^2 e^{-y^2/2}\Phi(cy)dy
=\frac{\sqrt{2\pi}}{2}\int_0^\infty y^2 \frac{1}{\sqrt{2\pi}}e^{-y^2/2}dy=\frac{\sqrt{2\pi}}{4}
\end{align}

Therefore, for general $c$, we have
\begin{align}\notag
&\int_0^\infty y^2 e^{-y^2/2}\Phi(cy)dy
=1_{c\geq0}\sqrt{\frac{\pi}{2}}-\sqrt{\frac{1}{2\pi}}\left(\tan^{-1}\frac{1}{c}-\frac{c}{c^2+1}\right)
\end{align}
Note that we follow the convention that $\tan^{-1}\frac{1}{0} = \tan^{-1}\frac{1}{0+}=\frac{\pi}{2}$.

\section{Proof of Theorem 2}

Firstly, it is obvious that ${E}\left(({sgn}(x_j)^2 y_j)^2\right) = {E}\left(y_j^2\right) = 1$, and ${E}\left(({sgn}(x_j) y_j)^4\right) = {E}\left(y_j^4\right) = 3$. Because $(x_j,x_j)$ is bi-variate normal, we have $x_j|y_j \sim N\left(\rho y_j,\ \ (1-\rho^2)\right)$ and
\begin{align}\notag
&{E}\left({sgn}(x_j) y_j)\right)
={E}\left(y_j{E}\left({sgn}(x_j) |y_j\right)\right)\\\notag
=&{E}\left(y_j{Pr}\left(x_j|y_j\geq0\right) -y_j{Pr}\left(x_j|y_j<0\right) \right)\\\notag
=&{E}\left(y_j\left(1-2\Phi\left(\frac{-\rho y_j}{\sqrt{1-\rho^2}}\right)\right)\right)\\\notag
=&{E}\left(y_j\left(2\Phi\left(\frac{\rho y_j}{\sqrt{1-\rho^2}}\right)-1\right)\right)\\\notag
=&2\int_{-\infty}^\infty t\phi(t)\Phi\left(\frac{\rho t}{\sqrt{1-\rho^2}}\right)dt \\\notag
=&4\int_{0}^\infty t\phi(t)\Phi\left(\frac{\rho t}{\sqrt{1-\rho^2}}\right)dt -2\int_0^\infty t\phi(t)dt\\\notag
=&4\frac{1+\rho}{2}\frac{1}{\sqrt{2\pi}}-2\frac{1}{\sqrt{2\pi}}=\sqrt{\frac{2}{\pi}}\rho,\hspace{0.2in} \text{using result from Lemma 1}
\end{align}
Similarly
\begin{align}\notag
&{E}\left({sgn}(x_j) y_j^3)\right)\\\notag
=&4\int_{0}^\infty t^3\phi(t)\Phi\left(\frac{\rho t}{\sqrt{1-\rho^2}}\right)dt - 2 \int_{0}^\infty t^3\phi(t)dt\\\notag
=&\frac{1}{\sqrt{2\pi}}\left(6\rho-2\rho^3\right),\hspace{0.3in} \text{using result from Lemma 1}
\end{align}

\section{Proof of Theorem 3}

\ \ First, we denote  $Z_k = \frac{\sum_{j=1}^k sgn(x_j) y_j}{\sqrt{k}\sqrt{\sum_{j=1}^k y_j^2}}$.  As $k\rightarrow\infty$, we have
\begin{align}\notag
&\frac{1}{k}\sum_{j=1}^k y_j^2 \rightarrow E\left(y_j^2\right) = 1, \ \ a.s. \hspace{0.4in}
Z_k = \frac{\frac{1}{k}\sum_{j=1}^k {sgn}(x_j) y_j}{\sqrt{\frac{1}{k}k}\sqrt{\frac{1}{k}\sum_{j=1}^k y_j^2}} \rightarrow \sqrt{\frac{2}{\pi}}\rho= g, \ \ a.s.
\end{align}
We express the deviation $Z_k - g$ as
\begin{align}\notag
&Z_k - g
 = \frac{\frac{1}{k}\sum_{j=1}^k {sgn}(x_j) y_j - g + g}{\sqrt{\frac{1}{k}\sum_{j=1}^k y_j^2}} - g\\\notag
=& \frac{\frac{1}{k}\sum_{j=1}^k {sgn}(x_j) y_j - g }{\sqrt{\frac{1}{k}\sum_{j=1}^k y_j^2}} + g \frac{1-\sqrt{\frac{1}{k}\sum_{j=1}^k y_j^2} }{\sqrt{\frac{1}{k}\sum_{j=1}^k y_j^2}}\\\notag
=& \frac{1}{k}\sum_{j=1}^k {sgn}(x_j) y_j - g + g \frac{1-\frac{1}{k}\sum_{j=1}^k y_j^2}{2} + O_P(1/k)
\end{align}
Thus, to analyze the asymptotic variance, it suffices to study:
\begin{align}\notag
&{E}\left( {sgn}(x)y - g + g  \frac{1-y^2}{2}\right)^2= {E}\left({sgn}(x)y-g(1+y^2)/2\right)^2\\\notag
=&{E}(y^2)+g^2{E}(1+y^4+2y^2)/4  - g {E}({sgn}(x)(y+y^3))\\\notag
=&1+g^2(1+3+2)/4- g {E}({sgn}(x)(y+y^3))\\\notag
=&1+3/2g^2-g^2-gg_3
=1+g^2/2-gg_3=1-\frac{1}{\pi}\left(5\rho^2-2\rho^4\right)
\end{align}
where  we recall
\begin{align}\notag
g_3=&{E}\left({sgn}(x) y^3)\right)=\frac{1}{\sqrt{2\pi}}\left(6\rho-2\rho^3\right)
\end{align}

\section{Proof of Theorem 4}

\begin{align}\notag
&{E}\left(y_- 1_{x\geq0} + y_+1_{x<0}\right)
=2\int_0^\infty \int_{-\infty}^0 y \phi(x,y;\rho) dx dy\\\notag
=&2\int_{0}^{\infty}\int_{-\infty}^0\frac{1}{2\pi\sqrt{1-\rho^2}}ye^{-\frac{x^2-2\rho xy+y^2}{2(1-\rho^2)}}dxdy \\\notag
=&2\int_{0}^{\infty}\int_0^{\infty}\frac{1}{2\pi\sqrt{1-\rho^2}}ye^{-\frac{x^2+2\rho xy+y^2}{2(1-\rho^2)}}dxdy \\\notag
=&\frac{2}{2\pi\sqrt{1-\rho^2}}\int_0^\infty y e^{-\frac{y^2}{2}}\int_{0}^{\infty}e^{-\frac{(x+\rho y)^2}{2(1-\rho^2)}}dxdy\\\notag
=&\frac{2}{2\pi}\int_0^\infty y e^{-\frac{y^2}{2}}\int_{\frac{\rho y}{\sqrt{1-\rho^2}}}^{\infty}e^{-\frac{x^2}{2}}dxdy\\\notag
=&\frac{2}{\sqrt{2\pi}}\int_0^\infty y e^{-y^2/2} \Phi\left(-\frac{\rho y}{\sqrt{1-\rho^2}}\right) dy\\\notag
=&\frac{2}{\sqrt{2\pi}}\frac{1-\rho}{2} =\frac{1-\rho}{\sqrt{2\pi}}
\end{align}
\begin{align}\notag
&{E}\left(y_- 1_{x\geq0} + y_+1_{x<0}\right)^2
=2\int_0^\infty\int_{-\infty}^0y^2\phi(x,y;\rho)dx\\\notag
=&\frac{2}{{\sqrt{2\pi}}}\int_0^\infty y^2e^{-\frac{y^2}{2}} \Phi\left({\frac{-\rho y}{\sqrt{1-\rho^2}}}\right)dy\\\notag
=&\frac{2}{{\sqrt{2\pi}}}\left(1_{\rho<0}\sqrt{\frac{\pi}{2}}+\sqrt{\frac{1}{2\pi}}\left(\tan^{-1}\left(\frac{\sqrt{1-\rho^2}}{\rho}\right)+\rho\sqrt{1-\rho^2}\right)\right)\\\notag
=&1_{\rho<0}+\frac{1}{\pi}\left(\tan^{-1}\left(\frac{\sqrt{1-\rho^2}}{\rho}\right)-\rho\sqrt{1-\rho^2}\right)
\end{align}

Similarly, we can prove
\begin{align}\notag
&{E}\left(y_- 1_{x<0} + y_+1_{x\geq0}\right)=\frac{1+\rho}{\sqrt{2\pi}}\\\notag
&{E}\left(y_- 1_{x<0} + y_+1_{x\geq0}\right)^2
=1_{\rho\geq0}-\frac{1}{\pi}\left(\tan^{-1}\left(\frac{\sqrt{1-\rho^2}}{\rho}\right)-\rho\sqrt{1-\rho^2}\right)\hspace{1in}\hfill\Box
\end{align}

\newpage

\section{Proof of Theorem 5}

\ \ \ Firstly, it is easy to see that, as $k\rightarrow\infty$, we have
\begin{align}\notag
&\frac{\frac{1}{k}\sum_{j=1}^k y_{j-}1_{x_j\geq0} + y_{j+}1_{x_j<0}}{\sqrt{\frac{1}{k}k}\sqrt{\frac{1}{k}\sum_{j=1}^k y_j^2}} \rightarrow \frac{1-\rho}{\sqrt{2\pi}}=s, \ \ a.s.
\end{align}
To analyze the asymptotic variance, it suffices to study:
\begin{align}\notag
&{E}\left( \left\{y_{-}1_{x>0} + y_{+}1_{x<0}\right\} - s + s  \frac{1-y^2}{2}\right)^2\\\notag
=& {E}\left(\left\{y_{-}1_{x>0} + y_{+}1_{x<0}\right\}-s(1+y^2)/2\right)^2\\\notag
=&{E}\left\{y_{-}1_{x>0} + y_{+}1_{x<0}\right\}^2+s^2{E}(1+y^4+2y^2)/4 
 - s {E}(\left\{y_{-}1_{x>0} + y_{+}1_{x<0}\right\}(1+y^2))\\\notag
=&1_{\rho<0}+\frac{1}{\pi}\left(\tan^{-1}\left(\frac{\sqrt{1-\rho^2}}{\rho}\right)-\rho\sqrt{1-\rho^2}\right)
+s^2\left(1+3+2\right)/4-s^2
-s\frac{1}{\sqrt{2\pi}}\left(2-3\rho+\rho^3\right)\\\notag
=&\left[1_{\rho<0}+\frac{1}{\pi}\left(\tan^{-1}\left(\frac{\sqrt{1-\rho^2}}{\rho}\right)-\rho\sqrt{1-\rho^2}\right)
-\frac{(1-\rho)^2}{2\pi}\right]
-\frac{(1-\rho)^2}{{4\pi}}\left(1-2\rho-2\rho^2\right)
\end{align}
where we have used the previous results
\begin{align}\notag
&{E}\left(y_- 1_{x\geq0} + y_+1_{x<0}\right)=\frac{1-\rho}{\sqrt{2\pi}}\\\notag
&{E}\left(y_- 1_{x\geq0} + y_+1_{x<0}\right)^2
=1_{\rho<0}+\frac{1}{\pi}\left(\tan^{-1}\left(\frac{\sqrt{1-\rho^2}}{\rho}\right)-\rho\sqrt{1-\rho^2}\right)\\\notag
&{E}(y^2\left\{y_{-}1_{x\geq0} + y_{+}1_{x<0}\right\}) =2{E}\left(y_{+}^31_{x<0}\right)
=\frac{1}{\sqrt{2\pi}}\left(2-3\rho+\rho^3\right)\hspace{1in}\hfill
\end{align}

\section{Proof of Lemma 2}

\begin{align}\notag
&V_1 = \cos^{-1}\rho\left(\pi-\cos^{-1}\rho\right)(1-\rho^2)\\\notag
&V_g = \frac{\pi}{2}-\rho^2\\\notag
&V_{g,n} = V_g - \rho^2\left(3/2-\rho^2\right)\\\notag
&V_s={2\pi}
\left[1_{\rho<0}+\frac{1}{\pi}\left(\tan^{-1}\left(\frac{\sqrt{1-\rho^2}}{\rho}\right)-\rho\sqrt{1-\rho^2}\right)-\frac{(1-\rho)^2}{2\pi}\right]\\\notag
&V_{s,n}=V_s -\frac{(1-\rho)^2}{{4\pi}}\left(1-2\rho-2\rho^2\right)
\end{align}

 Let $t = \cos^{-1}\rho$, i.e., $\rho = \cos t$. When $\rho\rightarrow1$ (i.e., $t\rightarrow0$), we have $\rho = 1-\frac{t^2}{2} + O\left(t^4\right)$, i.e., $t=\cos^{-1}\rho \approx \sqrt{2(1-\rho)}$. When $\rho\rightarrow-1$ (i.e., $t\rightarrow \pi$), we have $\rho = \cos t = -\cos(\pi -t) = -1+\frac{(\pi-t)^2}{2}+O\left(\pi -t\right)^4$, i.e., $t =\cos^{-1}\rho \approx \pi-\sqrt{2(1+\rho)}$. Combining the results, we have
\begin{align}\notag
V_1 =& \cos^{-1}\rho\left(\pi-\cos^{-1}\rho\right)(1-\rho^2) \\\notag
=&2\sqrt{2}\pi\left(1-|\rho|\right)^{3/2} + o\left(\left(1-|\rho|\right)^{3/2}\right), \ \ \text{as } |\rho|  \rightarrow 1
\end{align}

When $\rho=0$, we have
\begin{align}\notag
\frac{1}{V_m}=E\left\{
\frac{1}{(1-\rho^2)^{3}} \frac{\phi^2\left(\frac{\rho}{\sqrt{1-\rho^2}}{sgn}(x_j) y_j\right)
}{\Phi^2\left(\frac{\rho}{\sqrt{1-\rho^2}}{sgn}(x_j) y_j\right)} y_j^2\right\}
=E\left\{\frac{1/(2\pi)}{1/4} y_j^2\right\} = \frac{2}{\pi}
\end{align}
\begin{align}\notag
V_1 = \frac{\pi^2}{4},\hspace{0.2in} V_m= V_g = V_{g,n} = \frac{\pi}{2}, \hspace{0.2in} V_{s} = \pi -1,\hspace{0.2in} V_{s,n} = \pi-\frac{3}{2}
\end{align}
\begin{align}\notag
&\frac{V_m}{V_1}=\frac{V_g}{V_1} =\frac{ V_{g,n}}{V_1} = \frac{2}{\pi}\approx 0.6366,\ \ \
\frac{V_{s}}{V_1} = \frac{4}{\pi} -\frac{4}{\pi^2}\approx 0.8680,\ \ \
\frac{V_{s,n}}{V_1} = \frac{4}{\pi}-\frac{6}{\pi^2} \approx 0.6653
\end{align}

Consider $\rho\rightarrow 1$ and let $\Delta = 1-\rho$. We have  already shown that $V_1 =\pi\sqrt{2\Delta}2\Delta+o\left(\Delta^{3/2}\right)$. Moreover,
\begin{align}\notag
V_s=&2\left(\tan^{-1}\left(\frac{\sqrt{1-\rho^2}}{\rho}\right)-\rho\sqrt{1-\rho^2}\right)-(1-\rho)^2\\\notag
 =& 2\tan^{-1}\left(\sqrt{2\Delta-\Delta^2}\left(1+\Delta+O\left(\Delta^2\right)\right)\right)-2(1-\Delta)\sqrt{2\Delta-\Delta^2}-\Delta^2\\\notag
 =& 2\left(\sqrt{2\Delta-\Delta^2}\left(1+\Delta\right)-2\sqrt{2}\Delta^{3/2}/3\right)-2\sqrt{2\Delta}(1-\Delta)\sqrt{1-\Delta/2}+o\left(\Delta^{3/2}\right)\\\notag
 =&2\sqrt{2\Delta}\left(\left(1+\frac{3}{4}\Delta\right)-\Delta\frac{2}{3}-\left(1-\frac{5}{4}\Delta\right)\right)+o\left(\Delta^{3/2}\right)\\\notag
 =&8\sqrt{2\Delta}\Delta/3 +o\left(\Delta^{3/2}\right)\\\notag
V_{s,n}=&V_s-\frac{\Delta^2}{2}\left(1-2\rho-2\rho^2\right) = 8\sqrt{2\Delta}\Delta/3 +o\left(\Delta^{3/2}\right)
\end{align}
 Thus, $\frac{V_s}{V_1} = \frac{V_{s,n}}{V_1}=\frac{4}{3\pi}\approx 0.4244 $, as $\rho\rightarrow1$.

\newpage\clearpage

{
\bibliographystyle{plain}
\bibliography{../bib/IEEEabrv,../bib/mybibfile}

\begin{thebibliography}{14}
\providecommand{\natexlab}[1]{#1}
\providecommand{\url}[1]{\texttt{#1}}
\expandafter\ifx\csname urlstyle\endcsname\relax
  \providecommand{\doi}[1]{doi: #1}\else
  \providecommand{\doi}{doi: \begingroup \urlstyle{rm}\Url}\fi

\bibitem[Anderson(2003)]{Book:Anderson03}
Theodore~W. Anderson.
\newblock \emph{An Introduction to Multivariate Statistical Analysis}.
\newblock John Wiley \& Sons, Hoboken, New Jersey, third edition, 2003.

\bibitem[Charikar(2002)]{Proc:Charikar}
Moses~S. Charikar.
\newblock Similarity estimation techniques from rounding algorithms.
\newblock In \emph{STOC}, pages 380--388, Montreal, Canada, 2002.

\bibitem[Dasgupta(1999)]{Proc:Dasgupta_FOCS99}
Sanjoy Dasgupta.
\newblock Learning mixtures of gaussians.
\newblock In \emph{FOCS}, pages 634--644, New York, 1999.

\bibitem[Datar et~al.(2004)Datar, Immorlica, Indyk, and
  Mirrokn]{Proc:Datar_SCG04}
Mayur Datar, Nicole Immorlica, Piotr Indyk, and Vahab~S. Mirrokn.
\newblock Locality-sensitive hashing scheme based on $p$-stable distributions.
\newblock In \emph{SCG}, pages 253 -- 262, Brooklyn, NY, 2004.

\bibitem[Goemans and Williamson(1995)]{Article:Goemans_JACM95}
Michel~X. Goemans and David~P. Williamson.
\newblock Improved approximation algorithms for maximum cut and satisfiability
  problems using semidefinite programming.
\newblock \emph{Journal of ACM}, 42\penalty0 (6):\penalty0 1115--1145, 1995.

\bibitem[Gradshteyn and Ryzhik(1994)]{Book:Gradshteyn_94}
Izrail~S. Gradshteyn and Iosif~M. Ryzhik.
\newblock \emph{Table of Integrals, Series, and Products}.
\newblock Academic Press, New York, fifth edition, 1994.

\bibitem[Grimes and O'Brien(2008)]{Proc:Grimes_WWW08}
Carrie Grimes and Sean O'Brien.
\newblock Microscale evolution of web pages.
\newblock In \emph{WWW}, pages 1149--1150, 2008.

\bibitem[Henzinger(2006)]{Proc:Henzinger_SIGIR06}
Monika~Rauch Henzinger.
\newblock Finding near-duplicate web pages: a large-scale evaluation of
  algorithms.
\newblock In \emph{SIGIR}, pages 284--291, 2006.

\bibitem[Johnson and Lindenstrauss(1984)]{Article:JL84}
William~B. Johnson and Joram Lindenstrauss.
\newblock Extensions of \text{Lipschitz} mapping into \text{Hilbert} space.
\newblock \emph{Contemporary Mathematics}, 26:\penalty0 189--206, 1984.

\bibitem[Lehmann and Casella(1998)]{Book:Lehmann_Casella}
Erich~L. Lehmann and George Casella.
\newblock \emph{Theory of Point Estimation}.
\newblock Springer, New York, NY, second edition, 1998.

\bibitem[Li et~al.(2006)Li, Hastie, and Church]{Proc:Li_Hastie_Church_COLT06}
Ping Li, Trevor~J. Hastie, and Kenneth~W. Church.
\newblock Improving random projections using marginal information.
\newblock In \emph{COLT}, pages 635--649, Pittsburgh, PA, 2006.

\bibitem[Manku et~al.(2007)Manku, Jain, and Sarma]{Proc:Manku_WWW07}
Gurmeet~Singh Manku, Arvind Jain, and Anish~Das Sarma.
\newblock {D}etecting {N}ear-{D}uplicates for {W}eb-{C}rawling.
\newblock In \emph{WWW}, Banff, Alberta, Canada, 2007.

\bibitem[Papadimitriou et~al.(1998)Papadimitriou, Raghavan, Tamaki, and
  Vempala]{Proc:Papadimitriou_PODS98}
Christos~H. Papadimitriou, Prabhakar Raghavan, Hisao Tamaki, and Santosh
  Vempala.
\newblock Latent semantic indexing: A probabilistic analysis.
\newblock In \emph{PODS}, pages 159--168, Seattle,WA, 1998.

\bibitem[Vempala(2004)]{Book:Vempala}
Santosh Vempala.
\newblock \emph{The Random Projection Method}.
\newblock American Mathematical Society, Providence, RI, 2004.

\end{thebibliography}
}

\end{document}